\RequirePackage{fix-cm}
\documentclass[smallextended]{svjour3}       %
\smartqed  %
\usepackage{graphicx}

\usepackage{natbib}

\usepackage{aas_macros}

\usepackage[finalnew]{trackchanges}  %

\addeditor{MH}
\addeditor{AS}
\addeditor{PG}
\addeditor{R1}
\addeditor{R2}
\addeditor{Editor}

\usepackage{epsfig}
\usepackage{amsmath}
\usepackage{amssymb}

\usepackage[colorlinks=true,citecolor=blue,urlcolor=blue,breaklinks]{hyperref}

\newcommand{\dd}{\mathrm{d}}
\newcommand{\slope}{q}
\newcommand{\vnabla}{\bm{\nabla}}
\newcommand{\bcdot}{\bm{\cdot}}
\newcommand{\vektor}[1]{\bm{\mathit{#1}}}
\newcommand{\tensor}[1]{\bm{\mathsf{#1}}}
\newcommand{\cren}{e}
\newcommand{\rkl}[1]{\left(#1\right)}
\newcommand{\ekl}[1]{\left[#1\right]}

\newcommand{\skl}[1]{\left\langle#1\right\rangle}

\newcommand{\disfunc}{\mathcal{N}}

\newcommand{\curlyn}{\mathcal{N}}

\def\Gadget2{\rm{\textsc{Gadget\thinspace 2}\ }}

\def\g{{\rm\thinspace g}}

\def\kpc{{\rm\thinspace kpc}}

\def\Msun{\hbox{$\mathrm{\thinspace M_{\odot}}$}}
\def\pc{{\rm\thinspace pc}}

\def\s{{\rm\thinspace s}}

\def\Myr{{\rm\thinspace Myr}}
\def\Gyr{{\rm\thinspace Gyr}}

\newcommand{\h}{\frac{1}{2}}
\newcommand{\imh}{{i-\h}}
\newcommand{\iph}{{i+\h}}
\newcommand{\qi}{{q_i}}

\newcommand{\si}{{s_i}}

\newcommand{\simo}{{s_{i-1}}}
\newcommand{\tha}{\frac{3}{2}} %
\newcommand{\imth}{{i-\tha}}

\newcommand{\ncr}{{n_{\textrm{cr}}}}
\newcommand{\ncrLR}{{n_{\textrm{cr}}^{\mathrm{LR}}}}
\newcommand{\ecr}{{e_{\textrm{cr}}}}
\newcommand{\ecrLR}{{e_{\textrm{cr}}^{\mathrm{LR}}}}
\newcommand{\Pcr}{{P_{\textrm{cr}}}}

\newcommand{\upw}{{\tiny \textrm{u}}}
\newcommand{\pupw}{p_\upw}

\newcommand{\sign}{{\rm sign}}
\newcommand{\indx}[1]{\textrm{\scriptsize #1} }

\newcommand{\q}{\frac{1}{4}}

\newcommand{\xpt}{}
\newcommand{\xp}{}
\newcommand{\xt}{}
\newcommand{\p}{}

\newcommand{\bm}[1]{\mbox{{\boldmath $#1$}}}
\newcommand{\der}[3]{\frac{d^{#3} {#1}}{d {#2}^{#3}}}

\newcommand{\pder}[3]{\frac{{\partial}^{#3} {#1}}{{\partial} {#2}^{#3}}}
\newcommand{\cm}{\mbox{cm}}

\newcommand{\km}{\mbox{km}}

\newcommand{\muG}{\mu{\mbox{G}}}

\newcommand{\K}{\mbox{K}}

\def\asr{Adv.\ Spa.\ Res.}  %
\def\aap{Astron.\ Astrophys.}

\def\apss{Astrophys.\ Sp.\ Sci.}        %

\def\app{Astropart.\ Phys.}
\def\apj{Astrophys.\ J.}          %
\def\apjl{Astrophys.\ J.}  %
\def\apjs{Astrophys.\ J.\ Suppl.} %
\def\araa{Annu.\ Rev.\ Astron.\ Astrophys.}
\def\arnps{Annu.\ Rev.\ Nuclear and Particle Sci.} %

\def\icrc{Int.\ Cosmic Ray Conf.}
\def\jcap{JCAP}%

\def\mnras{MNRAS}                    %

\def\nat{Nature}
\def\pasj{Publ.\ Astron.\ Soc.\ Jpn.}

\def\prl{Phys.\ Rev.\ Lett.}      %
\def\prd{Phys.\ Rev.\ D.}             %
\def\pr{Phys.\ Rev.}              %
\def\rmp{Rev.\ Mod. Phys.}          %

\def\ssr{Space Sci.\ Rev.} 
\def\gray{$\gamma$-ray\ }
\def\grays{$\gamma$-rays\ }

\newcommand{\Dzz}{D_{zz}}

\def\Ekin{E_{kin}}

\def\densityunits{cm$^{-3}$\ MeV$^{-1}$} 
\def\fluxunits   {cm$^{-2}$        sr$^{-1}$ s$^{-1}\ $(MeV/nucleon)$^{-1}$}

\journalname{Living Reviews in Computational Astrophysics}
\begin{document}

\title{Simulations of cosmic ray propagation}

\author{Micha\l \ Hanasz  \and Andrew W. Strong \and Philipp Girichidis
}

\institute{M. Hanasz\at
              Institute of Astronomy, Nicolaus Copernicus University, ul. Grudziadzka 5, PL-87-100 Toru\'n\\
              Phone.: +48-56-611-30-55, 
              \email{mhanasz@umk.pl}           %
           \and
           A. Strong \at
              Max-Planck-Institut f\"ur extraterrestrische Physik, 85748 Garching, Germany\\
\email{aws@mpe.mpg.de}
\and
           P. Girichidis \at
              Leibniz-Institut f\"{u}r Astrophysik (AIP), An der Sternwarte 16, 14482 Potsdam, Germany\\
\email{pgirichidis@aip.de}
             }

\date{Received: 6 July 2020 / Accepted: 20 April 2021}

\maketitle

\begin{abstract}

We review numerical methods for simulations of cosmic ray (CR) propagation on galactic and larger scales. We present the development of algorithms designed for phenomenological and self-consistent models of CR propagation in kinetic description based on numerical solutions of the Fokker-Planck equation. The phenomenological models assume a stationary structure of the galactic interstellar medium and incorporate diffusion of particles in physical and momentum space together with advection, spallation, production of secondaries and various radiation mechanisms. The self-consistent propagation models of CRs include the dynamical coupling of the CR population to the thermal plasma. The CR transport equation is discretized and solved numerically together with the set of magneto-hydrodynamic (MHD) equations in various approaches treating the CR population as a separate relativistic fluid within the two-fluid approach or as a spectrally resolved population of particles evolving in physical and momentum space. The relevant processes incorporated in self-consistent models include advection, diffusion and streaming well as adiabatic compression and several radiative loss mechanisms.
We discuss applications of the numerical models for the interpretation of CR data collected by various instruments. We present example models of astrophysical processes influencing galactic evolution such as galactic winds, the amplification of large-scale magnetic fields and instabilities of the interstellar medium.

\end{abstract}

\noindent

\setcounter{tocdepth}{3}
\tableofcontents

\section{Introduction} \label{sec:intro}

Cosmic rays (CRs) are charged particles with non-thermal energy distributions \citep{StrongMoskalenkoPtuskin2007, GrenierBlackStrong2015, GabiciEtAl2019}. There are both hadronic and leptonic CRs among which protons and electrons are the most abundant particles. In the hadronic component the composition resembles approximately the element distribution in the universe \citep[e.g.][]{BlasiAmato2012a,GaisserStanevTilav2013}, so protons and helium nuclei are by far most abundant and account for most of the energy stored in high energy particles. Heavier elements and in particular unstable isotopes are rare but provide valuable information on the dynamics of CRs and serve as clocks for CR acceleration and transport. CR electrons are usually negligible concerning the dynamics but provide important information on the magnetic field via loss processes that produce radio and synchrotron radiation \citep[e.g.][]{Gaisser1991}. Being high-energy charged particles, CRs mainly interact with the gas via the magnetic field \citep[e.g.][]{Zweibel2013}. The transport processes and the dynamical interaction with the gas are thus tightly coupled to plasma processes.

The integrated energy in hadronic CRs is large enough to have a dynamical impact. In the interstellar medium (ISM) the CR proton energy is comparable to the thermal and kinetic counterpart \citep[e.g.][]{Ferriere2001}. On galactic scales there is evidence that CR protons are an important agent in driving galactic winds \citep[see, e.g.][]{NaabOstriker2017,Zweibel2017}. In addition they might have an impact on the distribution of gas in the galactic disc and thus alter the star formation process on molecular cloud scales, even though their dynamical impact on molecular clouds is expected to be much weaker compared to other drivers in the ISM like radiation or supernovae. In the densest part of cloud cores, into which CR protons are able to penetrate they ionize and heat the gas \citep[e.g.][]{PadovaniEtAl2020}. If CRs are efficiently coupled to the gas or penetrate into regions dense enough such that direct particle-particle collisions become relevent, they provide a temperature floor, which directly impacts the fragmentation scale and the seeds of star formation.

\subsection{Origin of CRs}

Most of the CRs are believed to be produced in shocks via diffusive shock acceleration (DSA). Early theoretical models \citep{Krymskii1977, AxfordLeerSkadron1977, Bell1978, BlandfordOstriker1978} as well as recent numerical simulations \citep[e.g.][]{CaprioliSpitkovsky2014a} confirm this paradigm, see also \citet{MarcowithEtAl2020} for a recent review. In the Milky Way and most of the star forming galaxies supernova remnants (SNR) are by far the most abundant strong shocks that provide the conditions to accelerate CRs up to an energy of approximately $10^{15}\,\mathrm{eV}$, so the particles up to that energy are likely to be of Galactic origin \citep{AckermannEtAl2013}. CRs are however observed up to energies of $10^{20}\,\mathrm{eV}$, which cannot be explained by SNe. Instead, Active Galactic Nuclei (AGN) are an alternative candidate to provide sufficiently energetic environments. In addition, the gyro-radii of those ulta-high energy CRs are larger than the disk of the Milky Way. If coming from a local source within our Galaxy, the individual CRs could therefore be traced back to their origin with an anisotropic distribution on the sky. However, this is not the case and those ultra high energy CRs are thus of extragalactic origin \citep{KoteraSilk2016}.

The spectral distribution of CRs shows a peak or a flattening at a few GeV followed by a power law with remarkably little scatter. Up to energies of $10^{15}\,\mathrm{eV}$ a spectrum with $E^{-2.7}$ is observed. Above that energy the scaling steepens, which leads to naming the kink at $10^{15}\,\mathrm{eV}$ the "knee". At $10^{20}\,\mathrm{eV}$ the spectrum flattens again, which is called the "ankle". At the highest energies the spectrum is less clear, mainly due to the small statistics.

Due to the steep spectrum, most of the energy can be located in a narrow energy range of from 1 to 10 GeV per nucleon. Models accounting for the dynamical impact of CRs therefore focus on GeV protons.

\subsection{Numerical approaches of CR transport and scope of this review}

CRs span a large energetic range from particle energies just exceeding the thermal distribution ($\sim$ MeV) up to the GZK cutoff \citep[e.g.][]{KoteraOlinto2011, AmatoBlasi2018, GabiciEtAl2019}. Over this range in momentum the dynamical range of the particle distribution function spans approximately 10 orders of magnitude. Consequently, the effective (numerical) models that describe CRs and the interactions with their environment differ. First, we need to distinguish between a \emph{macroscopic} and a \emph{microscopic} perspective, where we start with the macroscopic models.

For low-energy CRs with small gyro-radii we are often interested in the overall energy of CRs rather than the individual particle trajectories \citep[e.g.][]{PadovaniEtAl2020}. On the other hand, for ultra-high-energy CRs with gyro-radii comparable to entire galactic discs the particles' trajectory is of greater interest \citep{KoteraOlinto2011}. Equally important to the energy per particle is the fraction of CR energy compared to the other energies like the magnetic, thermal, radiation or kinetic one for the coupling between the CRs and their environment. If the integrated CR energy is much smaller that the other components, CRs can be treated as tracers without dynamical back-reaction. Depending on the system under consideration the effective transport of the CR tracers can differ from advection with the gas, to diffusion and to free streaming.

This is basically always the case for CR electrons, so CRs can be treated as tracer particles or passive fluid. For the hadronic component we can identify three main regimes, which we outline in more detail in Section~\ref{sec:basics-approaches}. The low-energy CRs do not significantly contribute to the dynamics via their total pressure, so again a passive tracer fluid or tracer particles are appropriate. The regime around a GeV contains comparable energy densities to other components in galaxies, so the back-reaction onto the dynamics needs to be included. There are several models of different complexity, which we outline in this review, ranging from simply adding another pressure to the hydrodynamical equations up to a more complex way of including plasma processes between CR particles and the magnetic waves. The very steep spectrum results again in an overall negligible integrated energy above $\sim\mathrm{100\,GeV}$. The passive treatment is then again appropriate. However, the large gyro-radii combined with the low number of particles favours the treatment as passive individual particles and their trajectories.

In the microscopic CR models the interaction of \emph{individual} particles with the magnetic field is of interest in order to understand plasma processes. This approach is a different numerical class, known as particle-in-cell methods, where the Lorentz force for each particle is computed and the particle distribution function is actually sampled by a large number of CRs. The complexity of this plasma approach exceeds the scope of this review and we would like to highlight the review by \citet{MarcowithEtAl2020}.

The field of CR physics is large ranging from their acceleration at small spatial scales of individual shock fronts up to the trajectory of the highest energy CRs and their specific origin from extra-galactic sources. Depending on the individual setups and environmental conditions, different models are chosen and combined. Among the three main types of models (particle-in-cell, test particle models and an effective fluid description) we mainly cover the last one in different applications. %

In this review we distinguish between the following numerical models:
\begin{itemize}
\item We consider CR propagation described by the canonical Fokker-Planck equation in a static magnetic field, referred to as 'phenomenological models', which are discussed in Sections~\ref{sect:galprop} and \ref{sect:other_codes}.
\item In the second part we discuss models considering joint solutions of the Fokker-Planck equation for CRs combined with the equations of MHD. These models are referred to as 'selfconsistent models' \citep[][]{StrongMoskalenkoPtuskin2007} and are discussed in Sections~\ref{sect:selfc_eqns}-\ref{sect:diff_models}. They can be further distinguished in two ways.
\begin{itemize}
\item The first one covers a dynamical coupling of the CRs to the MHD system by simply including an additional pressure term related to cosmic rays. The interactions between CRs and magnetic fields are included as subgrid models with effective coupling coefficients, i.e. without including the streaming process explicitly in the system. This class of models however, has recently been extended to include spectrally resolved CRs.
\item The latest approach also considers CR-generated Alfv\'en waves together with CR scattering off self-excited waves (streaming) which additionally couple CRs to the energy equation of the thermal plasma through dissipation of Alfv\'en waves (cosmic ray heating). 
\end{itemize}
\end{itemize}

\subsection{Basics and Approaches} 
\label{sec:basics-approaches}

We begin with some basic concepts of CR  propagation, and   relating these to observational data. 
Most of our  knowledge of CR propagation  comes {\it via} secondary CRs, with additional information from \grays and synchrotron radiation.
It is instructive to point out why secondary nuclei  are an ideal probe of CR propagation:
the fact that the primary nuclei are measured (albeit only near the solar position) means that the
secondary production functions can be computed from primary spectra, cross-sections and interstellar gas densities;
the secondaries can then be `propagated' and compared with observations.

Since it became known  that CRs fill the Galaxy it has been clear that nuclear interactions imply that 
their composition contains information on their propagation
\citep{1950PhRv...80..943B}. %
 An important breakthrough was the advent of satellite measurements of isotopic Li, Be, B in the 1970's
\citep{1975ApJ...201L.145G}. %
Since then the subject has expanded continuously  with models of increasing degrees of sophistication.
The observation that the  composition of CRs is different from that of solar,
in that rare solar-system nuclei like Boron are abundant in CR, proves the importance of propagation in the interstellar medium. Simplified estimates of the lifetime time of CRs in the interstellar medium conclude that there is a canonical column density of a few `few $\g \ \cm^{-2}$' of traversed material before the CRs interact with the gas and lose their energy (see section~\ref{sec:cr-clocks-motivation}).

\subsubsection{Particle vs. kinetic vs. fluid approach}

\begin{figure}
    \centering
    \includegraphics[width=\textwidth]{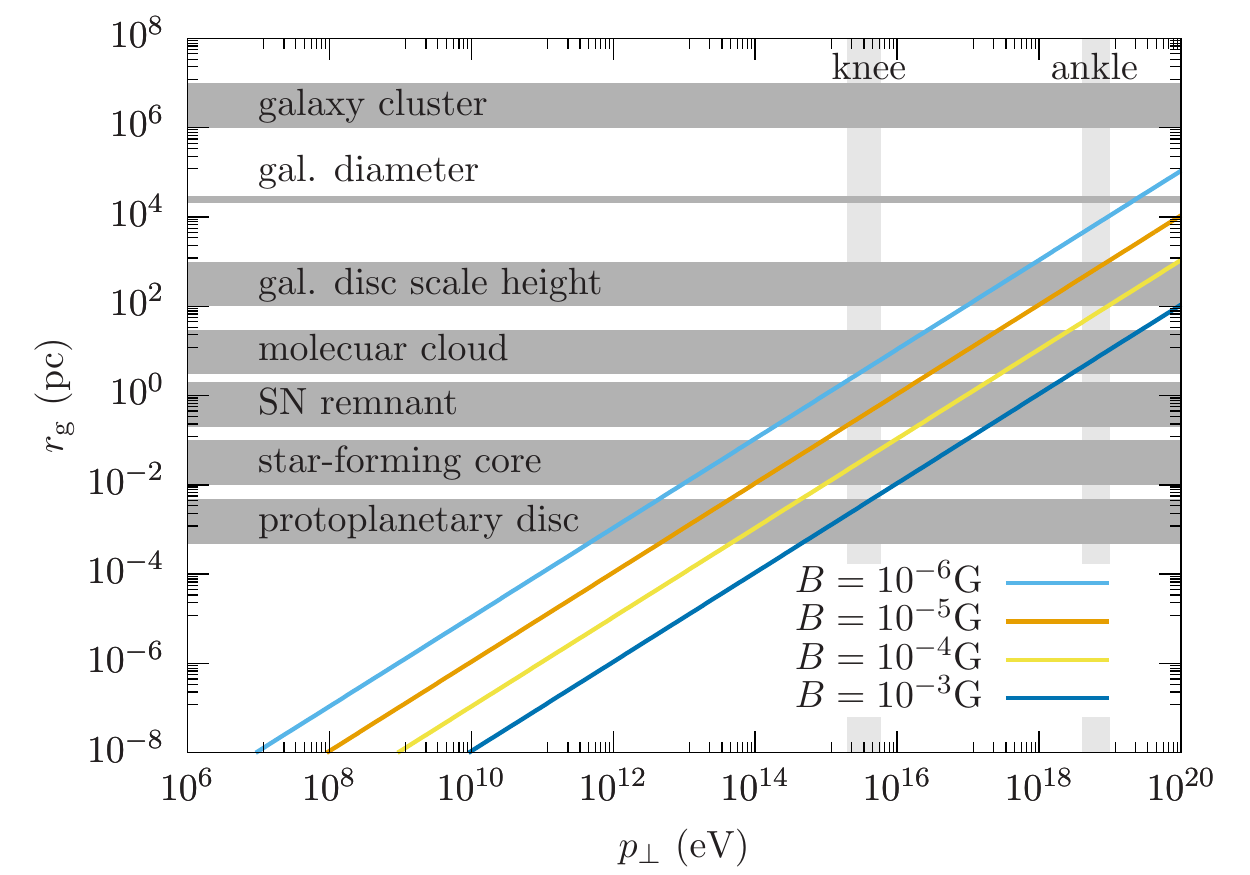}
    \caption{Gyro-radius of CRs as a function of particle momentum. For gyro-radii much smaller than the system under consideration a fluid approximation seems reasonable.}
    \label{fig:gyro-radius}
\end{figure}

One main distinction of CR transport is the difference between particle and fluid. If the mean free path of CRs is small compared to the characteristic simulated scales, i.e. CRs scatter efficiently, they are often treated as a fluid. Contrary, if the mean free path is comparable or larger than the system under consideration the trajectory of individual CRs need to be considered. Which approach is more appropriate depends on the spatial scales, the scattering frequency, and the energy of the CRs. To get an order of magnitude estimate of the typical scales we can investigate the gyro-radius
\begin{equation}
  r_\mathrm{g} = \frac{p_{\perp}} {|q|B},
\end{equation}
where $p_\perp$ is the momentum perpendicular to the magnetic field line, $|q|$ is the absolute value of the charge and $B$ is the magnetic field strength. Fig.~\ref{fig:gyro-radius} shows the gyro-radius of CR protons for different magnetic field strengths as a function of CR momentum together with typical sizes of astrophysical systems. The plot illustrates that for low-energy CRs the gyro-radii are perceptibly smaller than typical astrophysical objects. We would like to highlight that the gyro radius is not a definite measure for how the CRs need to be treated, but it nicely illustrates the separation of scales.

The confinement of CRs is more complicated and depends also on the total energy in CRs compared to the energy of the background system, i.e. whether CRs can be treated as tracer particles moving through the medium or whether CR themselves modify the magnetic field and gas properties. %
We can broadly separate three different regimes.    

\paragraph{CR momenta $\ll$GeV/c}
The spectrum at low CR energies is not well constrained due to solar modulation \cite[e.g.][]{Webber1998,Potgieter2013}. The best data of the CR spectrum is based on Voyager \cite[e.g.][]{CummingsEtAl2016} and AMS-02 \citep[e.g.][]{AguilarEtAl2014a,AguilarEtAl2014b} measurements, which is still in our very local astrophysical neighbourhood. At low energies the cross section of CRs with gas atoms and molecules increases, which leads to efficient losses (Coulomb losses, ionisation losses). On typical scales above a few parsec the gas scatters efficiently and the integrated energy of CRs is low enough such that those CRs are often treated as a diffusing fluid. On scales of star forming regions and protoplanetary discs the individual trajectories might be of interest.

\paragraph{CR momenta between 0.1 and $\sim$10 GeV/c}
This regime of the CR spectrum is the most difficult one because the total integrated CR energy is dominated by the contribution of the GeV particles. With energy densities exceeding the magnetic one, CRs do not just scatter off a background system but heavily influence the magnetic field and -- connected via MHD -- the local turbulent and thermal properties. Typically the scattering is very efficient, which allows a fluid approach. However, the simple diffusion approximation is likely to be violated. The resulting transfer of energy and the CR losses are complicated interplay between Coulomb and ionisation losses, hadronic losses and the interactions between CRs and the magnetic field.

\paragraph{CR momenta above $\sim$10 GeV/c}
At higher momenta each CR is much more energetic than other particles in the universe. However, the spectrum is so steep that the total energy stored in those CRs is subdominant, so they are dynamically irrelevant. The low scattering efficiency makes those CRs a reliable probe of their travel through the magnetised universe. Therefore, in this energy regime CRs are often treated by following individual particles along their trajectories.

\subsubsection{Grey vs. spectrally resolved CR fluids}

In order to account for the global effects of CRs in numerical models one often uses a grey approach, i.e. CR properties integrated over the full momentum range of the distribution function. In this simplified approach the CRs are treated as a fluid with global effective transport properties and interaction efficiencies. As in thermal gas dynamics an underlying spectrum needs to be assumed. Spectrally resolved methods allow for a more accurate treatment of the cooling.

\subsection{The canonical CR Propagation equation} 

\newcommand{\Dpp}{D_{pp}}
\newcommand{\Dxx}{D_{xx}}
\newcommand{\ddp}{{\partial\over\partial p}}

The canonical CR propagation equation can be written as \citep{1971ApJ...170..265S,1975MNRAS.172..557S, 1975MNRAS.173..245S,1975MNRAS.173..255S}

\begin{eqnarray}
\pder{f}{t}{} +\bm{u} \cdot \nabla f   & = & \frac{1}{3} (\nabla \cdot \bm{u}) p \pder{f}{p}{} + \nabla ( \tensor{D} \nabla f) \nonumber \\
                                                       && + \frac{1}{p^2} \pder{}{p}{} \left[  p^2 b_l\xp f +D_{pp} \pder{f}{p}{}    \right]  
                                                       + q(\vektor r, p, t), 
    \label{eq:cr_transp}
\end{eqnarray}

\noindent
where $f=f(\vektor r,p,t)$ is the CR density per unit momentum $p$ at position $\vektor r$ and at time $t$, $f(p)\,dp = 4\pi p^2 f(\vektor p)\,dp $ in terms of phase-space density $f(\vektor p)$, $b_l$  represents mechanical and adiabatic losses and $q(\vektor r, p, t)$ denotes the source terms.
The effective CR advection velocity is
\begin{equation}
\vektor{u}=\vektor{v}_{0} + \vektor{v}_s,
\label{eq:cr_advection}
\end{equation}
where  $v_0$ is the advection velocity of thermal plasma and
\begin{equation}
\vektor{v}_s =  - \vektor{b} v_{\mathrm{A}} \sign(\vektor{b}\cdot \nabla \ecr)
\label{eq:stream_speed}
\end{equation}
accounts for scattering of CR particles off self-excited Alfv\'en waves with $\vektor{b}$ denoting the unit vector aligned with magnetic field, and $v_A$ is the Alfv\'en speed.

The  waves are excited due to a resonant coupling with the streaming CR population. The condition for resonance is that the Doppler-shifted wave frequency $\omega - k v$ is an integral multiple  of the cyclotron frequency $n \Omega$ \citep{1969apj...156..445k}. For $n=0$  the gyroresonance occurs for waves at the same phase velocity as the particle velocity. 
If the drift velocity of CRs exceeds the Alfv\'en speed, the instability occurs for the waves propagating in the same direction.
Cosmic ray interacting with the wave traveling in the same direction are mainly scattered in pitch angle and give a small amount of energy to the wave. The effect is known as the streaming instability. Although the instability generates directly forward waves propagating along magnetic field in the same direction as cosmic rays, wave-wave interactions imply the existence of backward waves, propagating in the opposite direction \citep{1972Ap&SS..16..465C,1975MNRAS.172..557S}.

 The spatial diffusion coefficient is generally anisotropic and thus described by a diffusion tensor 
 \begin{equation}
 \tensor{D}=D_{xx} \vektor{b}\vektor{b},
 \end{equation}
with '$\vektor{b}\vektor{b}$' meaning the dyadic product of vector $\vektor{b}$.
In one dimension or for isotropic diffusion this reduces to a diffusion coefficient, $D_{xx}$, which is given by
\begin{equation}
D_{xx}=v^{2}\left\langle\frac{1-\mu^{2}}{2\left(\nu_{+}+\nu_{-}\right)}\right\rangle,
\label{eq:d_xx}
\end{equation}
where  $\nu_{\pm}$ means the collision frequency against forward (+) and backward (-) waves, $v$ is particle velocity 
and $\mu = \vektor{p} \cdot \vektor{b}/ p$ denotes the particle pitch-angle cosine.

Diffusive reacceleration is described as diffusion in momentum space determined by the coefficient 
\begin{equation}
D_{pp}=4 \gamma^{2} m^{2} v_{\mathrm{A}}^{2}\left\langle\frac{1-\mu^{2}}{2} \frac{\nu_{+} \nu_{-}}{\nu_{+}+\nu_{-}}\right\rangle,
\label{eq:d_pp}
\end{equation}
where $m$ is the particle mass and $\gamma$ is its Lorentz factor.
 
CR sources are usually assumed
to be concentrated near the Galactic disk and to have a radial distribution like for example supernova remnants (SNR).
A source injection spectrum and its isotopic composition are required; the composition is usually initially based on primordial solar
but can be determined iteratively from the CR data themselves for later comparison with solar.
The spallation part of $q(\vektor r, p,t ) $ depends on all progenitor species and their energy-dependent cross-sections,
 and the gas density $\rho(\vektor r)$; it is generally assumed that the spallation products have the same kinetic energy per nucleon as the progenitor.
K-electron capture  and electron stripping can be included  via the time scale for loss by fragmentation $\tau_f$ and $q$.
 $\Dxx$   is in general a function of  $(\vektor r, \beta, p/Z)$ where $\beta=v/c$ and $Z$ is the charge, and $p/Z$ determines the gyroradius in a given magnetic field;
 $\Dxx$ may be  isotropic, or more realistically anisotropic,
and may be influenced by the CR themselves (e.g. in  wave-damping  models).
The coefficient $\Dpp$\ is related to  $\Dxx$ by  $\Dpp\Dxx\propto p^2$, with the proportionality constant depending
 on the  theory of stochastic reacceleration
\citep{1990acr..book.....B,1994ApJ...431..705S} %
 as described in Section 2.5.
 $\vektor v$ is a function of   $\vektor r$ and $t$.
The term in $\vnabla \cdot \vektor v$ represents  adiabatic momentum  gain or loss in the
 non-uniform flow of gas with a frozen-in magnetic field whose inhomogeneities scatter the CR.
$\tau_f$ depends on the total spallation cross-section and    $\rho(\vektor r)$.
Observationally, the density $\rho(\vektor r)$ can be based on surveys of atomic and molecular gas, but can also incorporate 
small-scale variations such as the region of low gas density  surrounding the Sun. In hydrodynamical simulations, the density is naturally included in every computational cell.
This equation only treats continuous momentum-loss; catastrophic losses can be included via  $\tau_f$ and $q$.
CR electrons, positrons and antiprotons propagation constitute just special cases of this equation, differing only 
in their energy losses and production rates.

The {\it boundary conditions} depend on the model; often $f=0$ is assumed at the `halo boundary' 
 where particles escape into intergalactic space, but
this obviously just an approximation (since the intergalactic flux is not zero) which can be relaxed for  models
with a physical treatment of the boundary.

Equation~(\ref{eq:cr_transp}) is a time-dependent equation; often the steady-state solution is required, which can be
obtained either by setting $\partial f / \partial t = 0$ or following the time dependence
until a steady state is reached; the latter procedure is much easier to implement numerically.
Depending on the model setup the time-dependence of $q$ can be neglected unless effects of nearby recent sources  or the stochastic nature of sources are being studied. In hydrodynamical models, local sources are easily incorporated in regions of star formation or locally identitfied shocks.
By starting with the solution for the heaviest primaries and using this source term to compute the spallation source
for their products, the complete system can be solved including secondaries, tertiaries etc.
Then the CR spectra at the solar position can be compared with direct observations, including solar modulation if required.

Source abundances are determined iteratively, comparing propagation calculations with data;
for nuclei with very small source abundances, the source values are masked by secondaries and cross-section uncertainties  and are therefore hard to determine.
\citet{1998SSRv...86..239W} %
 gives a ranking from `easy' to `impossible' for the possibility of getting the source abundances using  Advanced Composition Explorer (ACE) data. 
A review of the high-precision abundances from ACE  is in
\citet{2001SSRv...99...15W} %
and for Ulysses in
\citet{2001SSRv...99...41C}. %
For a useful summary of the various astrophysical abundances relevant to interpreting CR abundances see 
 \citet{2005ApJ...634..351B}. %

\subsubsection{Streaming and diffusion} \label{sect:streaming}

 Equation~(\ref{eq:cr_transp}) contains the term representing advection of CR population at the velocity $\bm{u}=\bm{v}_0+ \bm{v}_s$, given by formula~(\ref{eq:cr_advection}). This includes the streaming velocity along the direction pointing down the CR energy density gradient. This term originates from the consideration of the cyclotron resonance streaming instability, discovered by \citep{1969apj...156..445k}, which sets in when a population of CRs move with a bulk speed greater than the Alfv\'en speed $v_{A}$.  Even a slight anisotropy of CRs, which naturally arises in the presence of sources, causes  unstable growth of the waves due to momentum transfer from the CRs to waves via pitch-angle scattering. 
 The  streaming  CRs  transfer their energy to Alfv\'en waves and subsequently scatter off self-generated waves \citep[see e.g.][for a detailed introduction]{2013MNRAS.434.2209W}.  Streaming and diffusion are considered respectively as the first and second order effects in expansion of the distribution function in powers of inverse wave particle scattering frequency $\nu$  \citep[e.g.][]{1971ApJ...170..265S,WienerPfrommerOh2017}.
 
The waves are subject to damping  due to ion-neutral  friction,  due to wave-particle interactions between thermal ions and the low-frequency beat waves formed by two interfering  Alfv\'en waves (non-linear Landau damping) and due to the cascade of waves to smaller scales (turbulent damping)  \citep[see][for a brief summary on this topic]{2019MNRAS.485.2977T}. The system reaches  equilibrium when the wave damping rate becomes equal to their growth rate due to the streaming instability.  The scattering leads to isotropisation of the CR distribution in the reference frame co-moving with the waves, implying that the CRs bulk velocity becomes equal to the Alfv\'en velocity plus a local fluid velocity.  

Scattering of CR particles off Alfv\'en waves manifests itself as diffusion process in phase space, with spatial and momentum diffusion coefficients given  respectively by formulae  (\ref{eq:d_xx}) and (\ref{eq:d_pp}). Diffusion of CRs  is intrinsically related to the streaming process which depends on local plasma conditions.  In a more general case diffusion might be related also to an extrinsic turbulence
driven by sources other than CRs. The effective diffusion coefficient depends on the dominant wave damping mechanism. %

Wave damping means dissipation of wave energy and heating of the interstellar medium.
The heating rate due to dissipation of Alfv\'en waves is 
\begin{equation}
\Gamma_{\textrm{ wave}} = - \bm{v}_s \cdot \nabla P_{cr}. 
\label{eq:cr_heating_rate}
\end{equation}
If the coupling of CRs is reduced, due to wave damping mechanisms, the effective propagation speed of the CR population might be higher than the Alfv\'en speed, however, it is argued  that the heating rate is still given by the expression~(\ref{eq:cr_heating_rate}), because all momentum and energy transfer between the CRs and gas is mediated by hydromagnetic waves which propagate at the speed $\bm{v}_A$ \citep[e.g.][]{Zweibel2013}.

The concept of CR diffusion explains why energetic charged particles
have highly isotropic distributions and why they are retained well in the
Galaxy. The Galactic magnetic field which tangles the trajectories of
particles plays a crucial role in this process.  Typical values of the
diffusion coefficient found from  fitting to CR  data is $\Dxx\sim(3 - 5)\times10^{28}$ cm$^{2}$ s$^{-1}$ at an energy 
of $\sim$1~GeV per nucleon and it increases with magnetic rigidity as $R^{0.3}-R^{0.6}$ in different
versions of the empirical diffusion model of CR propagation.  Here, the magnetic rigidity is defined as $R=pc/Ze$ with the momentum $p$ and the charge $Ze$. In a given magnetic field configuration, particles with the same rigidity follow the same trajectory.

On the microscopic level the diffusion of CR  results from 
particle scattering off random MHD waves and discontinuities. The effective
``collision integral'' for charged energetic particles moving in a magnetic
field with small random fluctuations $\delta B\ll B$  is given by  the
standard quasi-linear theory of plasma turbulence \citep{Kennel66}.
The wave-particle interaction is of resonant character so that an energetic
particle is predominantly scattered by those irregularities of magnetic field
which have their projection of the wave vector on the average magnetic field
direction equal to $k_{\parallel}=\pm s/\left(  r_{\mathrm{g}}\mu\right)  $,
where $\mu$ is the particle pitch angle. The integers $s=0,1,2...$ correspond
to  cyclotron resonances of different order. The efficiency of scattering
depends on the polarization of the waves and on their distribution in
$\vektor{k}$-space. The first-order resonance $s=1$\ is the most important for
the isotropic and also for the one-dimensional distribution of random MHD
waves along the average magnetic field. In some cases -- for the calculation of
scattering at small $\mu$ and for the calculation of perpendicular diffusion --
the broadening of resonances and magnetic mirroring effects should be taken
into account. The resulting spatial diffusion is strongly anisotropic locally
and goes predominantly along the magnetic field lines. However, strong
fluctuations of the magnetic field on large scales of $L\sim100$ pc, where the
strength of the random field is several times higher than the average field
strength, lead to the isotropization of global CR diffusion in the
Galaxy. The rigorous treatment of this effect is not trivial, since the
field is almost static and the strictly one-dimensional diffusion along the
magnetic field lines does not lead to non-zero diffusion perpendicular to
$\vektor{B}$, see \cite{2002PhRvD..65b3002C}%
 and the references therein.

Following several detailed reviews of the theory of CR  diffusion
\citep{1971RvGSP...9...27J,Toptygin85,1990acr..book.....B,2002cra..book.....S} %
the diffusion coefficient  at $r_{\mathrm{g}}<L$\ can be roughly
estimated as $\Dxx\approx\left(  \delta B_{\mathrm{res}}/B\right)  ^{-2}%
vr_{\mathrm{g}}/3$, where $\delta B_{\mathrm{res}}$ is the amplitude of the random
field at the resonant wave number $k_{\mathrm{res}}=1/r_{g}$. The spectral
energy density of interstellar turbulence has a power law form
$w(k)dk\sim k^{-2+a}dk$, $a=1/3$ over a wide range of wave numbers
$1/(10^{20}$ \textrm{cm}$)<k<1/(10^{8}$ \textrm{cm}$)$, see
\citet{2004ARA&A..42..211E},
and the strength of the random field at the main scale is $\delta B\approx5$ $\mu
$G. This gives an estimate of the diffusion coefficient $\Dxx\approx$
$2\times10^{27}\beta R_{_{\mathrm{GV}}}^{1/3}$ cm$^{2}$ s$^{-1}$ for all CR
particles with magnetic rigidities $R<10^{8}$ GV, in a fair agreement with the
empirical diffusion model (the version with distributed reacceleration). The
scaling law $\Dxx\sim R^{1/3}$ is determined by the value of the exponent
$a=1/3$, typical for a Kolmogorov spectrum. Theoretically
\citep{1995ApJ...438..763G} %
 the
Kolmogorov type spectrum might refer only to some part of the MHD turbulence
which  includes the (Alfv\'{e}nic) structures strongly elongated along the
 magnetic-field direction and which are not able to provide the significant scattering
and required diffusion of cosmic rays. In parallel, the more isotropic (fast
magnetosonic) part of the turbulence, with a smaller value of random field at
the main scale and with the exponent $a=1/2$ typical for the Kraichnan type turbulence
spectrum, may exist in the interstellar medium 
\citep{2004ApJ...614..757Y}. %
 The Kraichnan spectrum
gives a scaling $\Dxx\sim R^{1/2}$ which is close to the high-energy asymptotic form
of the diffusion coefficient obtained in the `plain diffusion' version of
the empirical propagation model. Thus the approach based on  kinetic theory
gives a proper estimate of the diffusion coefficient and predicts a power-law
dependence of diffusion on magnetic rigidity, but the  determination of the actual diffusion coefficient has to make use of fitting to
models of CR  propagation in the Galaxy.

\subsubsection{Convection/Advection}

The transport of CRs is a combination of the locally advective contribution, where CRs are advected with the flow, plus the diffusive and/or streaming transport relative to the gas. Depending on the system under consideration convection and advection habe been distinguished. From a hydrodynamical perspective the CRs are advected with the flow. Historically, global models of galaxies do not resolve the gas flow but cover the turbulent motions in the galaxy and the perturbations globally as a convective system. In the literature we therefore find both terminologies.

While the most frequently considered mode of CR transport is diffusion, 
the existence of galactic  winds in many galaxies suggests that advective transport
should also be important.
Winds are common in galaxies and can be CR-driven \citep[e.g.][]{NaabOstriker2017}%
 CRs also play a dynamical role in galactic halos
\citep{1991A&A...245...79B,1993A&A...269...54B}. %
Convection not only transports CR, it can also produce adiabatic energy losses as the wind speed increases away from the disk.
Convection was first considered by 
\cite{1976ApJ...208..900J} %
and followed up by
\cite{1977ApJ...215..677O,1977ApJ...215..685O,1978ApJ...222.1097J,1979ApJ...229..747J,1993A&A...267..372B}. %
Both  1-zone and 2-zone models have been studied: 
a 1-zone model has convection and diffusion everywhere, a 2-zone model  has diffusion alone up to some distance from the plane, and diffusion plus convection beyond.

For {\it one-zone} diffusion/convection models
a good diagnostic  is the energy-dependence of the secondary-to-primary ratio: 
a purely convective transport would have no energy dependence (apart from the velocity-dependence of the reaction rate), contrary to what is observed.
If the diffusion rate decreases with decreasing energy,  any convection will eventually take over
and cause the secondary-to-primary ratio to flatten at low energy: this is observed but 
convection  does not reproduce e.g. the 
 Boron-to-Carbon ratio  (B/C) very well \citep{1998ApJ...509..212S}. %

\citet{1997A&A...321..434P} %
studied a self-consistent  {\it two-zone} model with a wind driven by CR and thermal gas in a rotating Galaxy.
The CR propagation is entirely diffusive in a zone $|z|< 1$ kpc, and diffusive-convective outside.
CR reaching the convective zone do not return, so it acts as a halo boundary with  height varying with energy and Galactocentric radius.
It is possible to explain the energy-dependence of the secondary-to-primary ratio with this model,
and it is also claimed to be consistent with radioactive isotopes.
The effect of a Galactic wind on the radial CR gradient has been investigated 
\citep{2002A&A...385..216B}; %
they constructed a self-consistent model with the wind driven by CR, and with anisotropic diffusion.
The convective velocities involved in the outer zone are large ($\sim$100 km s$^{-1}$) but this model is still
consistent with radioactive CR nuclei which set a much lower limit
\citep{1998ApJ...509..212S}, %
since this limit is only applicable in the inner zone. 
Observational support of such models would require direct evidence for a Galactic wind in the halo.

\subsubsection{Reacceleration}

In addition to spatial diffusion, the scattering of CR particles
on randomly moving MHD waves leads to stochastic
acceleration which is described in the transport equation as diffusion in
momentum space with some diffusion coefficient $D_{pp}$. One can estimate it
as $D_{pp}=p^{2}V_{\mathrm{a}}^{2}/\left(  9\Dxx\right)  $ where the Alfv\'en
velocity $V_{\mathrm{a}}$ is introduced as a characteristic velocity of
weak disturbances propagating in a magnetic field, see
\citep{1990acr..book.....B,2002cra..book.....S}
for rigorous formulas.

\section{Early models}
\subsection{CR propagation  - basic relations and motivations}
\label{sec:cr-clocks-motivation}
Assuming that CRs are accelerated from the ISM of our Galaxy, we expect a similar composition of CRs and the thermal gas in the ISM. However, their observed abundances show a relative over-abundance of the light elements Li, Be, and B, which must be produced during their transport through the ISM. These elements are produced in spallation processes and are thus secondary particles. A classical measure of the ratio of stable secondaries to primaries is the B/C ratio, which depends on particle energy, but can be approximated to zeroth order to be $\sim0.3$. The grammage is the amount of material that the CRs have to pass through before they interact with an atom in the ISM,
\begin{equation}
\chi = \int\rho(l) dl.
\end{equation}
The grammage can be related to the spallation products by comparing the travelled depth with  the mean mass per particle in the ISM $m_\mathrm{ISM}$ and the typical cross section for the spallation of carbon to boron,
\begin{equation}
\frac{\mathrm{B}}{\mathrm{C}}\sim\frac{\chi}{m_\mathrm{ISM}/\sigma_{\mathrm{C}\rightarrow\mathrm{B}}} \sim 0.3.
\end{equation}

Using typical numbers for the ISM we find a grammage of $\sim 10\,\mathrm{g\,cm}^{-2}$. At a disc surface density of approximately $\Sigma_\mathrm{gas}\sim10\,\mathrm{M}_\odot\,\mathrm{pc}^{-2} = 2\times10^{-3}\,\mathrm{g\,cm}^{-2}$ the CR must cross the disc $10^3$ times, assuming that the grammage is accumulated by travelling through the disc. The associated travel or residence time, which can also be understood as the lifetime of a CR in the Galaxy  or its escape time from the disc, is given by
\begin{equation}
t_\mathrm{res}=\frac{\chi}{\Sigma_\mathrm{gas}}\,\frac{h}{v} \sim 10^6\,\mathrm{yr}
\end{equation}
for a gas scale height of $h\sim100\,\mathrm{pc}$ and a CR velocity $v=c$.

The linear distance that a CR travels during the residence time is $l=t_\mathrm{res}c\sim500\,\mathrm{kpc}$, so much larger than galactic scales and the CRs need to be confined to the Galaxy. In order to estimate escape timescales of the CRs unstable secondaries provide a valuable tool. The longest lived and best measured unstable isotope is $^{10}$Be. Its decay time is $t_\mathrm{decay}=1.39\,\mathrm{Myr}$, which is of the order of the residence time. The cross sections for the spallation of carbon into stable ($^{9}$Be) and unstable ($^{10}$Be) beryllium are similar, $\sigma_{\mathrm{C}\rightarrow\mathrm{^{10}Be}}\sim\sigma_{\mathrm{C}\rightarrow\mathrm{^9Be}}$. This means that an initial abundance ratio ($^{10}$Be/$^{9}$Be) of unity at production decreases over time by $t_\mathrm{decay}(^{10}\mathrm{Be})/t_\mathrm{res}$. Measurements of this ratio reveal a residence time of $10-20\,\mathrm{Myr}$, so larger than the residence time of CRs.

A simple transport equation along the vertical dimension reads
\begin{equation}
\frac{\partial N}{\partial t} = Q_0(p)\delta(z) + \frac{\partial}{\partial z}\left[D \frac{\partial N}{\partial z}\right].
\end{equation}
Here, $N$ is the number density of CRs, $Q_0$ is the CR source function, $D$ is the diffusion coefficient, and $\delta$ is the Dirac-$\delta$ distribution. By assuming steady state and an injection of CRs close to the midplane at $z=0$, we can reduce the equation to
\begin{equation}
Q_0(p)\delta(z) = D\frac{\partial^2 N}{\partial z^2}
\end{equation}
and find for $z>0$
\begin{equation}
\frac{\partial N}{\partial z} = \mathrm{const.} \Leftrightarrow N(z) = N_0\left(1-\frac{z}{H}\right),
\end{equation}
where $H$ is the size of the halo, which is poorly constrained to a value of approximately $5\,\mathrm{kpc}$. The total grammage $\chi=\overline{\rho}t_\mathrm{res} c$ will be reached by moving though the average density of the total volume (disc plus halo), $\overline{\rho}=\mu m_\mathrm{p}n_\mathrm{ISM} h/H$, with a disc scale height of $h=100\,\mathrm{pc}$, an average mean molecular weight of $\mu=1.4$, and an average ISM density of $1\,\mathrm{cm}^{-3}$. The diffusion coefficient can be estimated to be
\begin{align}
D &= \frac{H^2}{t_\mathrm{res}}\\
&= \frac{H\mu m_\mathrm{p}n_\mathrm{ISM} h c}{\chi}\\
&\sim 3\times10^{28} \left(\frac{H}{5\,\mathrm{kpc}}\right)\left(\frac{\chi}{10\,\mathrm{g\,cm}^{-2}}\right)^{-1}\mathrm{cm}^2\,\mathrm{s}^{-1}
\end{align}

Even though these estimates are very simplified, they provide two valuable features of GeV CRs that remain valid even with much more complex assumptions. The first is that the CRs are distributed relatively smoothly through the ISM. Locally, the CR density varies, but much less than the gas structures, so molecular clouds are located in an almost uniform sea of GeV CRs. The second is that due to frequent scattering, the CR distribution is locally isotropic.

\subsection{Weighted Slabs and Leaky Boxes}

The closely related  leaky-box and weighted slab formalisms have provided the basis for most of the literature interpreting CR data.

In the leaky-box model, the diffusion and convection terms are approximated by the leakage term with some characteristic escape time of CR from the Galaxy. The escape time
 $\tau_{esc}$ may be a function of particle energy (momentum), charge, and mass number if needed, but it does not depend on the spatial coordinates. There are two cases when the leaky box equations can be obtained as a correct approximation to the diffusion model: 1) the model with fast CR diffusion in the Galaxy and particle reflection at the CR halo boundaries with some probability to escape
\citep{1964ocr..book.....G}, %
2) the formulae for CR         density in the Galactic disk in the flat halo model $(z_h \ll R)$ with thin source and gas disks ($z_{gas} \ll z_h)$ 
which are formally equivalent to the leaky-box model formulae in the case when stable nuclei are considered
\citep{1976RvMP...48..161G}. %
 The nuclear fragmentation is actually determined not by the escape time $\tau_
{esc}$ but rather by the escape length in g cm$^{-2}$: $x = v\rho\tau_{esc}$ ,
where $\rho$ is the average gas density of interstellar gas in a galaxy with the volume of the cosmic ray halo included.

The solution of a system of coupled transport equations for all isotopes involved in the process of nuclear fragmentation is required for studying CR propagation. A powerful method, the weighted-slab technique, which consists of splitting  the problem into astrophysical and nuclear parts was suggested for this problem
\citep{1960ICRC....3..220D,1964ocr..book.....G} %
 before the modern computer epoch. The nuclear fragmentation problem is solved in terms of the slab model wherein the CR beam is allowed to traverse a thickness $x$ of the interstellar gas and these solutions are integrated over all values of $x$ weighted with a distribution function $G(x)$  derived from an astrophysical propagation model. In its standard realization
\citep{1981ApJ...247..362P,1987ApJS...64..269G} %
 the weighted-slab method breaks down for low energy CRs where one has strong energy dependence of nuclear cross sections, strong energy losses, and energy dependent diffusion. Furthermore, if the diffusion coefficient depends on the nuclear species the method has rather significant errors. After some modification
 \citep{1996ApJ...465..972P} %
 the weighted-slab method becomes rigorous for the important special case of separable dependence of the diffusion coefficient on particle energy (or rigidity) and position with no convective transport. The modified weighted-slab method was applied to a few simple diffusion models in
\citet{2001ApJ...547..264J,2001AdSpR..27..737J}. %
 The weighted-slab method can also be applied to the solution of the leaky-box equations. It can  easily be shown that the leaky-box model has an exponential distribution of path lengths $G(x) \propto \exp(-x/X)$ with the mean grammage equal to the escape length $X$.

In a purely empirical approach, one can try to determine the shape of the distribution function $G(x)$ which best fits the data on abundances of stable primary and secondary nuclei 
\citep{1970ARNPS..20..323S}.  %
 It has been established that the shape of $G(x)$ is close to  exponential: $G(x) \propto \exp(-x/X(R,\beta))$, and this justifies the use of the leaky-box model in this case.
 There are various calculations of  $G(x)$ 
\citep{1998ApJ...505..266S,2000AIPC..528..421D,2001ApJ...547..264J,2001AdSpR..27..737J}.  %

The possible existence of truncation, a deficit at small path lengths (below a few g cm$^{-2}$ at energies near 1 GeV/n),
relative to an  exponential path-length distribution,      
    has been discussed for decades 
\citep{1970ARNPS..20..323S,1987ApJS...64..269G,1993ApJ...402..188W,1996A&A...316..555D}.  %
The problem was not solved mainly because of cross-sectional uncertainties. In a consistent theory of CR diffusion and nuclear fragmentation in the cloudy interstellar medium, the truncation  occurs naturally if some fraction of CR sources resides inside  dense giant molecular  clouds
\citep{1990AA...237..445P}. %

For radioactive nuclei, the classical approach is to compute the `surviving fraction' which is the ratio of the observed abundance
to that expected in the case of no decay.
Often the result is given in the form of an effective mean gas density, to be compared with the average density in the Galaxy,
 but this density should not be taken at face value.  The surviving fraction can better be related to physical parameters 
\citep{1998A&A...337..859P}. %
None of these methods can face the complexities of propagation of CR electrons and positrons with their large energy and spatially dependent energy losses.

\subsection{Explicit models} 

Finally the mathematical effort required to put the 3-D Galaxy into a 1-D formalism becomes
overwhelming, and it seems better to work in physical space from the beginning: this
 approach is intuitively simple and easy to interpret.
We can call these `explicit models'.
The explicit solution approach including secondaries was
pioneered by
\citep{1976RvMP...48..161G}  %
 and  applied to newer  data by
\cite{1992ApJ...390...96W,1993A&A...267..372B} %
 with analytical solutions for 2D diffusion-convection models with a cosmic-ray source distribution,
which however had many restrictive approximations to make them tractable (no energy losses, simple gas model).
More recently  a semi-empirical model which is 2D and includes energy-losses and reacceleration
has been developed
\citep{2001ApJ...555..585M,2002A&A...394.1039M}. %
This is  a closed-form solution expressed as a Green's function to be integrated over the sources.
It incorporates a radial CR source distribution, but the gas model is a simple constant density within the disk.
\cite{2004ApJ...609..173T} %
 give an analytical solution for the time-dependent case with a generalized gas distribution.

A `myriad sources model'
\citep{2003ApJ...582..330H},  %
  which is actually
a Green's function method without energy losses, yields similar results to 
\cite{2004ApJ...613..962S} %
for  the diffusion coefficient and   halo size.

The most advanced explicit solutions to date are the fully numerical models  described in other sections.
Even this has limitations in treating some aspects (e.g. when particle trajectories become important at high energies)
so one might ask whether a fully Monte-Carlo approach (as is commonly done for energies $>10^{15}$ eV)
would not be better in the future, given increasing computing power.
This would allow effects like field-line diffusion (important for propagation perpendicular to the Galactic plane)  to be explicitly included.
However it is still challenging: a GeV particle diffusing with a mean free path of 1 pc 
in a Galaxy with 4 kpc halo height
takes  $\sim$$(4000/1)^2 \approx 10^7 $ scatterings to leave the Galaxy,
which would even now need supercomputers to obtain adequate statistics.
 Hence we expect a numerical solution of the propagation equations to remain an important approach
for the foreseeable future.

\section{Phenomenological models}

\noindent

\subsection{ GALPROP } \label{sect:galprop}

The GALPROP project
\citep{1998ApJ...509..212S} %
  was invented with the following aims:

1.  to enable simultaneous predictions of all relevant observations including CR nuclei, electrons and positrons, \grays and synchrotron radiation,

2. to overcome the limitations of analytical and semi-analytical methods, taking advantage of advances in computing power,  as CR, \gray and other data become more accurate,

3.  to incorporate the best current information on Galactic structure and CR source distributions,

4. to provide a {\it publicly-available}  code as a basis for further expansion.

The first point was the real driving factor, the idea being that all data relate to the same system, the Galaxy, and one cannot for example allow
a model which fits CR secondary/primary ratios while not fitting \grays or not being compatible with the known interstellar gas distribution.
There are so many simultaneous constraints, and that to find one model satisfying all of them is a challenge,  which in fact has not been met up to now.
GALPROP has been adopted as the standard for diffuse Galactic \gray emission  for NASA's Fermi-LAT \gray observatory.

We give a very brief summary of GALPROP; for details we refer the reader to the relevant papers
\citep{1998ApJ...509..212S,1998ApJ...493..694M,2000ApJ...537..763S,2002ApJ...565..280M,2004ApJ...613..962S,2006ApJ...642..902P} %
and two Annual Reviews articles,
 \citep{StrongMoskalenkoPtuskin2007} %
with new developments described in
\citet{GrenierBlackStrong2015}%
.
Developments of GALPROP have continued. It is maintained as public software\footnote{see http://galprop.stanford.edu and https://gitlab.mpcdf.mpg.de/aws/galprop} which includes an Explanatory Supplement describing the method in full detail.
Recent developments are described in
\citep{2017ApJ...846...67P,2019ApJ...887..250P,2020ApJ...889..167B}. %

The CR propagation equation  is solved numerically on a spatial grid, either in 2D with cylindrical symmetry in the Galaxy or in full 3D.
The boundaries of the model in Galactocentric radius and height above the disk, and the grid spacing, are user-definable.
 In addition there is a grid in momentum;
momentum (not e.g. kinetic energy) is used because it is the natural quantity for propagation.
Parameters for all processes in equation (\ref{eq:cr_transp}) can be controlled with input parameters.
The distribution of CR sources can be freely chosen, typically to represent supernova remnants (SNR).
Source spectral shape and isotopic composition (relative to protons) are input parameters.
Interstellar gas distributions are based on current HI and CO surveys, and the interstellar radiation field (ISRF), for lepton energy losses and inverse Compton scattering  is based on a separate detailed calculation. 
CR fragmentation and destruction cross-sections are based on extensive compilations and parameterizations
\citep{2004AdSpR..34.1288M}. %
The numerical solution proceeds in time until a steady-state is reached; a time-dependent solution is also an option.
Checks for convergence are implemented.
Starting with the heaviest  primary nucleus considered (e.g.\ $^{64}$Ni) the propagation solution is used to compute the source term for its spallation products,
which are then propagated in turn, and so on down to protons, secondary electrons and positrons, and antiprotons.
In this way secondaries, tertiaries etc. are included.
(Production of $^{10}$B via the $^{10}$Be-decay channel is  important and requires a second iteration of this procedure.)
 GALPROP includes K-capture and electron stripping
processes, where a nucleus with an electron (H-like) is considered a separate species because of the difference in the lifetime.
Since H-like atoms have only one K-shell electron,
the K-capture decay half-life has to be increased by a factor of 2
compared to the measured half-life value.

 Primary electrons  are treated separately.
Normalization of  protons, helium and electrons to experimental data is provided (all other isotopes  are determined by the source composition and propagation).
\grays and synchrotron are computed  using interstellar gas survey data (for pion-decay and bremsstrahlung)   and the ISRF model (for inverse Compton).
Spectra of all species on the chosen grid and the \gray and synchrotron skymaps  are output in a standard astronomical format  for comparison with data.
 Extensions to GALPROP includes non-linear wave damping
\citep{2006ApJ...642..902P}. %

We remark that
while GALPROP has the ambitious goal of being `realistic', it is obvious that any such model  can only be a crude approximation to reality.
Some known limitations are:
boundary condition (flux set to zero) at the halo boundary is not physical,
only  energies below $10^{15}$ eV are treated (no trajectory calculations), spatially uniform source abundances are assumed, though stochastic sources in space and time are also implemented.

CR propagation is traditionally treated as a spatially smooth,  steady-state process.
Because of the rapid diffusion and long containment time-scales in the Galaxy this is to first order
 a sufficient approximation,
but there are cases where it breaks down.
The rapid energy loss of electrons and positrons above about 100 GeV and the stochastic nature of their sources produces spatial and temporal
variations.
  Supernovae are stochastic events and each SNR source of CR accelerates for only  $10^4 - 10^5$ years, which leaves an
imprint on the distribution of electrons.
 This leads to large fluctuations in the CR electron/positron density at high energies, so that
the lepton spectrum measured near the Sun may not be typical
\citep{2004ApJ...613..962S}. %
These effects are  much smaller for nucleons since there are essentially no energy losses except ionization at low energies, but they are still included.
Such effects can influence the B/C ratio 
\citep{2004ApJ...609..173T,2005ApJ...619..314B}.  %
A recent time-dependent GALPROP model is described in \citet{2019ApJ...887..250P}. %

Here we give some technical details of the GALPROP package, taken from the
GALPROP Explanatory Supplement supplied with the code.
These can be compared with the other approaches described in this review.

{\it Transport Equation.}
GALPROP solves the transport equation with a given source distribution
and boundary conditions for all cosmic-ray species. This
includes Galactic wind (convection), diffusive reacceleration in the
interstellar medium, energy losses, nuclear fragmentation, and decay.
The numerical solution of the transport equation is based
on an   implicit second-order scheme \citep[e.g.][]{1992nrfa.book.....P}. 
The spatial boundary conditions assume either zero CR density at the boundaries or, more physically plausible, free particle escape at the boundaries. Since
we have a 3-dimensional $(R,z,p)$ or 4-dimentional $(x,y,z,p)$ problem
(spatial variables plus momentum)
we use ``operator splitting'' to handle the implicit solution.

The propagation equation is written in the form:
\begin{align}%
\label{A.1}
{\partial \disfunc \over \partial t} 
&= q(\vektor r, p) 
+ \vnabla \cdot ( \Dxx\vnabla\disfunc - \vektor V\disfunc )
+ \ddp\, p^2 \Dpp \ddp\, {1\over p^2}\, \disfunc \nonumber \\
&- {\partial\over\partial p} \left[\dot{p} \disfunc
- {p\over 3} \, (\vnabla \cdot \vektor V )\disfunc\right]
- {1\over\tau_f}\disfunc - {1\over\tau_r}\disfunc\ ,
\end{align}%
where $\disfunc=\disfunc (\vektor r,p,t)$ is the density per unit of total
particle momentum, $\disfunc(p)dp = 4\pi p^2 f(\vektor p)$ in terms of
phase-space density $f(\vektor p)$, $q(\vektor r, p)$ is the source term,
$\Dxx$ is the spatial diffusion coefficient, $\vektor V$ is the convection
velocity, reacceleration is described as diffusion in momentum space
and is determined by the coefficient $\Dpp$, $\dot{p}\equiv dp/dt$
is the momentum loss rate, $\tau_f$ is the time scale for
fragmentation, and $\tau_r$ is the time scale for the radioactive
decay.

 One can estimate 
 $D_{pp}=p^{2}V_{\mathrm{a}}^{2}/\left(  9 \Dxx\right)  $ where the Alfv\'en
velocity $V_{\mathrm{a}}$ is introduced as a characteristic velocity of
weak disturbances propagating in a magnetic field, see
\citep{1990acr..book.....B,2002cra..book.....S}
for rigorous formulas.

For a given halo size the diffusion coefficient as a function of momentum
and the reacceleration or convection parameters is determined by
boron-to-carbon ratio data.  The spatial diffusion coefficient
is taken as $\Dxx = \beta D_0(\rho/\rho_0)^{\delta}$ if necessary with
a break ($\delta=\delta_{1,2}$ below/above rigidity $\rho_0$),
where the factor $\beta$ ($= v/c$) is a
consequence of a random-walk process.
For the case of reacceleration
the momentum-space diffusion coefficient $D_{pp}$ is related to the
spatial coefficient $\Dxx$ %
where $\delta=1/3$ for a Kolmogorov spectrum of interstellar turbulences.  
The convection
velocity (in $z$-direction only) $V(z)$ is assumed to increase
linearly with distance from the plane ($dV/dz>0$ for all $z$);  this
implies a constant adiabatic energy loss.
Since the wind cannot blow in both directions at $z=0$ this formulation requires
a zero velocity there. A more general case where the wind starts at zero and reaches a constant value at a specified $z$ has therefore been
implemented using a tanh function.

The distribution of cosmic-ray sources is parameterized and usually chosen to follow the pulsar distribution from radio observations, since pulsars should be a good tracer of SNR, which are difficult to detect at large distances.  The injection spectrum
of nucleons is assumed to be a power law in momentum, $dq(p)/dp
\propto p^{-\gamma}$.
Energy losses  for nuclei by ionization and Coulomb
interactions are included, and for electrons by ionization, Coulomb
interactions, bremsstrahlung, inverse Compton, and synchrotron.
The code uses  cross-section
measurements and energy dependent fitting functions.

The code calculates the production and propagation of secondary antiprotons from $pp$ collisions.
Secondary positrons and electrons in cosmic rays are the final product
of decay of charged pions and kaons which in turn created in
collisions of cosmic-ray particles with gas. Pion production by
$pp,p{\textrm He},\textrm{He}p,\textrm{HeHe}$ collisions are included.

The nuclei are aligned on the same kinetic energy per nucleon $\Ekin$
since this simplifies the secondary-to-primary computation, where primaries 
produce secondaries of the same $\Ekin$.
However the basic CR density used has {\it units} of density per total momentum $p$ 
since this is natural for propagation.
The actual units used internally are ${c\over 4\pi} \curlyn(p)$, where $\curlyn(p) = d\curlyn/dp$ 
in units of \densityunits, i.e. $\curlyn = 4\pi p^2 f(p)$.

When the {\it flux} $I(\Ekin)$ in \fluxunits\ is necessary, it can be simply obtained from
\begin{equation}%
\label{eq.1}
I(\Ekin)= {\beta c\over 4\pi} {d\curlyn\over dp} {dp\over d\Ekin} = {c\over 4\pi} \curlyn(p) A,
\end{equation}%
where $A$ is the nucleus mass number.
This follows from $dp =  {A\over\beta} d\Ekin$.
The combined requirements of transport and fragmentation are thus elegantly met. 
The normal units for presentation of CR data are \fluxunits, and with this scheme
the conversion is trivial. The nucleus energy scales are logarithmic in $\Ekin$.

\medskip
{\it Numerical solution of the propagation equation.}
\medskip

\paragraph{Full explicit method.}

The diffusion, reacceleration, convection and loss terms in
eq.~(\ref{A.1}) can all be finite-differenced for each dimension ($R, z,
p$)  or  $(x, y, z, p)$  in the form
\begin{equation}%
{\partial \disfunc_i\over\partial t}
= {\disfunc^{t+\Delta t}_i -\disfunc^t_i \over \Delta t}
= {\alpha_1\disfunc^{t}_{i-1} -\alpha_2\disfunc^{t}_i 
+ \alpha_3\disfunc^{t}_{i+1}\over\Delta t}+ q_i \ ,
\end{equation}%
where all terms are functions of $(R, z, p)$ or  $(x, y, z, p)$.

This is the {\it fully time-explicit method} \citep{1992nrfa.book.....P} where the updating
scheme is
\begin{equation}%
\disfunc_i^{t+\Delta t} = \disfunc_i^t + \alpha_1 \disfunc_{i-1}^{t} - 
\alpha_2 \disfunc_i^{t} +\alpha_3 \disfunc_{i+1}^{t} + q_i
\Delta t \ .
\end{equation}%
which generalizes simply to any number of dimensions since all the quantities are known from the current step.
It gives  more accurate solutions, which tend to the exact solution according to the computed diagnostics,
but are not unconditionally stable (while Crank-Nicolson is). For this reason it is only applicable for short enough timesteps. Since no solution of matrix equations is required, this method is faster than Crank-Nicolson for the
same timesteps, and this compensates for the need for smaller steps.

\paragraph{Fully implicit method.}

The diffusion, reacceleration, convection and loss terms in
eq.~(\ref{A.1}) can all be finite-differenced for each dimension ($R, z,
p$)  or  $(x, y, z, p)$  in the form
\begin{equation}%
{\partial \disfunc_i\over\partial t}
= {\disfunc^{t+\Delta t}_i -\disfunc^t_i \over \Delta t}
= {\alpha_1\disfunc^{t+\Delta t}_{i-1} -\alpha_2\disfunc^{t+\Delta t}_i 
+ \alpha_3\disfunc^{t+\Delta t}_{i+1}\over\Delta t}+ q_i \ ,
\end{equation}%
where all terms are functions of $(R, z, p)$ or  $(x, y, z, p)$.

This is the {\it fully time-implicit method} %
 where the updating
scheme is
\begin{equation}%
\disfunc_i^{t+\Delta t} = \disfunc_i^t + \alpha_1 \disfunc_{i-1}^{t+\Delta t} - 
\alpha_2 \disfunc_i^{t+\Delta t} +\alpha_3 \disfunc_{i+1}^{t+\Delta t} + q_i
\Delta t \ .
\end{equation}%
This method is unconditionally stable for all $\alpha$ and $\Delta t$, but is only 1st-order accurate in time.

The tridiagonal system of equations
\begin{equation}%
- \alpha_1 \disfunc_{i-1}^{t+\Delta t} 
+ ( 1 + \alpha_2) \disfunc_i^{t+\Delta t}
- \alpha_3 \disfunc_{i+1}^{t+\Delta t} 
= \disfunc_i^t + q_i \Delta t ,
\end{equation}%
is solved for the $\disfunc_i^{t+\Delta t}$ by standard methods.
 Note that for energy losses we use `upwind'
differencing to enhance stability, which is possible since we have
only {\it loss} terms (adiabatic energy {\it gain} is not included
here).

\paragraph{Crank-Nicolson method}.

Alternatively, the  propagation equation  can be finite-differenced   in the form
\begin{align}%
{\partial \disfunc_i\over\partial t}
& = {\disfunc^{t+\Delta t}_i -\disfunc^t_i \over \Delta t} \nonumber \\
&= {\alpha_1\disfunc^{t+\Delta t}_{i-1} -\alpha_2\disfunc^{t+\Delta t}_i 
+ \alpha_3\disfunc^{t+\Delta t}_{i+1}\over2\Delta t}\nonumber\\
&+{\alpha_1\disfunc^{t}_{i-1} -\alpha_2\disfunc^{t}_i 
+ \alpha_3\disfunc^{t}_{i+1}\over2\Delta t}
+ q_i \ 
\end{align}%
This is the {\it Crank-Nicolson method} %
 where the updating
scheme is
\begin{align}%
\disfunc_i^{t+\Delta t} =
 \disfunc_i^t &+ {\alpha_1\over2} \disfunc_{i-1}^{t+\Delta t} - 
{\alpha_2\over2} \disfunc_i^{t+\Delta t} + {\alpha_3\over2} \disfunc_{i+1}^{t+\Delta t} \nonumber \\
 &+ {\alpha_1\over2} \disfunc_{i-1}^{t} -
{\alpha_2\over2} \disfunc_i^{t         } + {\alpha_3\over2} \disfunc_{i+1}^{t        }
 + q_i\Delta t \ .
\end{align}%
It  thus uses a combination of implicit and explicit terms, forming the time-average of the differentials.
Like the fully implicit method, this method is unconditionally stable for all $\alpha$ and $\Delta t$, but is 2nd-order accurate in time, so that 
larger time-steps are possible than with the 1st-order scheme.

The tridiagonal system of equations
\begin{align}%
&- {\alpha_1\over2} \disfunc_{i-1}^{t+\Delta t} 
+ ( 1 + {\alpha_2\over2}) \disfunc_i^{t+\Delta t}
- {\alpha_3\over2} \disfunc_{i+1}^{t+\Delta t} \nonumber \\
&= \disfunc_i^t + q_i \Delta t 
+ {\alpha_1\over2} \disfunc_{i-1}^{t} 
- {\alpha_2\over2} \disfunc_i^{t}
+ {\alpha_3\over2} \disfunc_{i+1}^{t} 
\end{align}%
or
\begin{align}%
&- {\alpha_1\over2} \disfunc_{i-1}^{t+\Delta t} 
+ ( 1 + {\alpha_2\over2}) \disfunc_i^{t+\Delta t}
- {\alpha_3\over2} \disfunc_{i+1}^{t+\Delta t} \nonumber \\ 
&= {\alpha_1\over2}       \disfunc_{i-1}^{t} 
+ (1 - {\alpha_2\over2}) \disfunc_i^{t}
+ {\alpha_3\over2}       \disfunc_{i+1}^{t} 
+ q_i \Delta t
\end{align}%
is again solved for the $\disfunc_i^{t+\Delta t}$ by the standard method. Note that the RHS has all known quantities from the current time-step.

The Crank-Nicolson method described above applies to a one-dimensional case; 
the application to 2 or 3 spatial and one momentum dimension requires a generalization.
A straightforward expansion to more dimensions implies solving large matrix equations (no longer tridiagonal);
instead the so-called ADI  (alternating direction implicit) method is used, in which the implicit solution is
applied to each dimension in turn. Each application uses just the operator for that dimension, so the tridiagonal
scheme can still be used.
This however is not completely valid since it solves a slightly different problem from that with the full operator;
however for small enough timesteps  the solution is accurate (see Section on Tests of GALPROP in the GALPROP Explanatory Supplement).

The {\it explicit} method, where the full operator can be used in each timestep without any overhead for solving matrix equations, is also useful
for obtaining an accurate solution at the end of a run. Although it is not unconditionally stable, this does not matter provided the timesteps are small enough, which is in any case required for the implicit methods to maximise their accuracy. A suitable mix of explicit and implicit methods to obtain an accurate solution with minimum computing requirements, is the goal.

For 2D,  three spatial boundary conditions 
\begin{equation}%
\disfunc(R,z_h,p) = \disfunc(R,-z_h,p) = \disfunc(R_h,z,p) = 0
\end{equation}%
may be imposed at each iteration. 
This is not physically expected, although it is a common conventional assumption.
More physically plausible is free escape at the boundaries, which is not the same. 
For this, it is sufficient simply to not impose the above conditions in the updating scheme, since the $\alpha_1, \alpha_3$ coefficients do not act outside the boundaries,
so there is  no diffusive or convective flux inwards at the boundaries. 
The resulting solutions  then have  $\disfunc>0$ at the boundaries.
In future more physical boundary conditions could be implemented, e.g. specifying the outward streaming velocity or the escape probability at the boundaries.
No boundary conditions are imposed or
required at $R$ = 0 or in $p$. 
 Grid intervals are typically $\Delta R
= 1$ kpc, $\Delta z = 0.1$ kpc; for $p$ a logarithmic scale with ratio
typically 1.2 is used.
  Although the model is symmetric around $z = 0$
the solution is generated for $-z_h < z < z_h$ since this is required
for the tridiagonal system to be valid.

For 3D,  the  spatial boundary conditions 
\begin{equation}%
  \disfunc(\pm x_h, y, z,p)  = \disfunc( x, \pm y_h, z,p) = \disfunc( x, y,\pm z_h,p)  = 0
\end{equation}%
may be imposed at each iteration, and  free escape at the boundary is an option as for 2D.
 Again no boundary conditions are imposed in $p$.  Grid intervals are typically $\Delta x= \Delta y = 0.5$ kpc, $\Delta z = 0.1$ kpc.

Since we have a 3-dimensional $(R,z,p)$ problem we use `operator
splitting' to handle the implicit solution, as follows. 
 We apply the
implicit updating scheme alternately for the operator in each dimension
in turn, keeping the other two coordinates fixed.
 The source and fragmentation, decay terms are used in every step,  so to account for the 3
substeps, ${1\over 3} q_i$ and $1\over3\tau$ are used instead of $q_i$,
$1/\tau$ for the source term and the fragmentation, decay terms respectively. 
 The coefficients for 3 spatial dimension are the same except that R is replaced by (x,y) and the finite differencing coefficients have the same form as for z, and  ${1\over 4} q_i$ and $1\over4\tau$ are used  for the source term and the fragmentation, decay terms respectively, to account for the 4 substeps.
 With this scheme the solution can be done via the tridiagonal solution for each dimension in turn, as described in 
 \citep{1992nrfa.book.....P}.
 The spatial 3D scheme is simpler than the 2D one since the diffusion operator is easier to formulate (x,y,z have the same form), and in addition it does not have the problem of the boundary condition at R=0.
In the case of anisotropic diffusion, $\Dzz$ is used in the $z$-direction.

Finally  all nuclei are normalized to the proton flux from the parameter file, using the relative abundances given as parameters.
The value of $\Ekin$ is taken as the reference value for the proton normalization.
All results are output to FITS files for  comparison with data and further analysis.

\subsection{Other codes} \label{sect:other_codes}

Following the success of the GALPROP approach described above, other projects with similar goals were started.  
Here we just mention these without details, which can be found in the papers.
The DRAGON project 
\citep{2014PhRvD..89h3007G} %
and references therein
 which extends the CR propagation to anisotropic and spatially-dependent  diffusion. 
DRAGON is also publicly available\footnote{at http://www.dragonproject.org}.
A more physical approach to diffusion with turbulence incorporated in DRAGON is given in
\cite{2014ApJ...782...36E}. %

USINE is a semi-analytical CR-propagation package, which has the advantage of speed for model parameter explorations, and which has been the basis of many recent investigations. 
USINE is publicly available\footnote{https://dmaurin.gitlab.io/USINE/}
as described by \citep{2020CoPhC.24706942M}.
It has been used for many years already
\citep{2011A&A...526A.101P} %
and references therein.
It has recently been used to study secondary antiprotons \citep{2020PhRvR...2b3022B} %
and the size of the Galactic CR halo \citep{2020arXiv200400441W}. %

The numerical packages mentioned have limitations in terms of both accuracy and speed, and hence the spatial resolution achievable. 
Hence their use has mainly been restricted to 2D models with cylindrical symmetry. A new approach is implemented in the PICARD model
\citep{2013arXiv1308.2829W,2014APh....55...37K}, %
 which is fully 3D in concept and has state-of-the art numerical techniques. This makes it possible to handle models with spiral structure at good (e.g. 10 pc) resolution with reasonable computer resources.

\section{Self-consistent models}

\subsection{The system of equations} \label{sect:selfc_eqns}
  
 To reduce the physical and computational complexity of the CR propagation problem numerous authors neglected the explicit consideration of the streaming process by eliminating the Alfv\'en wave-related component from the CR propagation speed $\bm{u}$ in equation~(\ref{eq:cr_transp}) and by defining the spatial and momentum diffusion coefficients as free parameters of the model. In phenomenological models the values of diffusion coefficient are deduced from secondary to primary CR abundances taken from observational data. Depending on particular needs Eqn.~\eqref{eq:cr_transp} might be integrated directly or with the aid of its number density or energy density moments. The latter quantities enable construction of numerical solvers in a conservative manner, using finite volume methods both in spatial and in momentum dimensions. In the following considerations we neglect particle acceleration processes, therefore we assume $D_{pp} = 0$. We also omit particle acceleration and radioactive decay.

\subsubsection{CR number density}

The number density of CR particles in an arbitrarily chosen range $[p_\mathrm{L}, p_\mathrm{R}]$ in momentum space is defined as
\begin{equation}
\ncrLR (\bm{x},t) \equiv \int_{p_\mathrm{L}}^{p_\mathrm{R}} 4\pi p^2 f(\bm{x},p,t) dp.
\end{equation}
We multiply Eq. (\ref{eq:cr_transp}) by $4\pi p^2$ and integrate over $p$
to get the evolution equation for the particle number density
\begin{align}
\pder{\ncrLR}{t}{} &= - \nabla\cdot (\bm{u} \ncrLR)
                      + \nabla \left(  \langle  \tensor{D}_n \rangle  \nabla \ncrLR\right) \nonumber \\
                      &+ \left[\left(\frac{1}{3}  (\nabla \cdot \bm{u} ) p  + b_l  \right ) 4\pi p^2 f\xpt\right ]_{p_\mathrm{L}}^{p_\mathrm{R}} +Q^{\mathrm{LR}},
\label{eqn:n}
\end{align}
where $Q^{\mathrm{LR}}$ is the spatial density of CR sources and $\langle\tensor{D}_n\rangle$
is the momentum-averaged spatial diffusion tensor of the particle number density (see also Section~\ref{sec:diffusion-spectral}).
In the limit of $p_\mathrm{L} = 0$ and $p_\mathrm{R}=\infty$ we get the conventional form of the diffusion-advection equation for CR particle number density
 \begin{equation}
\pder{\ncr\xt}{t}{} = - \nabla\cdot (\bm{u} \ncr\xt)
                      + \nabla \left(  \langle \tensor{D}^n \rangle  \nabla \ncr\xt \right) +Q\xpt.
\label{eq:n_all}
\end{equation}

\subsubsection{CR energy density}

The CR energy density in a section $({p_\mathrm{L}},{p_\mathrm{R}})$ of the momentum axis is defined as
\begin{equation}
\ecrLR(\bm{x},t) \equiv \int_{p_\mathrm{L}}^{p_\mathrm{R}} 4\pi p^2 T(p) f(\bm{x},p,t) dp.
\end{equation}
We integrate eq. (\ref{eq:cr_transp}) multiplied by $4\pi p^2 T\xpt$ over $p$,
where $T\xpt \equiv \sqrt{p^2 c^2 + m^2 c^4} - m c^2$ is  kinetic energy and $m$ is the rest-mass of CR protons, electrons or other particles. The resulting equation for the CR energy density reads
\begin{eqnarray}
\pder{\ecrLR}{t}{} & = & - \nabla\cdot (\bm{u} \ecrLR)  + \nabla \left(  \langle  D^e_{xx} \rangle\nabla \ecrLR \right)
                              + \left[\left(\frac{1}{3}  (\nabla \cdot \bm{u} ) p  + b_l  \right ) 4\pi p^2  f\xpt T\p \right ]_{p_\mathrm{L}}^{p_\mathrm{R}} \nonumber  \\
                    &&     -  \int_{p_\mathrm{L}}^{p_\mathrm{R}}   \left(\frac{1}{3}  (\nabla \cdot \bm{u} ) p  + b_l\xp  \right ) 4\pi p^2  f\xpt  \frac{c p }{\sqrt{p^2+m^2 c^2}} dp +S^{\mathrm{LR}} \ .
\label{eq:e}
\end{eqnarray}
where $\langle\tensor{D}_e\rangle$ denotes the momentum averaged diffusion coefficient (see Section~\ref{sec:diffusion-spectral}) of the CR energy and $S^{\mathrm{LR}}$ the sources of CR energy.
The CR pressure contribution from particles in the momentum range $ [{p_\mathrm{L}},{p_\mathrm{R}}]$  is
\begin{equation}
P_{cr}^{\mathrm{LR}} \equiv \frac{4\pi}{3}  \int_{p_\mathrm{L}}^{p_\mathrm{R}}  p^3 v\p f dp \ .
\end{equation}
In the limit of  $p_\mathrm{L} = 0$, $p_\mathrm{R}=\infty$ and $b_l=0$ we get the conventional form \cite[see eg.][]{1985A&A...151..151S} of the momentum-integrated diffusion-advection equation for the CR particle energy density
\begin{equation}
\pder{\ecr}{t}{} = - \nabla\cdot (\bm{u} \ecr)  + \nabla \left(  \langle \tensor{D}_e \rangle  \nabla \ecr \right)
                      -  P_{cr}   (\nabla \cdot \bm{u} )  +S \ ,
\label{eq:e_all}
\end{equation}
together with an adiabatic equation of state relating the momentum integrated CR  pressure  to CR energy density 
$P_{cr} = (\gamma_{cr} -1) \ecr$
with adiabatic index equal to $4/3$ for a relativistic gas or its  value is in the  range $[4/3,5/3]$ if the trans-relativistic population of CR particles is significant.

\subsection{Two-fluid diffusion-advection models} \label{subsect:two_fluid_models}

The development of self-consistent methods for the CR transport equation started with  theoretical work by \citet{1981ApJ...248..344D,1982A&A...111..317A} who integrated the kinetic equation (\ref{eq:cr_transp}) in  momentum space to get a single equation for the total CR pressure. They supplemented the resulting equation to the set of fluid equations for the conventional gas. The two components, thermal gas and the population of the nonthermal CR particles were coupled by  the CR pressure term. The resulting two-fluid system was applied in analytical  studies of hydrodynamic shock structure in the presence of CRs. These preliminary studies have shown that even for moderately strong shock waves most of the upstream energy flux in the background plasma is transferred to cosmic rays, and thus  demonstrated the importance of the CR induced dynamical effects in the system composed of CRs and thermal plasma.  \citet{1985A&A...151..151S} used the two-fluid system of equations  for  studies of the dynamics of interstellar gas in galactic disks.

\citet{1986MNRAS.223..353D} presented a first numerical algorithm to model the propagation of CRs together with thermal gas in 1-D. The algorithm, based on the solution of an appropriate Riemann problem (see Sect.~\ref{sect:Riemann_problem})
has been used to study the stability of  CR-modified shocks. %

\citet{1990ApJ...353..149K}  extended the two-fluid model of diffusive shock acceleration by inclusion of an in-situ CR injection at steady state shocks. 
In a companion paper \citet{1990ApJ...363..499J} presented the results of time-dependent numerical simulations of CR-mediated shocks based on the two-fluid model. By means of the piecewise parabolic method (PPM) \citet{1984JCoPh..54..174C} modelled the evolution of  plane parallel, piston-driven shocks and spherical adiabatic blast waves. They investigated  shocks that sweep up ambient cosmic rays as well as those that inject the cosmic rays directly. 
\citet{1993ApJ...402..560J} developed a time-dependent two fluid model  of CR acceleration during the impact of shocks on dense two-dimensional clouds. 

Diffusive shock acceleration of relativistic protons and their dynamical feedback on the background flow was included in the two-fluid model of \citet{1994ApJ...432..194J} who simulated the dynamical evolution of cosmic gas clouds moving supersonically through a uniform low-density medium. In their simulations more than 10\% of the initial kinetic energy of the flow was converted into a combination of thermal and CR energy. The fraction of energy going to CRs exceeded in some cases the energy transfered to gas.

\cite{1994ApJS...90..975F} carried out numerical simulations with a new  conservative total variation diminishing (TVD) MHD scheme based on a Godunov-type method %
by \cite{RyuJones1995,RyuJonesFrank1995}. %
They treated CRs as a massless, diffusive fluid governed by a conservation equation for the CR energy derived from the momentum-dependent diffusion-advection equation \citep{1975MNRAS.172..557S}.
The CR energy equation was solved using a second order monotonic advection and Crank-Nicolson scheme.
 The paper demonstrated  first results of time-dependent two-fluid CR modified shock simulations in one spatial dimension.

\cite{1995ApJ...441..629F} performed dynamical simulations of oblique MHD cosmic-ray (CR)-modified plane shock evolution. They solved the system of two-fluid CR-MHD equations and also a more complete system consisting of MHD equations and the momentum dependent diffusion-advection equation. 
The authors compare the results of two-fluid  and momentum-dependent diffusion-advection  approaches. 

A comprehensive discussion of the strength and weakness of the two-fluid model was presented by \cite{1997ApJ...476..875K}. %
They found a good agreement of the two-fluid model with the dynamical properties of the more detailed diffusion-advection results. They also found that the validity of the two-fluid formalism does not necessarily mean that steady state two-fluid models provide a reliable tool for predicting the efficiency of particle acceleration in real shocks.

\cite{2003A&A...412..331H} presented  a new numerical algorithm for two-fluid diffusion-advection propagation of CRs coupled dynamically with thermal plasma, with anisotropic, magnetic field-aligned diffusion.  The algorithm was implemented  into the MHD code ZEUS-3D \citep{StoneNorman1992a, StoneNorman1992b}. 
The paper  focussed on the  development of a stable algorithm for anisotropic diffusion of CRs along magnetic field vectors defined on a staggered mesh.
Details of the anisotropic diffusion algorithm are presented in Section~\ref{sect:aniso-diff}.
The  CR propagation algorithm, involving an explicit diffusion algorithm coupled to the MHD system, has been demonstrated to be stable within the standard CFL time-step restriction $\Delta t \leq 0.5 (\Delta x)^2/D $, where $D$ is the diffusion coefficient. The algorithm was applied in simulations of  the Parker instability triggered by cosmic rays injected by a SN remnant and subsequently in numerical experiments of  the CR-driven dynamo  process \citep{2004ApJ...605L..33H,2009ApJ...706L.155H} and in numerical models of CR-driven winds \citep{2013ApJ...777L..38H}.

\cite{2006MNRAS.373..643S} realized that the usual Fickian  approach to the diffusion, which assumes that the flux of the diffusive quantity is proportional to its instantaneous gradient, leads to several problems.  They noted that the anisotropic part of CR diffusion tensor becomes singular at magnetic X-points, leading to infinite CR propagation speeds, and consequently limits the timestep of an explicit time-stepping algorithm to zero.   In order to ensure finite propagation speed they modified the equation for the diffusive flux to a non-Fickian form motivated by the turbulent transport of passive scalars \citep{2003PhFl...15L..73B}. The resulting  equation for the diffusive quantity was reduced to the form of telegraph equation containing an extra second time-derivative, which acts as the displacement current in electrodynamics. The new term was included with an artificially reduced speed of light in order to reduce the propagation speed to numerically acceptable values.

\subsection{Dynamical CR MHD models}

\subsubsection{Riemann problem} \label{sect:Riemann_problem}

\begin{figure}
  \centering
  \includegraphics[height=2.5cm]{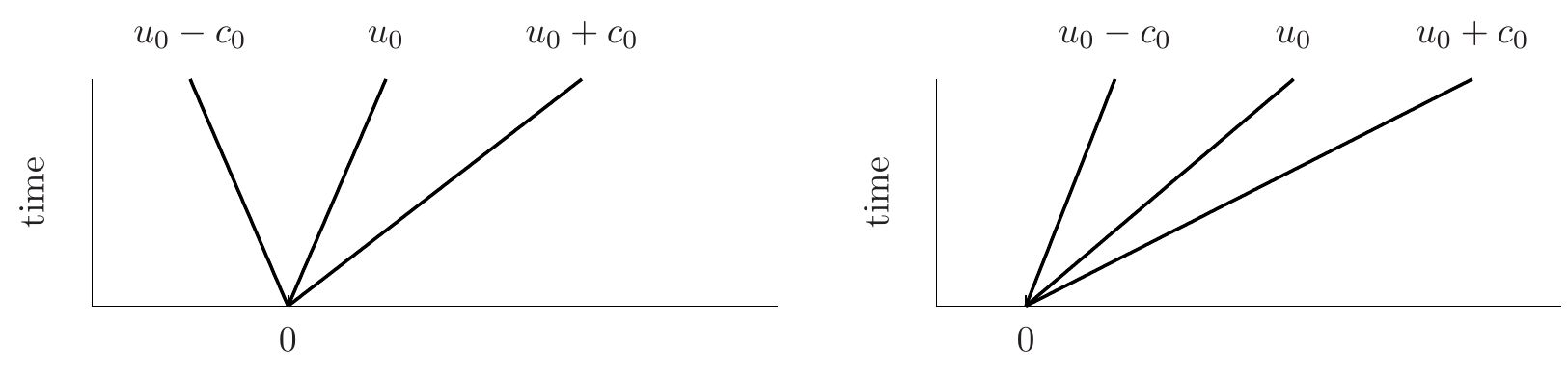}
  \caption{visualisation of the characteristics of the sound waves. The left side shows the subsonic case the right side the supersonic case.}
  \label{fig:characteristics-simple}
\end{figure}

The aim of this section is to introduce the Riemann problem as well as the principle of the Godunov method to clarify the terms we use in subsequent sections. The basis for the dynamical CR models are the fluid equations for the thermal gas, which can be expressed in \emph{primitive} ($\rho$, $u$, $P$) or \emph{conserved} quantities ($\rho$, $\rho u$, $\epsilon$). The differential equations of fluid dynamics in primitive form and one spatial dimension read 

\begin{align}
\rho_t+(\rho u)_x&=0,\\
u_t+uu_x+\frac{p_x}{\rho}&=0,\\
\epsilon_t+u\epsilon_x+\frac{p}{\rho}u_x&=0,
\end{align}
where we use the short notation $q_t\equiv\partial q/\partial t$. The three equations describe the conservation of mass, momentum and total energy. We define $e_\mathrm{tot}=\rho(\frac{1}{2}u^2+\epsilon)$, where $\epsilon=\frac{p}{(\gamma-1)\rho}$ and $\gamma=\frac{C_p}{C_v}$.
The equivalent set of HD equations in conservative form is
\begin{equation}
U_t+F(U)_x=0  .
\end{equation}
\begin{equation}
U=\left[
\begin{array}{c}
u_1\\u_2\\u_3
\end{array}\right]\equiv\left[
\begin{array}{c}
\rho\\ \rho u\\e_{tot}
\end{array}\right],\ \ F=\left[
\begin{array}{c}
f_1\\f_2\\f_3
\end{array}\right]\equiv\left[
\begin{array}{c}
\rho u\\
\rho u^2+p\\
u(e_{tot}+p)
\end{array}\right]  ,
\end{equation}
where  $U$ is the vector of conservative quantities,  $F$ is the vector of corresponding fluxes.
The equations of hydrodynamics are hyperbolic partial differential equations \citep[see, e.g.][]{Leveque2002, toro-2009, Balsara2017}, whose time dependent solutions are given by the so-called charateristics.
The characteristic curves are lines in space-lime along which $U$ stays constant. Their slope is given by eigenvalues of the Jacobian $|\partial F_i / \partial x_j |$. For the advection equation of quantity $q$ with constant velocity $u$ in one dimension,
\begin{equation}
  \frac{\partial q}{\partial t} + u_0 \frac{\partial q}{\partial x} = 0,
\end{equation}
the characteristcs are linear functions that simply translate the quantity $q$. More generally we find solutions by geometrical translation of the initial state to a given point along characteristic lines.
In the full set of equations we find that besides advection the system is influenced by the pressure in the gas and the corresponding sound waves $c_0 = \sqrt{\gamma P_{0} / \rho_{0}}$. The system yields three different characteristcs,
\begin{align}
  \lambda_{-1} &=u_{0}-\sqrt{\gamma P_{0} / \rho_{0}} = u_{0}-c_{0}  ,\\
  \lambda_{0} &=u_{0} , \\
  \lambda_{+1} &=u_{0}+\sqrt{\gamma P_{0} / \rho_{0}} = u_{0}+c_{0} ,
\end{align}
which are the ones for advection ($\lambda_{0}$), the advected fluid and a left-going sound wave ($\lambda_{-1}$) as well as the advection and a right-going sound wave ($\lambda_{+1}$). The characteristics of this problem are illustrated in Fig.~\ref{fig:characteristics-simple}.

In numerical setups we discretize the domain into cells $i$ and use the discretized form of conservation laws. Integration over a control volume - a finite section of space and time - leads to the evolution equation (Godunov method) 
\begin{equation}
U_i^{n+1}=U_i^n+\frac{\Delta t}{\Delta x}\left[F_{i-\frac{1}{2}}-F_{i+\frac{1}{2}}\right]  ,
\label{eq:cons-law}
\end{equation}
where $U_i^n$ are averaged conservative quantities over the cell volume and $F_{i\pm\frac{1}{2}}$ are the fluxes between cells averaged over the timestep.

\begin{figure}
    \centering
    \includegraphics[width=\textwidth]{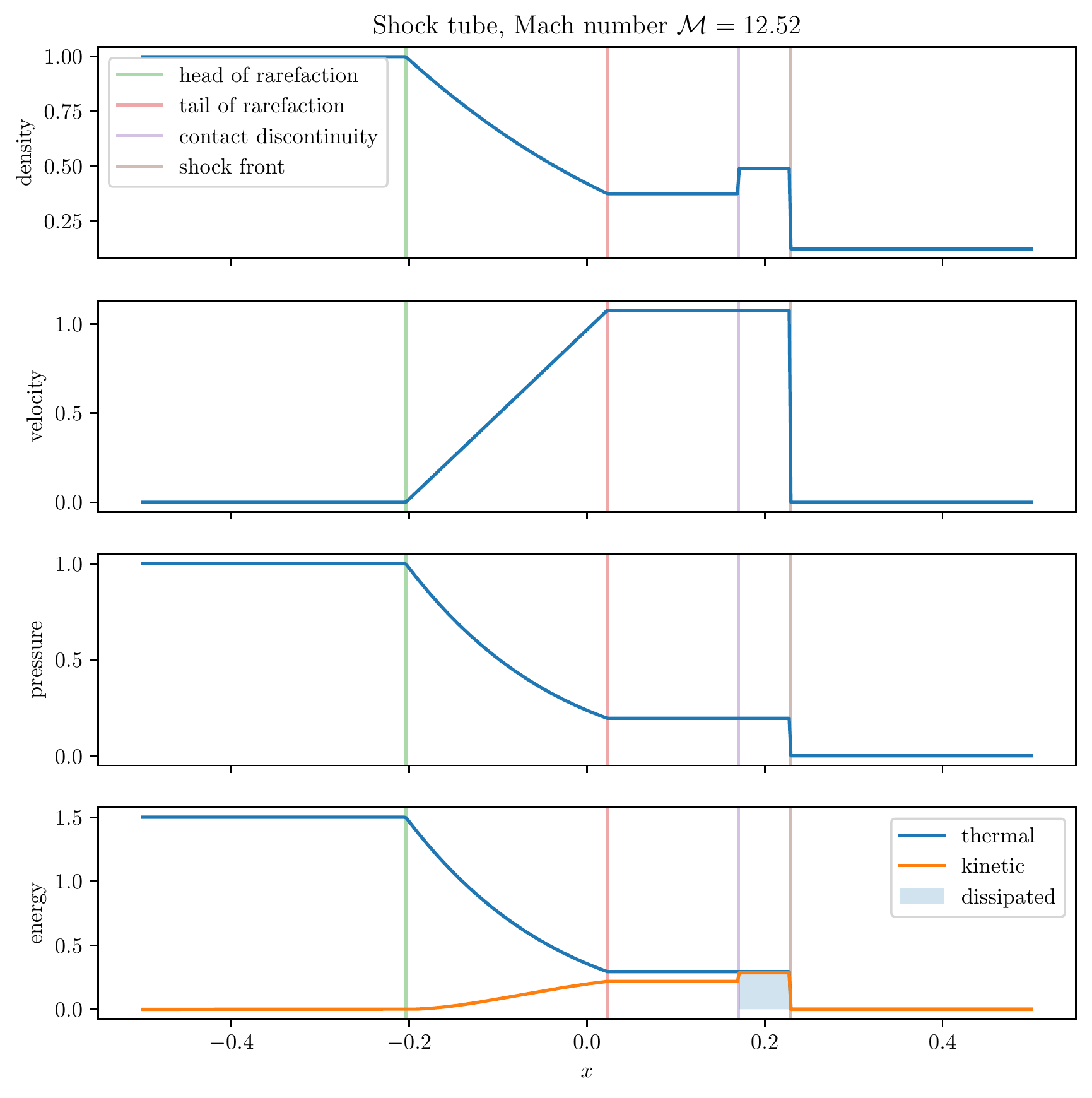}
    \caption{Shock tube solution at late time. One can clearly distringush between the unperturbed states at the very left and right, the rarafaction wave, the contact discontinuity and the shock front \citep{Leveque2002}.}
    \label{fig:shock-tube-late-time}
\end{figure}

The \emph{Riemann problem} describes an initial value problem using a discontinuous initial state $S$ with piecewise constant values on either side of the interface $x_0$
\begin{equation}
S = 
\begin{cases} 
  S_\mathrm{1} & x\leq x_0 \\
  S_\mathrm{2} & x>x_0,
\end{cases}
\end{equation}
The time evolution is described by three types of wave structures, namely \emph{shocks}, \emph{rarefaction waves} and \emph{contact discontinuities}. The solutions can be found by solving the so-called Rankine-Hugoniot jump conditions which represent conservation laws across discontinuities
\begin{align}
{[\rho v]} & =0  ,\\
{\left[\rho v^{2}+P\right]} & =0  ,\\
\left[\frac{1}{2} v^{2}+\epsilon+\frac{P}{\rho} \right] &= 0  ,
\end{align}
which is explicitly written as
\begin{align}
\rho_1 v_1 & = \rho_2 v_2\\
\rho_1 v_1^{2}+P_1 & =\rho_2 v_2^{2}+P_2 \\
\frac{1}{2} v_1^{2}+\epsilon_1+\frac{P_1}{\rho_1} &= \frac{1}{2} v_2^{2}+\epsilon_2+\frac{P_2}{\rho_2}
\end{align}
Using the Mach number
\begin{equation}
\mathcal{M}_{1} \equiv \frac{v_{1}}{c_{1}}=\sqrt{\frac{\rho_{1} v_{1}^{2}}{\gamma P_{1}}}=\sqrt{\frac{\bar{m} v_{1}^{2}}{\gamma k_{\mathrm{B}} T_{1}}}
\end{equation}
we can write the ratios of densities, pressures and temperatures as follows
\begin{align}
\frac{\rho_{2}}{\rho_{1}}&=\frac{v_{1}}{v_{2}}=\frac{(\gamma+1) \mathcal{M}_{1}^{2}}{(\gamma-1) \mathcal{M}_{1}^{2}+2} &&\stackrel{\gamma=1}{\longrightarrow} \mathcal{M}_{1}^{2} ,\\
\frac{P_{2}}{P_{1}}&=\frac{\rho_{2} k_{\mathrm{B}} T_{2}}{\rho_{1} k_{\mathrm{B}} T_{1}}=\frac{2 \gamma \mathcal{M}_{1}^{2}-(\gamma-1)}{\gamma+1} &&\stackrel{\gamma=1}{\longrightarrow} \mathcal{M}_{1}^{2} ,\\
\frac{T_{2}}{T_{1}}&=\frac{\left[(\gamma-1) \mathcal{M}_{1}^{2}+2\right]\left[2 \gamma \mathcal{M}_{1}^{2}-(\gamma-1)\right]}{(\gamma+1)^{2} \mathcal{M}_{1}^{2}} &&\stackrel{\gamma=1}{\longrightarrow} 1 .
\end{align}
Exact solutions can be found with some help of numerical root finding. In numerical practice it is computationally less expensive to compute approximate solutions which do not rely on root finding routines, but instead rely on purely algebraical methods, using explicitely Rankine-Hugoniot conditions (HLL). A particularly important setup is the Sod shock tube \citep{1978JCoPh..27....1S} with the initial conditions
\begin{equation}
\left(\begin{array}{c}
\rho_\mathrm{1} \\
P_\mathrm{1} \\
v_\mathrm{1}
\end{array}\right)=\left(\begin{array}{c}
1.0 \\
1.0 \\
0.0
\end{array}\right),\left(\begin{array}{c}
\rho_\mathrm{2} \\
P_\mathrm{2} \\
v_\mathrm{2}
\end{array}\right)=\left(\begin{array}{c}
0.125 \\
0.1 \\
0.0
\end{array}\right) .
\end{equation}
The most prominent example is a shock tube, with four important interfaces, which are illustrated in Fig~\ref{fig:shock-tube-late-time}. From left to right we highlight the head and the tail of the rarefaction wave, the contact discontinuity as well as the shock front moving to the right.

The system described so far does not include magnetic fields, which extend the set of characteristics by magnetic waves. Despite the importance of the magnetic field for CRs in general we refrain from deriving the additional equations here. This setup is for one simple thermal fluid. CRs can be included in three different manners into this setup. The first and most simple inclusion is simply the advection of the CR fluid with the gas velocity. The second possibility includes the CR pressure into the jump condition and accounts for the possibility of two different adiabatic indices for the different fluids as described in \citet{2006MNRAS.367..113P}. The third possibility does not just include CRs as an existing fluid, but includes the acceleration of CRs in a phenomenological way. A fraction of the dissipated energy at the shock is converted to CRs and treated as a source term in the shocked region. This approach is described in \citet{2017MNRAS.465.4500P}.

\subsubsection{Extensions of the Riemann problem including CRs}

To address the question of dynamical importance of shock-injected CRs during the structure formation in  the $\Lambda$CDM universe \cite{2006MNRAS.367..113P} derived a complete analytic solution of the Riemann shock-tube problem \citep{1978JCoPh..27....1S} for the medium composed of thermal gas and relativistic CR component in two-fluid approximation.   They applied their solution  in smooth particle hydrodynamics (SPH) cosmological simulations designed for studies of CR energy injection at cosmological shocks. 
  
\citet{2007JCoPh.227..776M} %
 added the number density moment of the CR propagation equation~(\ref{eq:cr_transp})  to the system of HD equations and performed characteristics analysis of the  system of  quasilinear equations describing the dynamical evolution of thermal gas 
 combined with a spectral evolution of a CR population. The study focussed on the hydrodynamical part of the momentum-dependent algorithm (see Sect. \ref{sect:theory-momentum-dependent} for details) for the evolution of CRs, introduced in the COSMOCR code \citep{2001CoPhC.141...17M}.
The CR population was approximated with a piecewise power-law distribution function. A set of  conservation laws for the number density  of CR particles in individual spectral bins, treated as passive scalars, has been supplemented to the system of conservation laws for thermal gas density, momentum and total (gas plus CR) energy density.
The exchange of energy between the thermal fluid and the CR components is  modeled with flux conserving methods both in configuration and in momentum space. The proposed numerical method is based on a combination of Glimm's method \citep{doi:10.1002/cpa.3160180408} preserving the discontinuous character of shocks and a higher order Godunov method \citep{toro-2009} in the smooth flows.

\citet{2016JPhCS.719a2021K} analyzed the CR-MHD equations and proposed a fully conservative approach to the system of equations consisting of the set of MHD equations and the equation describing the number density of CR particles in a two-fluid approach. 
They proposed a numerical scheme providing solutions that satisfy the Rankine-Hugoniot conditions. 
By using the CR concentration normalized to the thermal gas density $\chi = \rho_{\textrm cr}/\rho$ they have shown that
their conditions are  equivalent to those obtained by \cite{2006MNRAS.367..113P}. %
They derived the corresponding Roe-type MHD solver \citep{1981JCoPh..43..357R} for the full system of MHD equations  (with the total energy including the energy of thermal gas, cosmic rays and magnetic fields) supplemented  with the evolution equation for the CR concentration. 
They found that solutions obtained for different forms for the CR energy equations, with source terms ${\boldmath v} \cdot  \boldmath \nabla p_{\mathrm cr} $ \citep{2004ApJ...607..828K}  or $-P_{\textrm cr} \boldmath{\nabla} \cdot \boldmath{v}$ \citep{2003A&A...412..331H,2012ApJ...761..185Y,DuboisCommerconBeno2016}, differ from the Riemann solution between the shock front and the contact discontinuity. They find that the Rankine-Hugoniot relation is seriously violated when the CR pressure dominates in the post-shocked gas.

\cite{2019arXiv190607200G} %
found that the standard CR two-fluid model described in terms of three conservation laws (expressing conservation of mass, momentum and total energy) and one additional equation (for the CR pressure)  cannot be cast in a satisfactory conservative form. The presence of non-conservative terms with spatial derivatives in the model equations prevents a unique weak solution behind a shock. Nevertheless, all methods converge to a unique result  if the energy partition between the thermal and non-thermal fluids at the shock is prescribed a priori. This highlights the closure problem of the two-fluid equations at shocks.
They extended the analysis further and made a comprehensive comparison of different discretization strategies. They constructed a full classification of the available discretization options for the two-fluid cosmic-ray hydrodynamics by comparison of numerical aspects of the '$p\mathrm{d}v$' and '$v\mathrm{d}p$' methods applied for  different combinations of mathematically equivalent energy equations. They compared the cases of the gas energy equation with the CR energy equation, the total energy equation with the CR energy equation and total energy equation plus CR entropy equation in two variants including the operator split and unsplit method. After extensive numerical testing, they found that the numerical results are consistent for various reconstruction schemes only with the '$p\mathrm{d}v$' method in a fully unsplit fashion with $v$ and $p_{cr}$  %
computed from the HLL Riemann solver values of momentum and density states.

\subsubsection{CR injection in SN shocks}
A separate issue is how to properly inject CRs in astrophysical shocks. The standard approach is to follow the assumption of Diffusive Shock Acceleration theory (DSA), relating the injection efficiency of CRs at shocks to the Mach number of the shock \citep{2007APh....28..232K}. They find that the shock Mach number can be determined  using Rankine-Hugoniot boundary conditions across shocked cells \citep{2000ApJ...542..608M,2003ApJ...593..599R,2008ApJ...689.1063S,2009MNRAS.395.1333V}.
\citet{2012MNRAS.421.3375V} implement an algorithm for CR energy injection in the ENZO code and apply an approach based on the differences of the gas and CR pressure between cells.
Their algorithm  selects candidate shocked cells by requiring that the flow in the cell is converging: $\nabla v < 0$. Then the 3D distribution of cells around the candidate cell is analyzed along the three axes, and it is checked that the gas temperature $T$, and the gas pseudo-entropy, $S = p \rho^{-\gamma}$  change in the same direction, $\nabla S \cdot \nabla\rho > 0 $ \citep{ 2003ApJ...593..599R,2008ApJ...689.1063S,2009MNRAS.395.1333V}.
The gradient of the temperature then sets the candidate post-shock and pre-shock cells. The Mach number is then evaluated from the information of the pressure jump between cells given by the Rankine– Hugoniot condition
\begin{equation}
\frac{P_2}{P_1} = \frac{2 \gamma M^2 - \gamma +1}{\gamma +1}
\end{equation}

If the shock  Mach number is known the energy flux of shock accelerated protons can be computed as
\begin{equation}
\phi_{{\textrm CR}} = \eta(M) \frac{\rho_u c_s^3M^3}{2}, 
\end{equation}
where $\rho_u$ is the comoving upstream gas density and $\eta(M)$ is a function of $M$ \citep{2007APh....28..232K}.
The injected energy density of CRs within each cell results from integration of the CR energy flux over time step
\begin{equation}
e_{{\textrm CR}} = \frac{\phi_{{\textrm CR}}}{\rho} \frac{\mathrm{d}t}{\mathrm{d}x},
\end{equation}
where $\rho$ is gas density in the cell, $\mathrm{d}x$ is the cell size and $\mathrm{d}t$ is the timestep. The method has been further extended 
by inclusion of  the effects of the upstream CR-to-thermal ratio by \cite{2013ApJ...764...95K} and the magnetic obliquity by \cite{paisetal2018}.

\cite{2017MNRAS.465.4500P} presented new methods to integrate the CR evolution equations coupled to magneto-hydrodynamics on an unstructured moving mesh of the parallel AREPO code for cosmological simulations. They included an effective recipe for diffusive shock acceleration of CRs at resolved shocks based on \citet{SchaalSpringel2015} and at supernova remnants in the interstellar medium. CR losses are included in terms of Coulomb and hadronic interactions, with CR streaming neglected. The accuracy of their model was demonstrated using simulations of plane-parallel shock tubes that were compared to exact solutions of the Riemann shock tube problem for a composite of CRs and thermal gas with effective CR acceleration. Capabilities of the new algorithm have been verified in large scale cosmological simulations and in simulations of galaxy formation.

\subsection{Unified Alfv\'en wave regulated CR transport}

One of the first attempts to incorporate the Alfv\'en wave regulated CR transport introduced in Sect. \ref{sect:streaming}
in numerical solutions of CR propagation was undertaken  by \cite{1991A&A...245...79B} in studies of galactic wind driving mechanisms. Their system of equations, following the analytical work by \cite{1982A&A...116..191M}, included the set of stationary MHD equations extended with the CR propagation (energy balance) equation and with a separate equation for the energy density of the fluctuating field of Alfv\'en waves propagating down the CR pressure gradient.  A further extension of the model by \cite{2012A&A...540A..77D} discussed time dependent non-stationary solutions of galactic winds.

\cite{Sharma_2010} incorporated CR energy in the form derived by \cite{1982A&A...116..191M}
\begin{equation}
\pder{P_\mathrm{cr}}{t}{} +  \bm{\nabla} \cdot (\bm{v} P_\mathrm{cr}) + \bm{\nabla} \cdot \left( \frac{4}{3} P_\mathrm{cr} \bm{v_{s}} \right)   
= -P_\mathrm{cr} \frac{\bm \nabla \cdot \bm{v}}{3} - \frac{\left| \bm{v}_s \cdot \nabla P_\mathrm{cr} \right|}{3},  
\label{eq:streaming-sharma}
\end{equation}
where $P_\mathrm{cr} $ is the cosmic ray pressure,  $\bm{v}$ is the plasma velocity and  
\begin{equation}
\bm{v}_{s} = - \bm{v}_a \  \frac{ \bm{b}  \cdot \nabla P_\mathrm{cr}}{|{\bm b}\cdot \nabla \Pcr |}
\label{eq:stream_vel}
\end{equation} 
is the streaming velocity with $\bm{v}_a = \frac{ \bm{B}}{\sqrt{4\pi \rho}}$ representing Alfv\'en  velocity directed down the CR pressure gradient and ${\bm b} = {\bm B}/ | {\bm B} |$ is the magnetic unit vector. Equation (\ref{eq:streaming-sharma}) describes the streaming of  CR gas along the magnetic field direction down the gradient of its pressure.

Equation~(\ref{eq:streaming-sharma}) is challenging for numerical algorithms  because the standard upwind time-evolution algorithms with a CFL-limited time-step result in spurious oscillations at the grid scale. These oscillations arise due to unphysically large flux gradients near the minima and maxima and propagate through the computational domain. Using a hydro-based CFL ($\Delta t=C \Delta x/v_s$) leads to spurious oscillations which originate from the local extrema even for a smooth initial condition. Monotonicity can be achieved at high computational price for an upwind update by setting the timestep $\Delta t \propto \Delta x^3$. Numerical solutions of Eqn.~(\ref{eq:streaming-sharma}) are unstable even with implicit methods and  regularization of the equation by an additional smoothing term, resembling incorporation of explicit viscosity in Euler's equations is  necessary. 
\cite{Sharma_2010} proposed to replace the discontinuous streaming velocity in Equation.~(\ref{eq:streaming-sharma}) by its smooth approximation $\tanh(v_s/\epsilon)$, where $\epsilon$ is a small parameter. This operation introduces an additional diffusive term to the CR propagation equation with the diffusion coefficient  $v_s/\epsilon$ and numerical stability is achieved for the timestep limitation given by
\begin{equation}
\Delta t \leq \frac{\Delta x^2 \epsilon}{2 v_s}
\end{equation}
It has been demonstrated that stable numerical solutions are achievable due to the regularization.

\begin{figure}
\includegraphics[width=\columnwidth]{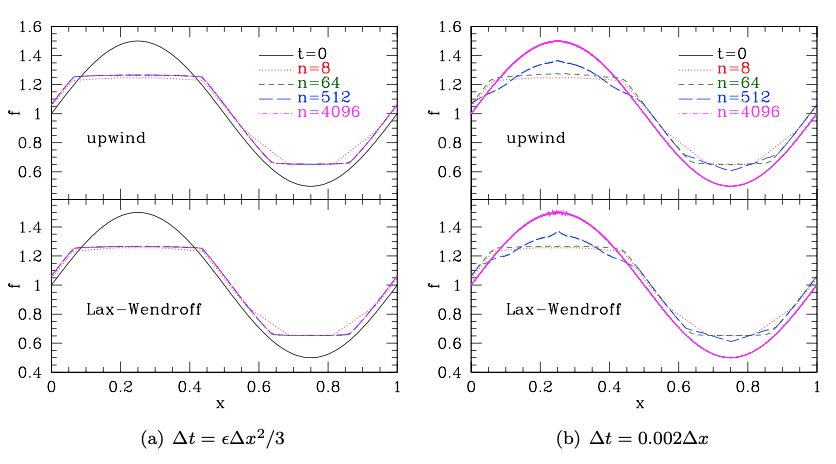}
\caption{The figure shows the flattened maxima typical for the  propagation of CRs down their gradient along magnetic field direction. Upper and lower panels show different integration schemes and different timestep choices. Different lines represent different resolutions of the 1D grid adopted for the tests. \citep[See][Fig 4.2 for details]{Sharma_2010}.
}
\label{fig:streaming-Sharma-2011}
\end{figure}

\cite{2012MNRAS.423.2374U} and later on \cite{2019A&A...631A.121D} showed that streaming can be recasted into a diffusion term and be solved with an implicit solver for diffusion. The  streaming  transport  term in the CR energy equation
$\bm{\nabla} \cdot [(\ecr + \Pcr)  {\bm v}_s]$ becomes 
\begin{equation}
\bm{\nabla} \cdot {\bm F}_{cr,s} = \bm{\nabla} \cdot  \left[- \frac{(\ecr + \Pcr)|B|}{{|\bm b}\cdot \nabla \ecr | \sqrt{4 \pi \rho}}   {\bm b} ({\bm b}\cdot \nabla \ecr )\right] = \bm{\nabla} \cdot  \left[-D_s {\bm b}({\bm b}\cdot \nabla \ecr) \right]
\end{equation}
and can be interpreted as anisotropic diffusive term, where $D_s$ is the diffusion coefficient corresponding to the streaming of CRs along magnetic field lines. The streaming-diffusion process has been modeled with an anisotropic implicit diffusion algorithm and is an interesting alternative to the one proposed by \cite{Sharma_2010}. Characteristic for the solutions obtained with the two different approaches are the flattened extrema apparent in numerical tests, such as those initiated with sinusoidal perturbations (see Fig~\ref{fig:streaming-Sharma-2011}).
 
To avoid the ad hoc smoothing \cite{2018ApJ...854....5J} proposed a new numerical algorithm based on the two-moment methods of radiative transfer under the reduced speed of light approximation. The system of equations for the propagation of CR energy density and flux reads

\begin{eqnarray}
\pder{\cren_\mathrm{cr}}{t}{} + \bm{\nabla} \cdot \bm{F}_\mathrm{cr} = (\bm{v} + \bm{v}_s) \cdot  \left(\bm{\nabla} \cdot \bm{P}_\mathrm{cr}  \right) + Q, \nonumber \\
\frac{1}{V_m^2} \pder{\bm{F}_\mathrm{cr}}{t}{} + \bm{\nabla} \cdot \bm{P}_\mathrm{cr} = - \sigma_\mathrm{cr} \cdot \left[ \bm{F}_\mathrm{cr} - v\cdot \left( E_\mathrm{cr} \bm{I} +   \bm{P}_\mathrm{cr} \right)  \right],
\label{eq:streaming-jiang-oh}
\end{eqnarray}
where $V_m$ is the maximum speed at which CRs can propagate which,  for practical reasons, is reduced with respect to the speed of light $c$. The authors demonstrated that their results are not sensitive to the exact value of $V_m$ as long as it is much larger than the maximum Alfv\'en and flow velocities in the simulations. On the other hand $V_m$ can be assumed much smaller than the speed of light, which is important for an effective mitigation of the CFL condition. 
Essential advantage of the method is that it ensures a stable evolution for isotropic and field aligned streaming and diffusion without any regularization procedure. The timestep given by the standard Courant condition  scales linearly with resolution. The full set of equations consists of the standard set of conservative MHD equations  extended with the set of the two Eqn.~(\ref{eq:streaming-jiang-oh}) and with additional CR-related source terms in the momentum and energy equations. The  difference with respect to the approach by  \cite{Sharma_2010} is that the interaction coefficient (inverse CR diffusion coefficient) $\sigma_\mathrm{cr}$ becomes zero when $\bm{\nabla} P_\mathrm{cr}$ approaches zero. In this regime the term $\frac{1}{V_m^2} \pder{\bm{F}_\mathrm{cr}}{t}{}$ becomes important, which is the main difference between these two approaches.
The flat extrema in the spatial CR distribution can be understood as the result of CR streaming outwards until the flat top develops. CRs cannot stream further to produce an inverted profile because that would require CRs to stream up their gradient.  An important element of the approach is the closure relation between the CR pressure tensor and the CR energy density. \cite{2018ApJ...854....5J} assume that $\bm{P}_\mathrm{cr} = \bm{I }\cren_\mathrm{cr}/3$. This assumption is correct only when CRs are strongly coupled to the gas, but in the case of strong wave dumping CRs may stream freely and the CR pressure can be very anisotropic. 
The algorithm has been implemented in the publicly available MHD code Athena++ \citep{2020arXiv200506651S}.

\cite{2019MNRAS.485.2977T} developed a new macroscopic transport theory, which includes both  diffusion and streaming of CRs  in the selfconfinement picture. They incorporated resonant excitation of  Alfv\'en waves through the gyroresonant instability by CRs streaming along magnetic field lines. CR scattering off these waves modulates the macroscopic CR transport. Their approach relies on the  system of three equations including an equation for the CR energy density $\cren_\mathrm{cr} $ and its  flux density  $F_\mathrm{cr} $  in the Eddington approximation of radiative transport and the equation describing the evolution of the energy density of Alfv\'en waves ${\cren_{a,\pm}} $ excited by the streaming CRs.  The $\pm$ signs denote co- and counter-propagating waves with respect to the large scale magnetic field. The Alfv\'en-wave regulated CR transport equations (eqs. (10-16) in \cite{2019MNRAS.485.2977T}) in the fluid (comoving) reference frame are 
\begin{eqnarray}
\lefteqn{\pder{{\cren_\mathrm{cr}}}{t}{} + \bm{\nabla} \cdot \left[ \bm{v} (\cren_\mathrm{cr} + P_\mathrm{cr}) + \bm{b} F_\mathrm{cr}   \right] =} \nonumber \\
&& \hspace {3cm} \bm{v} \cdot \bm{\nabla} P_\mathrm{cr}  - \bm{v}_a \cdot \bm{g}_\mathrm{gri,+} + \bm{v}_a \cdot \bm{g}_\mathrm{gri,-} ,\\
\lefteqn{\pder{(f_\mathrm{cr}/c^2)}{t}{} + \bm{\nabla} \cdot \left( \bm{v} F_\mathrm{cr}/c^2   \right) + \bm{b} \cdot \bm{\nabla} P_\mathrm{cr} =}\nonumber \\
&&  \hspace {3cm} - \left( \bm{b} \cdot \bm{\nabla} \bm{v}  \right) \cdot \left(   \bm{b} F_\mathrm{cr}/c^2   \right) 
-\bm{b} \cdot \left( \bm{g}_\mathrm{gri,+} + \bm{g}_\mathrm{gri,-}    \right) , \\
\lefteqn{ \pder{{e_{a,\pm}}}{t}{} + \bm{\nabla} \cdot \left[ \bm{v} ({e_{a,\pm}} + P_{a,\pm} ) \pm v_a \bm{b} {e_{a,\pm}}   \right] =} \nonumber \\
&& \hspace {3cm}  \bm{v} \cdot \nabla P_{a,\pm}  \pm \bm{v}_a \cdot \bm{g}_\mathrm{gri,\pm} - Q_{\pm}, 
 \end{eqnarray}
where  $\cren_\mathrm{cr}$ and  $F_\mathrm{cr}$, are measured with respect to the comoving (fluid) frame and the Alfv\'en wave energy density $\cren_{a,\pm}$ is measured in the lab frame, $\bm{b} = \bm{B}/B$ is the unit vector directed along magnetic field and
\begin{equation}
\bm{g}_{\textrm{gri,}\pm}  = \frac{\bm{b}}{3 \kappa_{\pm}} \left[F_\mathrm{cr} \mp v_a(\cren_\mathrm{cr} + P_\mathrm{cr} )\right] 
\end{equation}
are the Lorentz forces due to small-scale magnetic field fluctuations of Alfv\'en waves.

The subsystem is closed by the grey approximation for the CR diffusion coefficients linking the diffusion of CR energy density directly to the energy density of Alfv\'enic turbulence
\begin{equation}
\frac{1}{\kappa_\pm} = \frac{9\pi}{8} \frac{\Omega}{c^2} \frac{e_{a,\pm}}{2\epsilon_B} \left(1 + \frac{2 v_a}{c^2}  \right),
\end{equation}
where $\Omega = Ze B/(\gamma m c)$ is the relativistic gyrofrequency of a CR population with charge $Ze$ and Lorentz factor $\gamma$.

The scattering rate of CRs particles depends on the Alfv\'en wave energy density, so this energy density has to be dynamically evolved together with the propagation of CRs.  The transport of Alfv\'en  waves depends on several damping processes such as non-linear Landau damping, ion-neutral damping, turbulent and linear Landau damping as well as by sub-Alfv\'enically streaming CRs. %

The CR-Alfv\'{e}nic subsystem is coupled to the set of ideal MHD equations in a manner ensuring energy and momentum conservation in the non-relativistic limit of MHD
\begin{equation}
  \label{eq:conservation}
  \frac{\partial \vec{U}}{\partial t} + \vnabla\bcdot \tensor{F}
  = \vec{S} ,
\end{equation}
with the vector of conserved variables $\vec{U}$, the fluxes $\tensor{F}\equiv\tensor{F}(\vec{U})$, and the source terms $\vec{S}$. The three vectors can be described as  %
\citep[][]{2019MNRAS.485.2977T}
\begingroup
\renewcommand*{\arraystretch}{1.5}
\begin{eqnarray*}
  \label{eq:matrix}
  \vec{U} = \left( \begin{array}{c} \rho \\ \rho \vektor{v} \\ \cren \\ %
  \vektor{B} \end{array} \right),
  \ \ \ \
  \tensor{F} = \left( \begin{array}{c} 
      \rho \vektor{v} \\
      \rho \vektor{v} \vektor{v}^{\mathrm{T}} + P\bm{1} - \vektor{B} \vektor{B}^{\mathrm{T}}/(4\pi) \\
      (\cren + P) \vektor{v} - \vektor{B} \left( \vektor{v} \bcdot \vektor{B} \right)/(4\pi) \\
      \vektor{B} \vektor{v}^{\mathrm{T}} - \vektor{v} \vektor{B}^{\mathrm{T}}
    \end{array} \right),
\end{eqnarray*}
\begin{equation}
  \vec{S} = \left( 
    \begin{array}{c} 
      0 \\
      \mathbf{g} \\
      \mathbf{u \cdot g} + Q_{+} + Q_{-} \\ %
               \mathbf{0}
    \end{array} \right).
\end{equation}
Here, $\rho$ is the mass density, $\vektor{v}$ is the velocity, and  $Q_{+}$,  $Q_{-}$ are the source terms of thermal energy due to Alfv\'en wave energy losses. For the ion-neutral damping $Q_{\textrm{in}\pm} = \Gamma_{\textrm{in}} e_{a,\pm} $, where $\Gamma_{\textrm{in}}$ is the damping rate for Alfv\'en waves due to ion-neutral collisions.
For the nonlinear Landau damping $Q_{\textrm{nll}\pm} = \alpha e_{a,\pm}^2$, where the interaction coefficient $\alpha = \frac{\sqrt{\pi}}{8}\frac{v_{\textrm{th}}}{e_B} \langle k \rangle$ is a function of thermal velocity, magnetic field energy density and an averaged wavenumber.
The turbulent damping and linear Landau damping lead to $Q_{\textrm{turb+ll}\pm} = (\Gamma_{\textrm{turb}}+\Gamma_{\textrm{ll} }) e_{a,\pm}$ with turbulent damping rate given by $\Gamma_{\textrm{turb}} \simeq v_a \sqrt{k_{\parallel,\textrm{min}}  k_{\textrm{mhd,turb}}}$, 
where  $k_{\textrm{mhd,turb}}$ is the wavenumber at which the large scale MHD turbulence is driven and $k_{\parallel,\textrm{min} }\sim 1/r_L$,  $r_L$ is  the Larmor radius \citep[details see][and references therein]{2019MNRAS.485.2977T}.

Symbol  $\vektor{g}$ represents forces between CRs, Alfv\'en waves and the thermal gas:
\begin{equation}
\bm{g} = \bm{g}_{\textrm{Lorentz}} + \bm{g}_{\textrm{ponder}} + \bm{g}_{\textrm{gri,+}} + \bm{g}_{\textrm{gri,-}},
\end{equation}
where 
\begin{eqnarray}
\bm{g}_{\textrm{Lorentz}}  &=& - (\bm{1} - \bm {b \ b})  ) \cdot  \bm{\nabla}_\perp P_{\textrm{cr}},\\
\bm{g}_{\textrm{ponder}}  &=& - \bm{\nabla} (P_{\textrm{a},+} + P_{\textrm{a},-})
\end{eqnarray}
are the Lorentz force due to the large-scale magnetic field and the ponderomotive force respectively, and $P_{\textrm{a},+} + P_{\textrm{a},-} $ are the ponderomotive pressures.

  The thermal and CR energy densities are given by $\cren_\mathrm{th}$ and $\cren_\mathrm{cr}$, respectively. The total pressure $P$ and the total energy density excluding CRs are 
\begin{eqnarray}
  \label{eq:P_eps}
  P &=& P_\mathrm{th} + \frac{\vektor{B}^2}{2}, \\
  \cren &=& \cren_\mathrm{th} + \frac{\rho\vektor{v}^2}{2} + \frac{\vektor{B}^2}{2}.
\end{eqnarray}
The system of the equations is closed using an equation of state for both the thermal as well as the CR fluid,
\begin{eqnarray}
  \label{eq:EoS}
  P_\mathrm{th} &=& (\gamma_\mathrm{th}-1)\,\cren_\mathrm{th}, \\
  P_\mathrm{cr} &=& (\gamma_\mathrm{cr}-1)\,\cren_\mathrm{cr}.
\end{eqnarray}
The adiabatic indices are usually set to the canonical value of $\gamma_\mathrm{th}=5/3$ and to $\gamma_\mathrm{cr}=4/3$ implying a fully relativistic fluid.

\cite{2019MNRAS.485.2977T} presented  numerical solutions of the new CR-Alfv\'{e}n wave subsystem in one spatial dimension along the magnetic field. The  inclusion of the nonequilibrium kinetic   effects  in the Alfv\'{e}n wave propagation  advances significantly the  treatment of CR propagation  over the approaches assuming a fixed  Alfv\'en wave background. One of the tests reported by the authors exhibits the expected property of the system  that the non-linear Landau damping process reduces the wave energy, increases the CR flux density and makes the CR transport more diffusive. The formalism itself does not depend on free parameters, since the CR diffusion and the wave Landau damping coefficients are given by MHD quantities and the characteristic gyrofrequency of CR population in the grey approximation. The only tunable parameter -- the reduced speed of light -- is chosen so that the solution does not depend on its specific value. 

The aproach by \cite{2019MNRAS.485.2977T} recovers the equations of \cite{Sharma_2010} in the steady-state limit.
The major difference between the formalisms of  \cite{2018ApJ...854....5J} and \cite{2019MNRAS.485.2977T} is that the diffusion coefficient of CRs in the first case is reconstructed from the distribution of $\cren_\mathrm{cr}$
\begin{equation}
\kappa = \kappa_0 + \left|  \pder{p_\mathrm{cr}}{x}{}    \right|^{-1} v_a (\cren_\mathrm{cr} + P_\mathrm{cr}),
\end{equation}
while in the latter case it is related to the energy density of Alfv\'en waves and may be influenced by various damping processes, such as Landau damping, inaccessible directly in the former approach.
The results obtained with the method by \cite{2018ApJ...854....5J} are almost identical to those obtained with the \cite{2019MNRAS.485.2977T} 
in a test case  initiated with a Gaussian profile of CR energy density, however a similar test involving the Gaussian profile superposed to an initial uniform CR background, shown in Fig.~\ref{fig:tp19_fig6} indicates  advantages of  the separate treatment of  Alfve\'n wave component in the method proposed by \cite{2019MNRAS.485.2977T}.

\begin{figure}
\includegraphics[width=\textwidth]{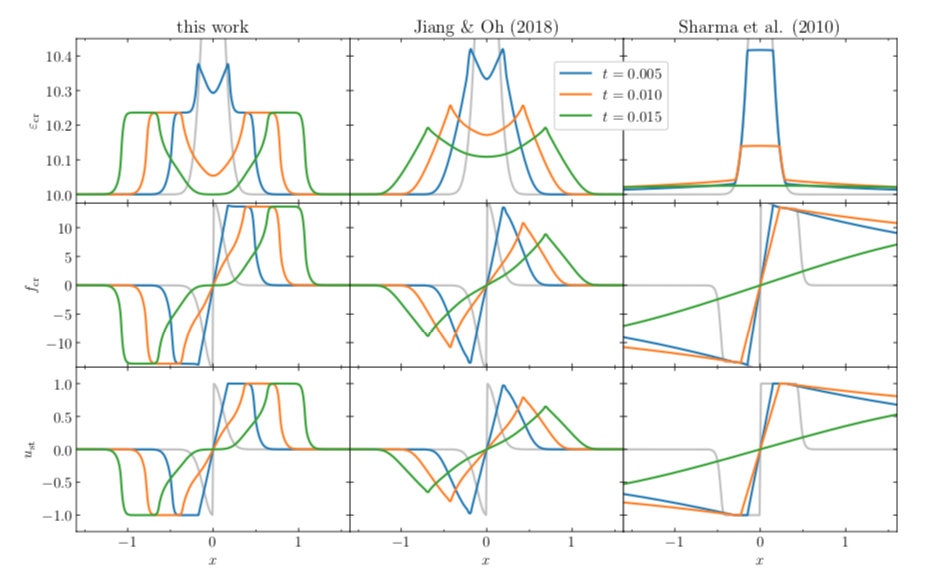}
\caption{Comparison of the time evolution of an isolated Gaussian profile of $\cren_\mathrm{cr}$ on top of a homogeneous CR distribution in the approaches by  \cite{2019MNRAS.485.2977T}, \cite{2018ApJ...854....5J} and by \citep{Sharma_2010}. Profiles of the CR energy distribution, energy flux and streaming spead reveal signifficant differences. }
\label{fig:tp19_fig6}
\end{figure}

Another interesting alternative for the direct incorporation of the streaming process on large galactic scales has been proposed by
\cite{2020MNRAS.493.2817K}
who develop theoretical models of CR propagation, relating CR diffusion to  the streaming process in magnetic field line random walk (FLRW). They describe the FLRW as the macroscopic diffusion coefficient to indicate that (1.) the averaging is done over scales larger than the coherence length of MHD turbulence, (2.) the process relies on the Alfv\'en wave regulated transport, which arises from the first-order term in distribution of pitch angles, while diffusion of CR relative to the bulk flow represents a second-order microscopic diffusion. They present an explicit formula for the CR parallel diffusion coefficient  (their Eqn. 20.) and show its dependence on energy (their Fig 2.) for three starbust galaxies and the Milky Way.  They apply the computed diffusion coefficient to a simple model of CR escape and loss, and show that the resulting $\gamma$-ray spectra are in good agreement with the observed spectra of the starbursts NGC 253, M82, and Arp 220. The model reproduces relatively hard GeV $\gamma$-ray spectra and softer TeV spectra without the need for any fine-tuning of advective escape times or the shape of the CR injection spectrum.

\subsection{Momentum-dependent diffusion-advection models} \label{sect:theory-momentum-dependent}

Up to now we have discussed numerical treatments of the diffusion-advection equation~(\ref{eq:cr_transp}) integrated over momentum space which leads us to the two-fluid approaches applied for the combination of a thermal fluid and CRs treated as a relativistic fluid.
Neglecting  the evolution of the whole CR population in momentum space is a significant simplification. 

However, the  Fokker-Planck equation~(\ref{eq:cr_transp}) can be also discretized and solved numerically in both spatial and momentum spaces to determine a time-evolution of the distribution function $f(\bm{x}, p)$ in a way similar to the approaches implemented in GALPROP \citep{1998ApJ...509..212S}, PICARD \citep{2014APh....55...37K}, DRAGON2 \citep{2017JCAP...02..015E}, but with  dynamical coupling to the thermal plasma taken into account.

Including the additional momentum dimensions in full generality becomes computationally demanding because the space of independent variables expands from three to six dimensions. A practical option is to reduce the problem to four dimensions by assuming that the distribution function is isotropic. %
A standard discretization with piecewise constant values for $f$ requires relatively high spectral resolution in order to obtain accurate results. Because of a typically high dynamical range of the power-low distribution function a few hundred bins are necessary to achieve a reasonable accuracy of the numerical integration \citep[see e.g.][]{2019MNRAS.488.2235W,2020MNRAS.491..993G}.

\citet{1987MNRAS.225..399F} and \citet{1987MNRAS.225..615B}  used partial differential equation solvers to investigate the time-dependent structure of plane parallel shocks propagating parallel to the large-scale magnetic field. 
They solved the CR transport equation~(\ref{eq:cr_transp}) in momentum space, together with the transport of gas and CRs in one spatial dimension. 
They used first-order Godunov methods to evolve the gas in a conservative manner together with a non-conservative treatment of the CR  population represented by the function $g=p^4 f$ on a uniform grid of the independent variable $y=\ln p$. The effectively smaller dynamical range of $g$ resulted  in a possibility to reduce resolution of the momentum grid. These authors applied a Lagrangian formalism for advection and diffusion of the CR spectrum and used the  Crank-Nicholson method. They  included injection of particles at the shock position and assumed a momentum-dependent CR diffusion coefficient.

Another way to reduce the computational cost is to consider CRs as a passive component needed only for observational diagnostics of the simulated object. In this case a smaller spatial resolution of synthetic images is usually sufficient, therefore the spectrum can be evolved for a relatively small set of Lagrangian particles. If  dynamical impact of CRs on the thermal fluid is to be taken into account, the spectrum is needed for each computational element of the spatial grid. This would imply the  need to process hundreds of momentum bins on multi-dimensional spatial grids leading to very high computational costs of numerical models.

To mitigate the computational demands several methods have been proposed to solve the Fokker-Planck equation using a piecewise power-law representation. The currently popular methods can be classified as one-moment and two-moment approaches, involving number density $\ncr$ and/or energy density $\ecr$ moments of the distribution function $f$ defined in Sect.~\ref{sect:selfc_eqns}.

\subsubsection{One-moment approaches}

To achieve a more accurate evolution of the spectrum with a lower number of momentum bins one needs an accurate representation of the spectrum within a single bin. Since the  observed spectra of CRs %
are power-laws to a good approximation, a natural choice is the numerical representation by the piecewise power-low distribution function.
\citet{1999ApJ...511..774J}  assumed a piece-wise power-law electron spectrum of the type
\begin{equation}
f(p) = f_{i-1/2} \left(\frac{p}{p_{i-1/2}}\right)^{-q_i},
\end{equation}
where  edges of $i$-th bin are placed at $p_{i-1/2}$ and the logarythmic slope $q_i$ is attributed to  bin centers. 
They used the number density as the dependent variable representing CRs
\begin{equation}
n_i = \int_{p_\imh}^{p_\iph} 4\pi p^2 f(p) dp 
\end{equation}
to obtain the number density moment of the CR discretized transport equation~(\ref{eq:cr_transp}).
They noted that only a few bins may capture the basic structure of the spectrum. They assumed a continuous spectrum at the bin boundaries. By imposing  the same values of the slope $q_i$ for two lowest bins in the log of momentum space one can solve, with the use of continuity assumption,  the system of equations recursively for the entire spectrum. \\

The method has been combined with the ideal MHD equations for an adiabatic plasma, solved using the ZEUS-3D Eulerian code \citep{StoneNorman1992a, StoneNorman1992b,1994ApJ...429..748J}.  The advection equation for the electron number density was solved by the standard fluid transport algorithm and the adiabatic compression term  was updated implicitly for the time-centered number density of electrons in each momentum bin. The CR electrons were treated as a passive population without a significant pressure affecting the thermal plasma component. The diffusion term in the CR propagation and the synchrotron losses were ignored. The code has been used to study the MHD evolution of a Type Ia supernova remnant with CR electrons and for SN interactions with clouds in two spatial dimensions. %

 Using the piecewise-powerlaw approximation within another code \cite{1999ApJ...512..105J}  solved the  equations of ideal nonrelativistic magnetohydrodynamics (MHD) in cylindrical coordinates.
 \citep{Tregillis_2001} used the method combined with 3D MHD simulations of electron transport in radio-galaxies.
 The code is an MHD extension of the \citet{Harten1983} conservative, second-order finite difference  total variation diminishing scheme, as detailed in \citet{RyuJones1995,RyuJonesFrank1995, RyuEtAl1998}. 
 The electron transport equation was derived from the general momentum-dependent diffusion advection equation~(\ref{eq:cr_transp}) with the effects of spatial diffusion neglected. To properly treat the particle density at shock discontinuities their algorithm evolved the normalized concentration of relativistic particles  $b_i = n_i/\rho$, where $\rho$ is the  gas density. \\

The one-moment piece-wise power-law method proposed by \citep{1999ApJ...511..774J} is relatively straightforward and computationally inexpensive, however, there is a fundamental problem with a continuous description of $f$ in particular for a locally varying spectrum.
Numerical tests \citep{2020MNRAS.491..993G}  show that if  energy is injected at one part of the spectrum, the continuity assumption enforces changes of the local slope across the entire spectrum. The resulting continuous representation then alternates between a concave and a convex spectrum.  Artificial oscillations superposed to the initially smooth spectrum have a tendency to grow towards high energy bins revealing a numerical instability of the scheme. To avoid these oscillations, the energy in all bins need to be adjusted. The problem is less severe for a  small number of bins, equal or less than 10.
An alternative for the continuous spectrum is the discontinuous modeling, which requires  to constrain two degrees of freedom.
Within the one-moment approach one may consider the additional restriction imposing a constant curvature of the spectrum 
through the relation $q_{i+1} - q_i = q_i - q_{i-1} $, which turns out to be suitable only for CR nucleons, but not for the fast cooling electrons \citep{2001CoPhC.141...17M}.

An interesting one-moment approach to the propagation of CR electrons,  not affected by the numerical instability described above, was proposed by \cite{2009ApJ...696.1142M} and implemented in SPEV code and followed by \cite{2018ApJ...865..144V} who incorporated their algorithm in the PLUTO code. These authors presented a hybrid framework  which describes the spectral evolution of highly energetic particles by means of (mesh-less) Lagrangian macro-particles embedded in a classical or relativistic MHD fluid. The main purpose of this work was the inclusion of sub-grid micro-physical processes at macroscopic astrophysical scales where the fluid approximation is adequate. The MHD equations were integrated by means of standard Godunov-type finite-volume schemes, while propagation of macro-particles representing CRs is described by the relativistic version of the cosmic-ray transport equation~(\ref{eq:cr_transp}).  In the numerical implementation the number density moment of the equation corresponding to Eqn.~(\ref{eqn:n})  was solved for the CR particle number density
 normalized to the fluid number density $\chi_p = n_p/n$ \citep{2018ApJ...865..144V}.
Spatial diffusion of CR particles as well as back-reaction from particle to the fluid were not included. 
The novel aspect of their approach is the Lagrangian approach in momentum space assuming that bin edge positions drift in accordance with the flow of particles in momentum space due to actual cooling and heating processes. Position changes of bin-edges were calculated by means of the method of characteristics in momentum space reducing the problem to the set of ordinary differential equations for each Lagrangian particle moving in physical coordinate space. The Lagrangian discretization in momentum space  implies that the particle number within each bins remains constant. %
 The power-law index $q$ needed for  computing of synchrotron emissivity was computed \emph{a posteriori}.

  \subsubsection{Two-moment approaches}
 
Another method proposed by  \citet{2001CoPhC.141...17M} binds the pair of distribution function parameters $(f_{i - 1/2}, q_i)$ to its two moments: number density $n_i$ and energy density $e_i$. In absence of energy loss processes (radiative and adiabatic losses) these two moments are conservative quantities and therefore they are suitable for an accurate evolution with conservative transport schemes in both the spatial and momentum domains. A summary of the two-moment piece-wise power-law method is presented in Sect.~\ref{sect:two-mom-miniati}. %
 
Using a similar approach  \citet{2005APh....24...75J} modelled the propagation of CRs including advection plus diffusion in space and advection in momentum.  The propagation equation does not include any strong cooling effects, such as the ionization or synchrotron cooling which act predominantly on one of the edges of the spectrum. Therefore the spectrum can be processed in an arbitrarily chosen range of momentum space, with the momentum boundary condition setting the particle number density to zero at the high energy end. At the low energy range of the spectrum the distribution was matched to the thermal particle distribution.
 The authors refer to the method as to “Coarse-Grained Momentum finite Volume” or “CGMV”. It extends the ideas introduced in \citet{1999ApJ...512..105J,2001CoPhC.141...17M}  for the case of CR transport across an Eulerian grid including advection and diffusion.

 \cite{2005APh....24...75J}  extended the CGMV method so that it can be applied to the treatment of fully nonlinear CR modified shocks.
The algorithm has been implemented in the CRASH code which is based on the high order Godunov-like shock tracking algorithm.
 The hydrodynamics routine in that code employs a nonlinear Riemann solver to follow shock discontinuities within the zones of an initially uniform grid. Thus, gas sub-shocks in CR-modified shocks remain discontinuous throughout a simulation, allowing CR transport to be modeled down almost to the scale of the physical shock thickness. CRASH also employs adaptive mesh refinement (AMR) around shocks in order to reduce the computational effort on the spatial grid. 

 With the aim of studying the Fermi bubbles (kpc-scale gamma-ray features centred on the Galactic Centre) \citet{2017ApJ...850....2Y} develop a new CRSPEC module for FLASH code to track the CR spectrum during MHD simulations. They construct a scenario for the Fermi bubbles resulting from AGN jet activity in the center of Milky Way. The specific features of the Fermi bubble's spectra including their spatial uniformity, shapes of the spectra and high energy cutoff at 110 GeV suggest a leptonic origin. The physical nature of the problem requires, therefore, a method to evolve the spectra of CR electrons dynamically together with the fluid-dynamical evolution of the expanding  bubble. The method they adopted to trace the CR electrons spectral evolution follows  the piece-wise power-law approach by \citet{2001CoPhC.141...17M}. They solve the two-moment set of equations for a population of CR electrons advected along trajectories of a set of tracer particles while CR diffusion neglected. They assumed a hydrodynamical evolution of the thermal component and  a fixed magnetic field taken from GALPROP models. In this sense their approach represents a intermediate step between the  CR propagation in a fixed plasma background and the self-consistent models. 

\cite{2019MNRAS.488.2235W} presented an efficient post-processing code for Cosmic Ray Electron Spectra that are evolved in Time (CREST) on Lagrangian tracer particles. The novel element of their method is the division of the overall spectrum into three parts treated with different methods. They note that for very low momenta the timescale related to the Coulomb cooling process becomes shorter than the typical time step of  MHD simulations. Similarly, for very high momenta the synchrotron and inverse Compton energy losses proceed on arbitrarily short timescales.  In order to efficiently calculate the CR electron spectrum with time steps similar to the MHD time step, they use analytical solutions for low and high momenta together with the fully numerical treatment for intermediate momenta. The situation is outlined in Figure \ref{fig:winner-etal2019-fig1}. 
\begin{figure}
\includegraphics[width=\textwidth]{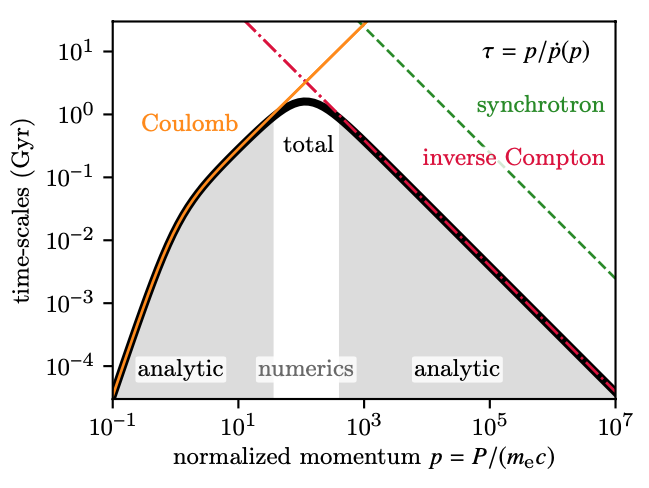}
\caption{Schematic view of characteristic time-scales for electron cooling illustrating the combination of analytical and numerical solutions proposed by \cite{2019MNRAS.488.2235W} The grey area shows the ranges where either Coulomb or inverse Compton plus synchrotron cooling dominate and where analytical solutions can be used. }
\label{fig:winner-etal2019-fig1}
\end{figure}
The analytical approach relies on the consideration of the momentum losses corresponding to the cooling terms in the CR transport equation~(\ref{eq:cr_transp}). The momentum losses integrated over a time step lead to the change of the  distribution function due to the cooling which shifts particles with from $p_\mathrm{ini}$ to $p$ within $\Delta t$ 
 \begin{equation}
f(p,t_0+\Delta t) = f(p_\mathrm{ini}(p,\Delta t), t_0) \frac{\dot{p}(p_\mathrm{ini}(p,\Delta t), t_0+\Delta t))}{\dot{p}(p,t_0)}.
\end{equation}
The cooled spectrum can be interpreted as a momentum shift of the initial spectrum at time $t_0$ multiplied with a momentum-dependent cooling factor. In the central part of the spectrum  the CR transport equation~(\ref{eq:cr_transp}) is solved with the aid of direct numerical integration relying on the operator split technique. The spectrum is calculated on a discrete momentum grid with piecewise constant values. A cubic interpolation function is used  to calculate the analytic solution after a time step $\Delta t$. 

The combination of analytical and numerical methods enables the integration of the spectra on reasonably large MHD time steps. The hybrid analytical-numerical method is implemented in the CREST module and coupled to the cosmological MHD code AREPO
\citep{Springel2010,2016MNRAS.455.1134P} with CR protons modeled as a relativistic fluid with a constant adiabatic index of 4/3 in a two-fluid approximation \citep{2017MNRAS.465.4500P}.
The coupling of CR propagation to the MHD system is done with the aid of Lagrangian tracer particles, which trace the velocity field \citep{2013MNRAS.435.1426G} and are passively advected with the gas. The CR electron transport equation is solved for each particle 
in a post-processing routine on every MHD time step. The hybrid technique enables modeling of the spectral CR electron evolution such as adiabatic expansion and compression, Coulomb losses, radiative losses: inverse Compton, bremsstrahlung and synchrotron processes, diffusive shock acceleration and re-acceleration, and Fermi-II reaccelleration. 
The authors demonstrate that the CR electron spectra are efficiently and accurately evolved in shock-tube and Sedov-Taylor blast wave simulations. This technique opens up the possibility to produce self-consistent synthetic observables of non-thermal emission processes in various astrophysical environments like SNR \citep{WinnerEtAl2020}.

A similar technique based on a combination of the numerical integration with the aid of piecewise power-law method and the analytical prescription of the Coulomb cooling has been applied by \cite{2020MNRAS.491..993G}. The analytical solution helps to overcome the excessive shortening of the integration timestep whenever the cooling timescale is significantly shorter than the hydrodynamical timescale. The numerical integration scheme incorporated the two-moment approach by \cite{2001CoPhC.141...17M}. The original algorithm  was modified in the part responsible for cooling losses, that were computed through a modification of inter-bin fluxes, rather than by inclusion of source terms (incorporated through the 'R'-term in the original approach, see section\ref{sec:algorithms}) to the evolution equation for the spectral energy density of CRs. The modification enables a more accurate computation of the spectral slope and is a less-diffusive alternative for the propagation of CR population in momentum space.
Tests of the algorithm  have confirmed that a combination of injection and cooling of CRs including Coulomb and hadronic losses shows a very good agreement with theoretical steady state spectra. 

Solutions of the CR transport equation~(\ref{eq:cr_transp}) including advection, diffusion and spectral evolution coupled with 
the MHD evolution of the thermal plasma turn out to be challenging. Up to now only  approaches restricted to a subset of physically relevant processes have been successful. The GALPROP-type codes include an extensive set of physical interactions of CR particles under the assumption of fixed magnetic field, the  CGMV method of \cite{2005APh....24...75J} is capable of modeling spectral evolution together with diffusion, advection and dynamical coupling of the CR population to the thermal plasma, but cannot cope with the fast cooling processes. Other implementations \citep{2001CoPhC.141...17M,2019MNRAS.488.2235W} engaging Lagrangian test particles include particle acceleration and various cooling processes, but ignore diffusion and dynamical coupling to thermal plasma.

Limitations of the original CGMV method become severe in the presence of fast cooling processes,
such as synchrotron emission, which  naturally tends to evacuate all particles from high energy bins, producing steep spectra near the upper cutoff. To avoid these limitations  \cite{2021ApJS..253...18O}  extended the CGMV method with movable boundaries, which change in response to CR momentum gains or losses near the extremes of the population distribution.  The extension relies on an operational definition of spectral cutoffs  which define spectral boundaries of the spectrum. 
The extension involves a special treatment of momentum bins containing spectral cutoffs. Contrary to the regular bins of fixed width, those bins have variable-width, and their outer edges coincide with spectral cutoffs. 

The CR spectra may have different cutoffs in each spatial grid cell, because local cooling conditions (magnetic and velocity fields) are generally different. Due to advective and diffusive propagation of the CRs across the spatial grid, different populations of particles mix in every cell. The essential part of the numerical problem is to estimate an effective cutoff for the mixture of different populations inflowing from neighboring cells with different cutoffs. The cutoff positions are estimated from the particle number density and energy density in the outer bins for an assumed value of an additional parameter $e_\mathrm{small}$ representing the smallest physically significant level of CR spectral energy density. Estimation of cutoff positions from the number density and energy density moments enables a numerically stable computation of power-law indexes in the active bins, especially those containing the cutoffs, what would not be possible in the case of fixed boundary conditions in momentum space.

The algorithm is designed to follow spectral evolution of  CRs  coupled with an MHD system on Eulerian grids and is particularly useful for modeling of CR electron population subject to strong synchrotron and inverse Compton cooling in the high-energy part of the spectrum. The algorithm is suitable for studies of CR electron  spectrum evolution in MHD simulations of the galactic interstellar medium.

\subsubsection{Advances in theory of CR propagation}

An important development has been to explicitly compute the trajectories of charged particles numerically in the magnetised ISM, thus going beyond the analytical approximations of e.g. the quasi-linear theory.
In principle this involves just the Lorentz force and Maxwell's equations, i.e. very basic physics.
This approach was pioneered by 
\cite{1994ApJ...430L.137G}, %
\cite{1999ApJ...520..204G}. %
The turbulent magnetic field is modelled as a superposition of modes with a power-law spectrum, using Alfv\'en waves.
The trajectories are computed for large numbers of particles, and then statistical properties--diffusion tensors etc--are calculated \citep[e.g.][]{RevilleEtAl2008}.
The diffusion coefficient perpendicular to the B-field is found to be much smaller than the parallel coefficient.
\cite{2002PhRvD..65b3002C} %
extended this approach to higher energies and a wider range of environments, including SNR, superbubbles and radio-galaxies.

\cite{2014ApJ...791...51D}  show that even simple magnetic nonuniformities combined with pitch angle scattering can enhance cross field line transport by several orders of magnitude, while pitch angle scattering is unnecessary for enhanced transport if the field is chaotic. Perpendicular  diffusion remains relatively small compared to parallel diffusion. 
\cite{2016MNRAS.457.3975S} study CR diffusion by means of particle simulations in turbulent magnetic fields. They obtain direct estimates of the diffusion tensor from test particle simulations in random magnetic fields, with the Larmor radius scale being fully resolved and find that diffusion coefficients obtained are consistent with existing transport theories that include the random walk of magnetic lines.
\cite{2017ApJ...839L..16S}  calculate cosmic-ray diffusivity in intermittent dynamo-generated magnetic fields using test particle simulations. The results are compared with those obtained from non-intermittent magnetic fields having identical power spectra. The presence of magnetic intermittency significantly enhances cosmic-ray diffusion over a wide range of particle energies. The authors demonstrate that the results can be interpreted in terms of a correlated random walk.
\cite{2018MNRAS.473.4544S} use test particle simulations, tracing the propagation of charged particles (protons) through a random magnetic field, to study the cosmic ray distribution at scales comparable to the correlation scale of the turbulent flow in the interstellar medium ($\sim100 \pc$ in spiral galaxies). They find that there is no spatial correlation between the cosmic ray number density and the magnetic field energy density.  Low-energy cosmic rays can become trapped between magnetic mirrors, whose location depends more on the structure of the field lines than on the field strength. These results are relevant for
interpreting synchrotron emission data which  often assume energy equipartition of CRs with the magnetic field energy. 

Meanwhile observation-based models of the Galactic magnetic fields and related theory are improving, and these can be input to  CR propagation studies, as addressed elsewhere in this review.

\subsection{Spatial diffusion of spectrally resolved CRs}
\label{sec:diffusion-spectral}

In contrast to advection, spatial diffusion includes an interaction of spatial and spectral changes. Being energy conserving, spatial diffusion itself does not transfer CRs in momentum space at one position in space, i.e., $\partial f/\partial p=0|_{x}$. But the transport of number and energy density to neighbouring cells depend on the spatial derivatives of $f$. Depending on the values in neighbouring fluid elements and combined with a different diffusion tensor for $n$ and $e$ will result in changes of both amplitude $f_{i-1/2}$ and slope $\slope_i$ within one spectral bin, and with that the other physical processes in the following evolution. We derive the terms for number and energy density separately. For the former one the spatial diffusion in momentum range $[p_1, p_2]$ is given by
\begin{align}
  \label{eq:ndiff}
  \partial_t n_\mathrm{diff} &= \int_{p_1}^{p_2} 4\pi\vnabla\bcdot(\tensor{D}\bcdot\vnabla f)p^2\,\dd p\notag\\
  &=4\pi\vnabla \bcdot\ekl{\int_{p_1}^{p_2} p^2 \tensor{D}\bcdot\vnabla f\,\dd p},
\end{align}
with the diffusion tensor $\tensor{D}$,
\begin{align}
  \tensor{D} &= \rkl{\begin{array}{ccc}
      D_{11} & D_{12} & D_{13}\\
      D_{21} & D_{22} & D_{23}\\
      D_{31} & D_{32} & D_{33}
  \end{array}}.
\end{align}
The components include the orientation of the magnetic fields. The components are chosen to be \citep{RyuEtAl2003},
\begin{equation}
  D_{ij} = D_{\perp}\delta_{ij} + (D_{\parallel} - D_{\perp})b_ib_j \label{eq:diff-tensor}
\end{equation}
with normalised components of the magnetic field, $b_i=B_i/|\vektor{B}|$. The dependency on momentum is expressed as
\begin{align}
\label{eq:diff-coeff-scaling}
  D_{\parallel}(p) &= D_{\parallel,10} \rkl{\frac{p}{p_{10}}}^{\alpha} ,\\
  D_{\perp}(p) &= D_{\perp,10} \rkl{\frac{p}{p_{10}}}^{\alpha},
\end{align}
with $p_{10} = 10\,\mathrm{GeV/c}$. Parallel and perpendicular diffusion coefficients with respect to the magnetic field are denoted as $D_{\parallel,10}$ and $D_{\perp,10}$. Equation~\eqref{eq:ndiff} can be solved directly by replacing the individual components yielding 
\begin{align}
  &\partial_t n_\mathrm{diff} =\notag\\
  &4\pi\,\vnabla\bcdot\,\int_{p_1}^{p_2}\,p^2\,\ekl{\begin{array}{c}
      D_{11}\partial_xf + D_{12}\partial_yf + D_{13}\partial_zf\\
      D_{21}\partial_xf + D_{22}\partial_yf + D_{23}\partial_zf\\
      D_{31}\partial_xf + D_{32}\partial_yf + D_{33}\partial_zf
  \end{array}}\,\dd p.
\end{align}
Here, $\partial_k=\partial/\partial k$ is the partial derivative with $k\in\{x, y, z\}$, so the spatial derivatives of $f$ need to be taken into account. Alternatively, we can write
\begin{align}
  \label{eq:ndiff_alt}
  \frac{\partial n_\mathrm{diff}}{\partial t} &=\int_{p_1}^{p_2} 4\pi\vnabla\bcdot(\tensor{D}_n\bcdot\vnabla f)p^2\,\dd p\notag\\
  &= \vnabla\bcdot\rkl{\skl{\tensor{D}_n}\bcdot\vnabla n_{cr}},
\end{align}
where the equation formally takes the form of a simple diffusion equation with only a modified diffusion tensor $\skl{\tensor{D}_n}$. The individual components of the diffusion tensor take the form
\begin{equation}
  \skl{D_{n,ij}} = \frac{\int_{p_1}^{p_2} p^2\,D_{ij}\,\partial_j f\,\dd p}{\int_{p_1}^{p_2} p^2 \partial_j f\,\dd p} .
\end{equation}
For $D_{ij}=D'_{ij}(p/p_{10})^\alpha$ we find
\begin{equation}
  \skl{D_{n,ij}} = \frac{D'_{ij}}{p_{10}^\alpha}\,\frac{\int_{p_1}^{p_2} p^{2+\alpha}\partial_j f\,\dd p}{\int_{p_1}^{p_2} p^2 \partial_j f\,\dd p} .
\end{equation}

In an analogous way we treat the energy density,
\begin{align}
  \label{eq:ediff}
  \partial_t e_\mathrm{diff} &= \int_{p_1}^{p_2} 4\pi\vnabla\bcdot(\tensor{D}\bcdot\vnabla f)p^2T(p)\,\dd p\notag\\
  &=4\pi\vnabla\bcdot \ekl{\int_{p_1}^{p_2} p^2T(p) \tensor{D}\bcdot\vnabla f\,\dd p},
\end{align}
with the modified diffusion coefficients
\begin{equation}
\skl{D_{e,ij}} = \frac{\int_{p_1}^{p_2} p^2T(p)\,D_{ij}\,\partial_j f\,\dd p}{\int_{p_1}^{p_2} p^2T(p) \partial_j f\,\dd p},
\end{equation}
which can be rewritten to include the scaling with momentum
\begin{equation}
\skl{D_{e,ij}} = \frac{D'_{ij}}{p_{10}^\alpha}\,\frac{\int_{p_1}^{p_2} p^{2+\alpha}T(p)\partial_j f\,\dd p}{\int_{p_1}^{p_2} p^2T(p) \partial_j f\,\dd p}.
\end{equation}

The full computation of the spectrally resolved diffusion involves many derivatives and introduces numerical errors. The usual approach of applying limiters is a very powerful method for conserved quantities. For a simple numerical representation of the particle distribution function, e.g. a piecewise constant representation, a limiter can simply be applied. But if more complex representations are chosen like piecewise powerlaws as in e.g. \citet{GirichidisEtAl2019}, the derivatives include the terms that scale with the spatial derivative of the amplitude as well as the derivative of the slope. Applying limiters to those individual terms are not simply applicable as the quantities are not conserved. A simple approximate solution to the full problem described above is derived in \cite{2005APh....24...75J}, where the number density weighted diffusion coefficient reads
\begin{equation}
D_{n_{i}}=\frac{\int_{p_{i}}^{p_{i+1}} p^{2} D \frac{\partial f}{\partial x} \mathrm{d} p}{\int_{p_{i}}^{p_{i+1}} p^{2} \frac{\partial f}{\partial x} \mathrm{d} p} \Rightarrow \frac{\int_{p_{i}}^{p_{i+1}} D f p^{2} \mathrm{d} p}{n_{i}} .
\end{equation}
The corresponding energy weighted counterpart can be expressed as
\begin{equation}
D_{g_{i}}=\frac{\int_{p_{i}}^{p_{i+1}} D \frac{\partial f}{\partial x} p^{3} \mathrm{d} p}{\int_{p_{i}}^{p_{i+1}} \frac{\partial f}{\partial x} p^{3} \mathrm{d} p} \Rightarrow \frac{\int_{p_{i}}^{p_{i+1}} D f p^{3} \mathrm{d} p}{g_{i}} ,
\end{equation}
with $g_{i}\left\langle D / p^{2}\right\rangle_{i}=\int_{p_{i}}^{p_{i+1}} p D f \mathrm{d} p$ .

\section{Numerical details}
\label{sec:algorithms}

\subsection{Numerical treatment of CR diffusion} \label{sect:diff_models}

Numerically solving diffusive CR transport faces two main challenges. First of all, the diffusion equation is an elliptic PDE, which means that the solution propagates infinitely fast, i.e. the domain of dependence is the entire domain. Ideally, one would like to solve the diffusion equation with an implicit algorithm including the entire domain. However, this is numerically demanding and approximate solutions are favoured. The solution using an explicit algorithm requires to satisfy the explicit von Neumann stability criterion with a maximum integration time step
\begin{equation}
\Delta t_\mathrm{max} = 0.5 \ C_{\indx{cr}} \frac{\Delta x^2}{D},
\end{equation}
where $\Delta x$ is the resolution of an individual cell and $D$ is the diffusion coefficient. The second complication is that the diffusion of CRs is highly anisotropic with the largest diffusion parallel to the magnetic field lines. The diffusion coefficient thus extends to a diffusion tensor, with coefficients describing the orientation of a flow with respect to the magnetic field orientation. For an explicit algorithm the time step criterion changes to 
\begin{equation}
  \Delta t_\mathrm{max} = 0.5 \ C_{\indx{cr}} \ 
  \frac{\min (\Delta x,\Delta y, \Delta z)^2}{D_{\parallel} + D_{\perp}}
\end{equation}

in the case of a dimensional split numerical scheme, where $D_\parallel$ and $D_\perp$ are the parallel and perpendicular diffusion coefficients, repsectively, and $C_{\indx{cr}} < 1 $ is the Courant number \citep[e.g.][]{2003A&A...412..331H}. 
In the case of an unsplit scheme the numerical coefficient changes to 0.3.
Whereas the implicit solution for isotropic diffusion can simply be computed using the Green's function, the anisotropic case does not have a simple form of the Green's function. Therefore, iterative methods are used.

\subsection{Anisotropic diffusion on a regular mesh} \label{sect:aniso-diff}

In this section we present details of the implementation of the anisotropic diffusion algorithm presented by \cite{2003A&A...412..331H}. 
The algorithm has been designed for the staggered mesh setup of the ZEUS-3D MHD code, with the MHD algorithm involving  the constraint transport (CT, \citealt{EvansHawley1988}) method to ensure the divergence-free evolution of the magnetic field. The staggered mesh locates the individual components of the magnetic field on different faces of grid cells: $B_x$ on $yz$ faces, $B_y$ on $xz$ faces and $B_z$ on $xy$. The CR energy density is located in cell centers and fluxes of CRs are located on cell faces. Different centering of the relevant quantities implies the need of  interpolation  ensuring numerical stability of the anisotropic diffusion algorithm.

The  CR diffusion tensor $\tensor{D}$ depends on the spatially variable magnetic field $\vektor{B}$.
Let us consider the diffusive part of the diffusion--advection equation~(\ref{eq:e_all}) with the diffusion tensor given by Eqn.~(\ref{eq:diff-tensor}):

\begin{equation}
   \partial_t e_{\indx{cr}}  + \vektor{\nabla} \cdot \vektor{F_{\indx{cr}}} = 0,\qquad 
    \vektor{F_{\indx{cr}}} = -\tensor{D} \vektor{\nabla} e_{\indx{cr}}. \label{cr-diff-eq} 
\end{equation} 
In the discrete representation,  the 3--dimensional conservation law reads
\begin{equation}
  \begin{array}{ll}
     e_{\indx{cr},i,j,k}^{n+1}=e_{\indx{cr},i,j,k}^n 
              &  -\frac{\Delta t}{\Delta x}
                 \left(F_{\indx{cr},i+\h,j,k}-F_{\indx{cr},i-\h,j,k} \right)\\
	      &  -\frac{\Delta t}{\Delta y}
                \left(F_{\indx{cr},i,j+\h,k}-F_{\indx{cr},i,j-\h,k} \right)       \\
	      &  -\frac{\Delta t}{\Delta z}
                \left(F_{\indx{cr},i,j,k+\h}-F_{\indx{cr},i,j,k-\h} \right) ,
      \end{array}
      \label{eqn:fluxes}
\end{equation}
where $e_{\indx{cr},i,j,k}^n $ and $e_{\indx{cr},i,j,k}^{n+1} $ are volume averaged CR energy densities in the cell  ${i,j,k}$, and $F_{\indx{cr},i-\frac{1}{2},j,k}$, $F_{\indx{cr},i+\frac{1}{2},j,k}$ are the CR fluxes through the left and right cell boundaries, in $x$-direction. The fluxes appearing in equation (\ref{eqn:fluxes}) should be understood as time-averaged fluxes through respective cell boundaries.
If we approximate these fluxes by their values computed at the time--level
$t^n$, then we obtain an explicit algorithm for the numerical
integration of the CR diffusion equation.
To compute  the diffusion tensor components at cell faces, we need
all components of the unit vector $\vektor{n}$, parallel to $\vektor{B}$ at each
cell face. We start with  computing the magnetic field aligned unit vectors at cell--faces perpendicular to the $x$--axis:    
\begin{equation}
\vektor{b}_{i-\h,j,k} = \frac{\vektor{\bar{B}}_{i-\h,j,k} }{|\vektor{\bar{B}}_{i-\h,j,k}| }, \end{equation}
and then do the same for the other faces.
The magnetic field component $B^x$ is already located at proper faces, but   the location of $B^y$ and  $B^z$ is different, thus an interpolation is necessary. The  linearly interpolated magnetic field vector located at $(x_{i-\h},y_j,z_k)$ is therefore      
\begin{eqnarray}
\vektor{\bar{B}}_{i-\h,j,k}=\left[B^x_{i-\h,  j,  k},     \frac{1}{4}\left(B^y_{i-1,j-\h,k}+B^y_{i,j-\h,k}+B^y_{i-1,j+\h,k}+B^y_{i,j+\h,k}\right)\right.,\; \\ \left.\frac{1}{4}\left(B^z_{i-1,j,k-\h}+B^z_{i,j,k-\h}+B^z_{i-1,j,k+\h}+B^z_{i,j,k+\h}\right)\right].\nonumber
\end{eqnarray}
The interpolation scheme for $B_y$, at the cell--face perpendicular to $x$--axis, is shown in Fig.~\ref{fig:b-avg}.

\begin{figure}
\centerline{\includegraphics[width=0.6\textwidth]{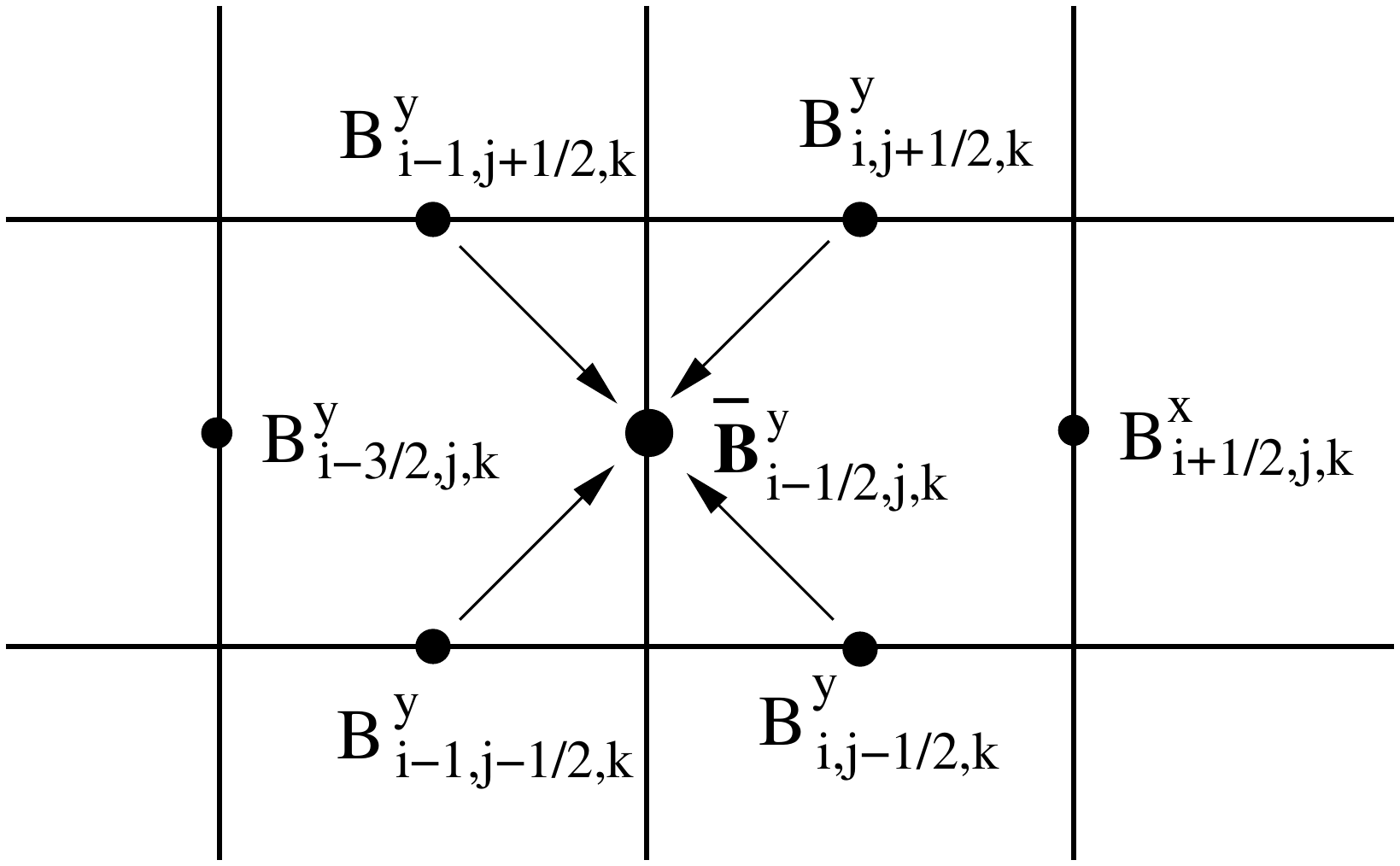}}
\caption{Interpolation of  magnetic--field components to cell--faces shown in 2D projection onto $xy$--plane.
}
\label{fig:b-avg}
\end{figure}

To compute  the diffusive fluxes of CRs across cell faces, one needs all components of the CR energy density gradient at each cell face.
The components of $\nabla {e_{\indx{cr}}}$, perpendicular to cell faces, are centered at cell--faces, e.g.:
\begin{equation}
\left(\partial_x e_{\indx{cr}} \right)_{(i-\h,j,k)} \simeq \frac{1}{\Delta x} (e_{{\indx{cr}},i,j,k} - e_{{\indx{cr}},i-1,j,k}),
\end{equation}
however, an interpolation is needed for the remaining components of CR energy--density gradient. 
The procedure follows in three steps, as depicted in Fig.~\ref{fig:ecr-grad}:

\begin{enumerate}
\item interpolation of $e_{\indx{cr}}$ to centers of cell faces (1) 
\item  computation of left-- and right finite differences of $e_{\indx{cr}}$, with respect to coordinates
parallel to cell faces, at positions (2) 
\begin{equation}
 \partial_{y,l} e_{\indx{cr}}                
 \equiv \frac{1}{2\Delta y}\left((e_{{\indx{cr}},i-1,j  ,k} {+} e_{{\indx{cr}},i ,j  ,k })  
               {-} (e_{{\indx{cr}},i-1,j-1,k} {+} e_{{\indx{cr}},i ,j-1,k})\right),
\end{equation}
\begin{equation}	   
\partial_{y,r} e_{\indx{cr}}                 
  \equiv \frac{1}{2\Delta y}\left((e_{{\indx{cr}},i-1,j+1,k} {+} e_{{\indx{cr}},i ,j+1,k})
                    {-}(e_{{\indx{cr}},i-1,j ,k}{+} e_{{\indx{cr}},i ,j  ,k})\right).  
\end{equation}
\item computation of face--centered  slopes at positions (3) monotonized with the aid of van Leer slope limiter 
\begin{equation}
\left(\partial_y {e_{\indx{cr}}}\right)_{(i-\h,j,k)}\simeq{\q\left(\partial_{y,l} e_{\indx{cr}}+\partial_{y,r} e_{\indx{cr}})  (1 + \sign(1, \partial_{y,l} e_{\indx{cr}} \ \partial_{y,r} e_{\indx{cr}})\right)}.
\end{equation}
\end{enumerate}
 We note that the  monotonization of \ $\nabla e_{\indx{cr}}$ components (step 3.), parallel to the considered cell faces, is essential for stability of the overall (CR+MHD) algorithm.
\begin{figure}
\centerline{\includegraphics[width=0.7\textwidth]{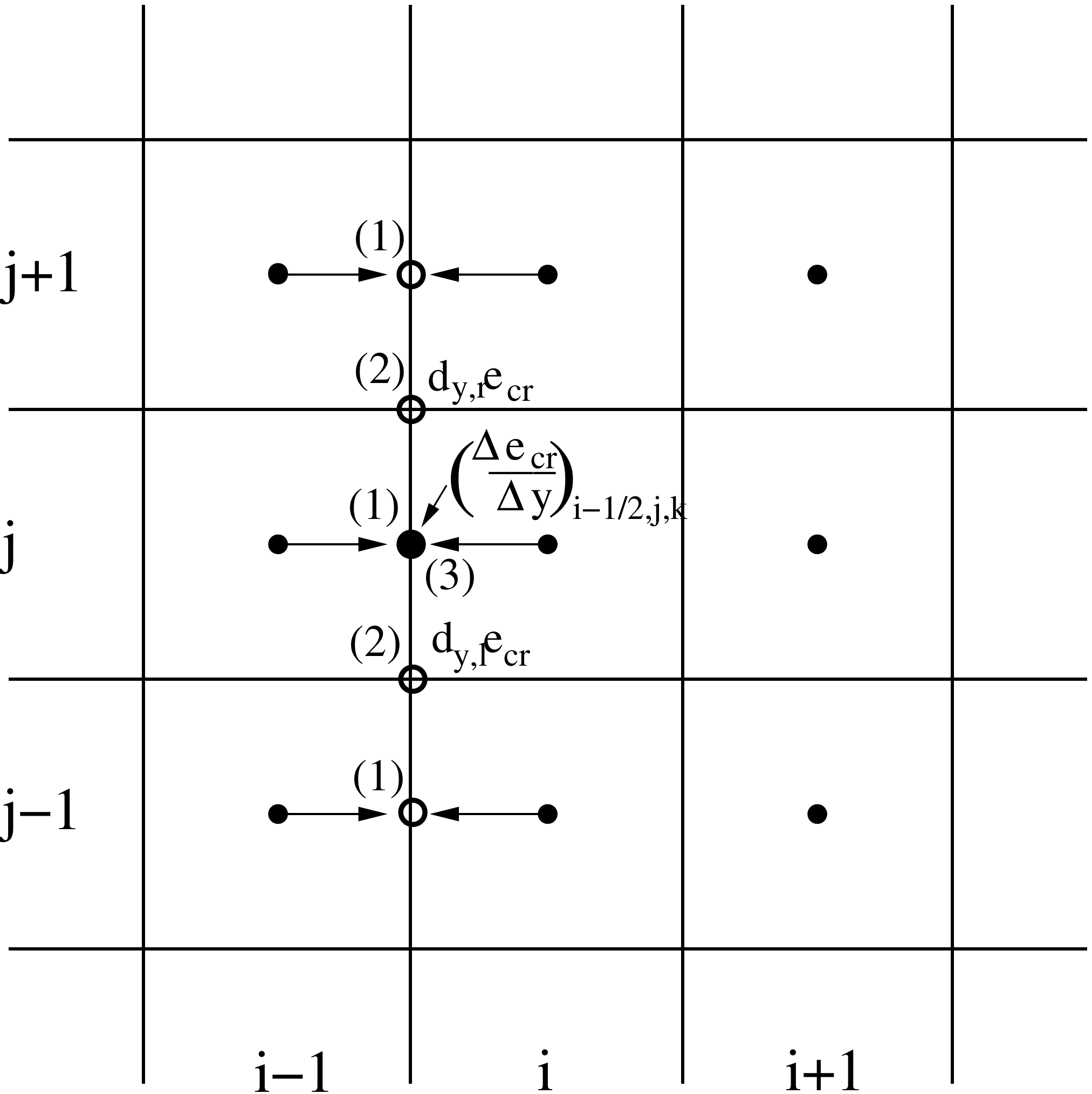}}
\caption{Computation of the monotonized $e_{\indx{cr}}$ gradient components, shown in 2D projection onto $xy$--plane.}
\label{fig:ecr-grad}
\end{figure}
The same procedure is applied to get the monotonized slope of $e_{\indx{cr}}$ in the $z$--direction.  
\begin{equation}
\left(\partial_z {e_{\indx{cr}}}\right)_{(i-\h,j,k)}\simeq\q\left(\partial_{z,l} e_{\indx{cr}}+\partial_{z,r} e_{\indx{cr}}) 
          (1 + \sign(1, \partial_{z,l} e_{\indx{cr}} \ \partial_{z,r} e_{\indx{cr}})\right).
\end{equation}
Computation of $(\vnabla e_{\indx{cr}})_{i,j-\h,k}$ and $(\vnabla e_{\indx{cr}})_{i,j,k-\h}$ follows in the analogous way.  

Another variant of transverse flux computation, designed to avoid negative CR energy densities when there is a large gradient, has been applied by  \cite{2012ApJ...761..185Y} following \cite{SharmaHammett2007}. Their implementation relies on a double use of one of the slope-limiters such as minmod, van Leer or monotonized central (MC) limiter. 

At this point we are ready to compute all CR--fluxes $\vektor{F}_{\indx{cr}} = -\tensor{D} \vnabla e_{\indx{cr}}$ at all cell--faces. We apply the directional--splitting technique to compute the total diffusive change of CR--energy density, in a single timestep.
Subsequent updates of CR--energy density proceed in  a directionally split manner:
\begin{eqnarray}
   e_{\indx{cr},i,j,k}^{n+b}=e_{\indx{cr},i,j,k}^{n+a}-\frac{\Delta t}{\Delta x} 
                \left[F_{\indx{cr},i+\h,j,k}-F_{\indx{cr},i-\h,j,k} \right]      \\ 
   e_{\indx{cr},i,j,k}^{n+c}=e_{\indx{cr},i,j,k}^{n+b}-\frac{\Delta t}{\Delta y} 
                \left[F_{\indx{cr},i,j+\h,k}-F_{\indx{cr},i,j-\h,k} \right]      \\  
   e_{\indx{cr},i,j,k}^{n+1}=e_{\indx{cr},i,j,k}^{n+c}-\frac{\Delta t}{\Delta z} 
                \left[F_{\indx{cr},i,j,k+\h}-F_{\indx{cr},i,j,k-\h} \right]      
\end{eqnarray}
where $e_{\indx{cr},i,j,k}^{n+a}$ is CR-energy density after the advection step, $e_{\indx{cr},i,j,k}^{n+b}$ and $e_{\indx{cr},i,j,k}^{n+}$ are CR-energy densities after these quantities have been updated in $x$ and $y$-directions, respectively

For more recent models of finite difference and finite volume algorithms for anisotropic diffusion we would like to point to \citet{vanEsEtAl2014} and \citet{vanEsEtAl2016}.

\subsubsection{A semi-implicit extension}

\citet{SharmaHammett2011} consider purely parallel diffusion along the magnetic field lines, i.e. $D_\perp=0$,
\begin{align}
\label{eq:aniso_cond}
\frac{\partial \cren}{\partial t} =& - \vnabla \cdot \vektor{q}, \\
\label{eq:heat_flux}
\vektor{q} =& - D_\parallel \vektor{b} (\vektor{b}\cdot \vnabla) \cren = - D_\parallel \vektor{b}
\nabla_\parallel e,
\end{align}
Defining the components of the diffusion operator on the right hand side
of Eq. (\ref{eq:aniso_cond}) as
\begin{equation}
\label{eq:operator}
\tensor{D}_{ij} = - \frac{\partial}{ \partial x_i} \left ( D_\parallel b_i b_j \frac{\partial}{\partial x_j} \right ) ,
\end{equation}
then the method
can be formulated as
\begin{equation}
\label{eq:op_2step}
\cren^{n+1} =
(1+\Delta t \tensor{D}_{yy})^{-1}
(1-\Delta t \tensor{D}_{yx})
(1+\Delta t \tensor{D}_{xx})^{-1}
(1 - \Delta t \tensor{D}_{xy}) \cren^{n}.
\end{equation}

An extension to adaptive mesh refinement using an implicit scheme was introduced by \citet{DuboisCommerconBeno2016} in the \textsc{Ramses} code \citep{Teyssier2002}. In their algorithm the general anisotropic diffusion is split into an isotropic component and a component parallel to the magnetic field,
\begin{equation}
\label{eq:conduction}\frac{\partial \cren} {\partial t}=-\vnabla\left(-D_{\parallel}\vektor{b} (\vektor{b}\vnabla) \cren \right)-\vnabla\left(-D_{\rm iso}\vnabla \cren \right).
\end{equation}
The updated CR energy in cell $i,j$ is given (in 2D) by
\begin{equation}
\label{eq:2dcond}\cren^{n+1}_{i,j}+\Delta t \frac{F^{n+1}_{i+\frac{1}{2},j}+F^{n+1}_{i,j+\frac{1}{2}} - F^{n+1}_{i-\frac{1}{2},j} - F^{n+1}_{i,j-\frac{1}{2}}} {\Delta x}= \cren^{n}_{i,j}\, ,
\end{equation}
for cell position $i,j$. These quantities are evaluated with the centred symmetric scheme proposed by~\cite{GuenterEtAl2005} for the anisotropic part of the flux. The anisotropic flux at cell interfaces $F^{\rm ani}_{i\pm1/2,j}$ and $F^{\rm ani}_{i,j\pm1/2}$ are evaluated from their cell corner fluxes $F^{\rm ani}_{i\pm1/2,j\pm1/2}$, thus
\begin{eqnarray}
F^{\rm ani}_{i+\frac{1}{2},j}&=&\frac {F^{\rm ani}_{i+\frac{1}{2},j-\frac{1}{2}} + F^{\rm ani}_{i+\frac{1}{2},j+\frac{1}{2}}} {2}\, , \nonumber \\
F^{\rm ani}_{i,j+\frac{1}{2}}&=&\frac {F^{\rm ani}_{i-\frac{1}{2},j+\frac{1}{2}} + F^{\rm ani}_{i+\frac{1}{2},j+\frac{1}{2}}} {2}\, . \nonumber
\end{eqnarray}
The anisotropic cell corner flux is
\begin{equation}
\label{eq:aniflux}F^{\rm ani}_{i+\frac{1}{2},j+\frac{1}{2}}=\bar \kappa_{\parallel} \bar b_{x} \left( \bar b_{x} \bar {\frac{\partial e} {\partial x}} + \bar b_{y} \bar{ \frac{ \partial e} {\partial y} }\right)\, ,
\end{equation}
where barred quantities are arithmetic averages over the cells connected to the corner.

The authors use the diffusion algorithm introduced in \citet{CommerconEtAl2014} including AMR and adaptive time-stepping on a level-by-level basis.
Each refinement level $\ell$ is evolved with a corresponding time step $\Delta t^\ell$, where each time step is twice as large for each coarser level $\ell-1$, $\Delta t^\ell =\Delta t^{\ell -1}/2$. This means that level $\ell$ evolves with two consecutive time steps before one time step of level $\ell-1$ is applied. Each level of refinement $\ell$ is connected to two types of non-uniform interfaces, namely the fine-to-coarse interface (between level $\ell$ and $\ell -1$) and the coarse-to-fine interface (between level $\ell$ and $\ell +1$). \citet{DuboisCommerconBeno2016} use Dirichlet boundary conditions, where the cell \emph{values} at level boundaries are imposed. In the other hand, for Neumann boundary conditions the \emph{fluxes} are imposed, which are needed to guarantee energy conservation similar to hydrodynamical solvers.

At the interface from fine-to-coarse, \citet{DuboisCommerconBeno2016} use the values of $\ell-1$ at time $n$ as an imposed boundary condition for level $\ell$. They use the minmod scheme~\citep{vanLeer1979} and interpolate the values of level $\ell-1$ on a finer virtual grid at level $\ell$ to determine the fine-to-coarse boundary.
For the coarse-to-fine interface they use values of $\ell+1$ at time $n+1$ as the imposed boundary conditions for level $\ell$. The value of the boundary coarse cell at level $\ell$ is restricted to the average value of the $2^{dim}$ cells of level $\ell+1$ to impose the coarse-to-fine boundary.
Eq. \ref{eq:2dcond} remains correct since the diffusion solver only deals with data estimated at the same level of refinement. The combination of Dirichlet boundary conditions at the level interfaces and the interpolation or restriction operations does not break the symmetry of matrix $A$. The imposed values of neighbouring cells at different levels of refinement enter the right-hand side (in vector $\vektor{c}$) of the above matrix system $A\vektor{x}=\vektor{c}$.

\subsubsection{Diffusion on a Voronoi mesh}

\citet{PakmorEtAl2016} implement the diffusion of the CR strictly along the magnetic field lines
\begin{equation}
\frac{\partial \cren_\mathrm{cr}}{\partial t} - \vnabla \bcdot \left[ D_\parallel \vektor{b} \left( \vektor{b} \bcdot \vnabla \cren_\mathrm{cr} \right) \right] = 0 ,
\end{equation} 
with the CR energy density $\cren_\mathrm{cr}$ and the parallel diffusion coefficient $D_\parallel$. Similar to \citet{GuenterEtAl2005} and \citet{SharmaHammett2007} they compute the gradient estimates at the centre of an interface between cells based on the gradient estimates at the corners. In three dimensions every corner of a Voronoi cell connects to four adjacent cells. The residual of the gradient fit reads
\begin{equation}
r_i = \phi \left( \vektor{c} \right) + \left( \vnabla \phi \right) \left( \vektor{s}_i - \vektor{c} \right) - \phi \left( \vektor{s}_i \right),
\end{equation}
with the unknown value at the corner of a cell $\phi\left( \vektor{c} \right)$, the gradient at the corner position $\vektor{c}$, $\vnabla\phi$, and the value at the centre of the cell $\vektor{s}_i$, $\phi(\vektor{s}_i)$. The residual can be rewritten in matrix form
\begin{equation}
\vektor{r} = \tensor{X}\vektor{q}-\vektor{Y},
\end{equation}
where $\vektor{r}$, $\vektor{q}$, and $\vektor{Y}$ are $N$-vectors ($N=4$), and $\tensor{X}$ is an $N\times N$ matrix. The individual components are defined as
\begin{eqnarray*}
(\vektor{q})_0 &=& \phi \left( \vektor{c} \right) ,\\
(\vektor{q})_{1..N-1} &=& \left( \vnabla \phi \right)_{0..N-2} ,\\
(\vektor{Y})_i &=& \phi \left( \vektor{s}_i \right) ,\\
(\tensor{X})_{i,0} &=& 1 ,\\
(\tensor{X})_{i,1..N-2} &=& \vektor{s}_{i,0..N-2} -  \vektor{c}_{0..N-2}.
\end{eqnarray*}
The residual is then minimized by solving
\begin{equation}
\left( \tensor{X}^T\ \tensor{X} \right) \vektor{q} = \tensor{X}^T\ \vektor{Y}.
\end{equation}
There is a unique solution for $\vektor{q}$ with zero residual because $\tensor{X}$ is a square matrix. In order to solve for $\vektor{q}$ a multiplication with $\left( \tensor{X}^T\ \tensor{X} \right)^{-1}$ from the left is used, which yields
\begin{equation}
\vektor{q} = \tensor{M} \ \vektor{Y},
\label{eq:gradcorner}
\end{equation}
with $\tensor{M} = \left( \tensor{X}^T\ \tensor{X} \right)^{-1} \ \tensor{X}^T$.
Here, $\tensor{M}$ only depends on the geometry of the mesh. The vector $\vektor{Y}$ only contains values at the cell centers. Therefore, $\tensor{M}$ only needs to be computed once every timestep as the geometry of the mesh does not change. One can obtain the value as well as the gradient of any quantity at a corner from a simple matrix-vector multiplication. Moreover, Eq.~(\ref{eq:gradcorner}) shows the linear dependence of the gradient at the corner on the values in the adjacent cells. This fact is needed for the implicit time integration.

The estimate of the gradient is given by
\begin{equation}
 \vnabla \cren_{\mathrm{cr}\mathrm{,n}} =  \frac{\cren_\mathrm{cr,L} - \cren_\mathrm{cr,R}}{ \left| \vektor{c}_\mathrm{L} - \vektor{c}_\mathrm{R} \right| } 
 \left(\frac{ \vektor{c}_\mathrm{L} - \vektor{c}_\mathrm{R} }{ \left| \vektor{c}_\mathrm{L} - \vektor{c}_\mathrm{R} \right| }\right) \bcdot\vektor{n}_\mathrm{face}.
\end{equation}

The solution is obtained by a semi-implicit time integration,
\begin{equation}
\begin{aligned}
\cren_{\mathrm{cr},i}^{\tilde{n}} = \cren_{\mathrm{cr},i}^{n} +  \frac{\Delta t}{V_i}  \sum_{j} D_{ij} \left( \vektor{b}_{ij} \bcdot\vnabla \cren_{\mathrm{CR,t},ij}^{n} \right) \vektor{b}_{ij} \bcdot\vektor{n}_{ij} A_{ij}.
\end{aligned}
\end{equation}

In a second step, the CR energy is advanced according to the fluxes associated with the normal component of the CR energy density gradients computed at the interfaces. This is done with an implicit backward Euler step:
\begin{equation}
\begin{aligned}
\cren_{\mathrm{cr},i}^{n+1} = \cren_{\mathrm{cr},i}^{\tilde{n}} +  \frac{\Delta t}{V_i}   \sum_{j} D_{ij} \left( \vektor{b}_{ij} \bcdot\vnabla \cren_{\mathrm{CR,n},ij}^{n+1} \right) \vektor{b}_{ij} \bcdot\vektor{n}_{ij} A_{ij}.
\end{aligned}
\end{equation}

The term $\vnabla \cren_{\mathrm{cr,n},ij}^{n+1}$ scales linearly with the CR energy densities. This yields a system of coupled linear equations, which can be solved using a matrix solver. \citet{PakmorEtAl2016} solve the linear system in a two-step procedure using the solvers from the \textsc{hypre} library \citep{FalgoutYang2002}.

\subsection{Numerical scheme for selfconsistent, spectral CR transport} \label{sect:two-mom-miniati}

Here we summarise the general framework formulated by  \citet{2001CoPhC.141...17M}, which we extend to middly relativistic and non-reativistic ranges of CR momenta.

\subsubsection{Evolution of the isotropic CR spectrum on the momentum grid}

We assume a piecewise power-law, isotropic (in momentum space)  distribution function
\begin{equation}
f(p) = f_{\imh} \left(\frac{p}{p_\imh}\right)^{-\qi}, \label{f(p)-powlaw-def:eqn}
\end{equation}
where the distribution function amplitudes $ f_{\imh}$ are defined on left edges of momentum bins $p_\imh$ and the spectral indices $\qi$ are attributed to bin interiors.

The number density of particles in a single momentum bin is
\begin{equation}
n_i = \int_{p_\imh}^{p_\iph} 4\pi p^2 f(p) dp = \int_{p_\imh}^{p_\iph} 4\pi p^2 f_\imh \left( \frac{p}{p_\imh }  \right)^{-\qi} dp,
\end{equation}

\begin{equation}
n_i = 
4\pi f_\imh p^3_\imh  \times \left\{ 
    \begin{array}{l l} 
            \frac{\left(\frac{p_\iph}{p_\imh}\right)^{3-\qi}-1}{3-\qi} & \quad \textrm{if $\qi \ne 3$ }   \\
            \ln \left(p_\iph/p_\imh\right) & \quad \textrm{if $\qi = 3$ }
   \end{array}
   \right. .
   \label{def:n_i}
\end{equation}
The particle density equation in the discretised form (spatial propagation terms and local sources are neglected here) reads
\begin{equation}
\pder{n_{i}}{t}{} = \left[b(p) 4\pi p^2 f\xpt\right ]_{p_\imh}^{p_\iph},
\label{eqn:n_i}
\end{equation}
where $b(p)$  is the loss term.
We integrate Eqn. (\ref{eqn:n_i}) over the timestep interval
\begin{equation}
  \int_t ^{t+\Delta t} \frac{\partial n_i}{\partial t} dt' =  - \left( \Delta n_\iph^{\Delta t} - \Delta n_\imh^{\Delta t} \right),
\label{eqn:n_i-int}
\end{equation}
where $\Delta n_\iph^{\Delta t}$ and  $\Delta n_\imh^{\Delta t}$ are  particle numbers  transferred  through bin  boundaries at  $p_{i \pm \h}$ during the time interval $\Delta t$. 
The discrete form of the particle number becomes
\begin{equation}
  n_i^{t+\Delta t}  =   n_i^t - \left( \Delta n_\iph^{\Delta t} - \Delta n_\imh^{\Delta t} \right). 
\end{equation}

For the kinetic energy density
\begin{equation}
T(p) = \sqrt{p^2 c^2 + m^2 c^4} - m c^2    \equiv g(p). \label{def:ekin}
\end{equation}
we shall approximate the function $g(p)$ by another piecewise powerlaw function defined for each momentum bin separately
\begin{equation}
g(p) \simeq  g_{\imh} \left(\frac{p}{p_\imh}\right)^{\si}, \label{eqn:g-powlaw}
\end{equation}
where the function amplitudes $ g_{\imh}$ are defined on left edges of momentum bins $p_\imh$ and the power indices $\si$ are attributed to bin interiors. The amplitudes $g_{\imh}$ are computed directly by evaluation of $g(p_{\imh})$ on bin edges and the corresponding slopes $\si$
are
\begin{equation}
\si =\frac{\log_{10} \left( \frac{g_{\iph}}{g_{\imh}} \right)}{\log_{10} \left( \frac{p_\iph}{p_\imh}\right) }.
\end{equation}
Thus, the general relation between energy and distribution function amplitudes is
\begin{equation}
e_i = 
4\pi f_\imh \g_\imh p^{3}_\imh \times \left\{ 
    \begin{array}{l l} 
             \frac{\left(\frac{p_\iph}{p_\imh}\right)^{3+\si-\qi}-1}{3+\si-\qi} & \quad \textrm{if $\qi \ne 3+\si$ }   \\
             \ln \left(p_\iph/p_\imh\right) & \quad \textrm{if $\qi = 3+\si$ }
   \end{array}
   \right. .
   \label{def:e_i}
\end{equation}
Energy equation in the discretised form reads
\begin{eqnarray}
  \frac{\partial e_i}{\partial t} & = & \left[ b(p) 4 \pi p^2 f(p) T(p) \right] ^{^{{p_\iph}}} _{_{p_\imh}} 
  - \int^{p_\iph}_{p_\imh} b(p)\frac{4 \pi c p^3 f(p)}{\sqrt{m^{2} c^2 + p^2}} dp. \label{e_i-def:eqn}
\end{eqnarray}
The energy loss term (the integral) on the right-hand side of  (\ref{e_i-def:eqn}) contains the derivative of kinetic energy
\begin{equation}
\der{T(p)}{p}{} = \frac{c p}{\sqrt{p^2 c^{2} + m^{2} c^4}}.
\end{equation}
Since $T(p)$ is approximated by a power-law function we can use Eqn. (\ref{eqn:g-powlaw}) to represent  its derivative
\begin{equation}
T'(p) = g'(p) \simeq g_{\imh} \frac{\si}{p_{\imh}}\left( \frac{p}{p_{\imh}}\right) ^{\si - 1}.  \label{eqn:ekin-deriv}
\end{equation}
We substitute the second right-hand side term in equation (\ref{e_i-def:eqn}), by  $e_i R_i$:
 \begin{equation}
  \frac{\partial e_i}{\partial t}  =  \left[ b(p) 4 \pi p^2 f(p) T(p) \right] ^{^{{p_\iph}}} _{_{p_\imh}} - e_i R_i , \label{eqn:e_i-new}
\end{equation}
where $R_{i}$ expresses the energy loss rate per unit energy
\begin{equation}
R_i = \frac{1}{e_{i}}  \int^{p_\iph}_{p_\imh} b(p)\frac{4 \pi c p^3 f(p)}{\sqrt{p^2 c^{2} + m^{2} c^4 }} dp. \label{def:Ri} 
\end{equation}
For adiabatic cooling 
\begin{equation}
b(p) = \frac{1}{3} (\nabla \cdot \bm{v})\; p \equiv  u_{d} \; p.
\end{equation}

To deal with the full (relativistic and non-relativistic) energy range we substitute  (\ref{eqn:ekin-deriv}) in (\ref{def:Ri})
\begin{eqnarray}
R_i &=& \frac{1}{e_{i}} \left[ \int^{p_\iph}_{p_\imh}  4 \pi  u_{d} p^3 f(p) g'(p) dp + \ldots \right]\\
      &=&   \frac{4\pi c u_{d}}{e_{i}} f_\imh g_\imh p^{3}_\imh \si \times \left\{ 
    \begin{array}{l l} 
             \frac{\left(\frac{p_\iph}{p_\imh}\right)^{3+\si-\qi}-1}{3+\si-\qi} & \quad \textrm{if $\qi \ne 3+\si$ }   \\
             \ln \left(p_\iph/p_\imh\right) & \quad \textrm{if $\qi = 3+\si$ }
   \end{array}
   \right. + \ldots \nonumber
\end{eqnarray}
where the dots stand for other cooling mechanisms.
By elimination of the energy density we find that 
\begin{equation}
R_{i} = u_{d} \si + \ldots
\end{equation}
We integrate now Eqn. (\ref{eqn:e_i-new}) over the timestep interval
\begin{equation}
  \int_t ^{t+\Delta t} \frac{\partial e_i}{\partial t} dt' =  - \left( \Delta e_\iph^{\Delta t} - \Delta e_\imh^{\Delta t} \right) - \int_t ^{t+\Delta t} e_i R_i dt',
\label{eqn:e_i-int}
\end{equation}
where $\Delta e_\iph^{\Delta t}$ and  $\Delta e_\imh^{\Delta t}$ are particle  energies  transfered  through bin  boundaries at  $p_{i \pm \h}$ during the time interval $\Delta t$. 

We assume that variations of $R_{i}$ over the timestep are negligible, thus we approximate the  integral on the r.h.s. of  (\ref{eqn:e_i-int}) as:
\begin{eqnarray}
  \int_t ^{t+\Delta t} e_i R_i dt' & = & \frac{1}{2} \Delta t \, R_i \left( e_i^{t+\Delta t} + e_i^t \right).
\end{eqnarray}
Hence
\begin{eqnarray}
  e_i ^{t+\Delta t} - e_i^t & = &  - \left( \Delta e_\iph^{\Delta t} - \Delta e_\imh^{\Delta t} \right) - \h \Delta t \, R_i \left( e_i^{t+\Delta t} + e_i^t \right), \nonumber \\
  e_i^{t+\Delta t} & = &  \frac{- \left( \Delta e_\iph^{\Delta t} - \Delta e_\imh^{\Delta t} \right) + e_i^t\left( 1 - \frac{R_i}{2}\Delta t \right)} { 1 + \h \Delta t \, R_i}.
\end{eqnarray}
The following part of this section will detail the calculation of the $\Delta n_{i-1/2}^{\Delta t}$ and 
$\Delta e_{i-1/2}^{\Delta t}$  terms, which involves fluxes of particles and their energies through bin boundaries.
The procedure relies by construction on the upwind computation of the fluxes in momentum space.

\paragraph{Computation of upstream momentum $p_{\textrm{u}}$.}

Particles loose or gain momentum due to physical effects underlying the source term $b(p)$ in Eqn. (\ref{e_i-def:eqn})
\begin{equation}
 b(p)  \equiv - \left(\der{p}{t}{}\right)_{\textrm{tot}} =   
 \frac{1}{3} \left( \nabla \cdot \bm{v}\right)  p + \ldots  = u_{d} \ p + \ldots.   
\label{p_upw-ad-der}
\end{equation}
We integrate (\ref{p_upw-ad-der}) for  the case of the adiabatic process
\begin{eqnarray}
    \int_{p(t)}^{p(t+\Delta t)} \frac{d p}{p} & = & - \int_{t}^{t+\Delta t} u_d \, dt, \nonumber \\
    \ln{p^{t+\Delta t}} & = & \ln{p^t} - u_d \Delta t \label{p_upw-ad-int}.
\end{eqnarray}
Our aim is to find the value of the momentum $\pupw$ at $t$ that becomes $\ln (p_\imh)$ at $t+\Delta t$
\begin{eqnarray}
  \ln{p_\imh} & = & \ln{\pupw} - u_d \, \Delta t, \nonumber 
\end{eqnarray}
and finally we find
\begin{eqnarray}
  \pupw &  = & p_\imh e^{u_d\Delta t}.
\end{eqnarray}
Taylor expansion up to the 1st  order in $\Delta t$ results in:
\begin{eqnarray}
  \pupw & \approx & p_\imh(1 + u_d \Delta t).
\end{eqnarray}
\paragraph{Computation of upwind quantities  $\Delta n^{\Delta t}_{\upw,  \imh } $. } \label{subsect:upfluxes}
In the case of cooling the number of particles transferred from $i$-th bin through the left bin-face is
\begin{eqnarray}
  \Delta n^{\Delta t}_{\upw,  \imh } & = & \int_{p_\imh}^{\pupw} 4\pi p^2 f(p) dp \nonumber \\
   & = &  4 \pi f_\imh p_\imh^3 \times \left\{ \begin{array}{l l l}
                    \;  \frac{  \left(\frac{\pupw}{p_{\imh}}\right)^{3-q_{i}}  - 1}{3 - q_i} \, &\textrm{if}  & q_i \neq 3  \vspace{7pt}\\
                      \ln{\left(\frac{\pupw}{p_\imh}\right)} \, &\textrm{if}  & q_i = 3                    
                    \end{array}
  \right. .
  \label{eqn:dn-cooling}
\end{eqnarray}
In the case of heating the the number of particles transferred from $i$-th bin through the right bin-face is
\begin{eqnarray}
  \Delta n^{\Delta t}_{\upw,  \imh } & = & \int_{\pupw}^{p_\imh} 4\pi p^2 f(p) dp \nonumber \\
 & = & 4 \pi f_\imth \, \pupw^3 \left( \frac{p_\imth}{\pupw} \right)^{q_{i-1}} 
               \times \left\{ \begin{array}{l l l}
                          \frac{\left(\frac{p_\imh}{\pupw}\right)^{3-q_{i-1}}-1}{3 - q_{i-1}} & \textrm{if}  &q_i \neq 3  \vspace{7pt}\\
                              \ln{\left(\frac{p_\imh}{\pupw}\right)} \, &\textrm{if}  & q_i = 3                    
                              \end{array}\right. .
 \label{eqn:dn-heating}
\end{eqnarray}

\paragraph{Computation of upwind quantities  $\Delta e^{\Delta t}_{\textrm{u}}$.}

In the case of cooling the the amount  of particle energy  transferred  from $i$-th bin through the left bin-face:
\begin{eqnarray}
  \Delta e^{\Delta t}_{\upw,  \imh } & = & \int_{p_\imh}^{\pupw} 4\pi p^2 f(p) T(p) dp \nonumber \\
   &\simeq & 4\pi f_\imh \g_\imh p^{3}_\imh \times \left\{ 
    \begin{array}{l l l} 
             \frac{\left(\frac{p_\upw}{p_\imh}\right)^{3+\si-\qi}-1}{3+\si-\qi} &  \textrm{if} & \qi \ne 3+\si   \\
             \ln \left(\frac{p_\upw}{p_\imh}\right) & \textrm{if} & \qi = 3+\si 
   \end{array}
   \right. .
\end{eqnarray}
In the case of heating  the amount  of particle energy  transferred from $i$-th bin through the right bin-face:
\begin{eqnarray}
  \Delta e^{\Delta t}_{\upw,  \imh } & = & \int_{\pupw}^{p_\imh} 4\pi  p^2 f(p) T(p) dp \nonumber \\
 & \simeq & 4 \pi f_\imth \g_{\imth}\, \pupw^{3} \left( \frac{p_\imth}{\pupw} \right)^{q_{i-1}-s_{i-1}} \\
               & &\times \left\{ \begin{array}{l l l}
                          \frac{\left(\frac{p_\imh}{\pupw}\right)^{3+\simo-q_{i-1}}-1}{3+\simo - q_{i-1}} &  \textrm{if} &q_{i-1} \neq 3 +\simo \vspace{7pt}\\
                              \ln{\left(\frac{p_\imh}{\pupw}\right)} \, &  \textrm{if} &q_{i-1} = 3  +\simo                  
                              \end{array}\right. . 
 \label{eqn:de-heating}
\end{eqnarray}

\subsubsection{Conversion between quantities} \label{sec:conv-en-fq}

The spectral behaviour is best computed using the Fokker-Planck equation using the amplitudes $f$ and the slopes $q$. For the coupling to hydrodynamical simulations is more convenient to use number density and energy density. Both sets of quantities ($f,q$ and $n,e$) contain equivalent information, so it is possible to use $f$ and $q$ for the spectral evolution and then convert to $n$ and $e$ for the coupled part to hydrodynamics. The computation of $n$ and $e$ based on $f$ and $q$ is straight forward. The backward conversion can be done by solving the ratio of $e_i/n_i$ for $q_i$. The amplitude $f_{i-1/2}$ cancels in the ratio and the solution for $q_i$ is unique.

\section{Astrophysical applications}\label{sec:applications}
\subsection{Modeling galactic CR propagation with GALPROP}

\subsubsection{Examples of GALPROP model predictions}

\begin{figure*}
  \centerline{\includegraphics[width=0.5\textwidth]{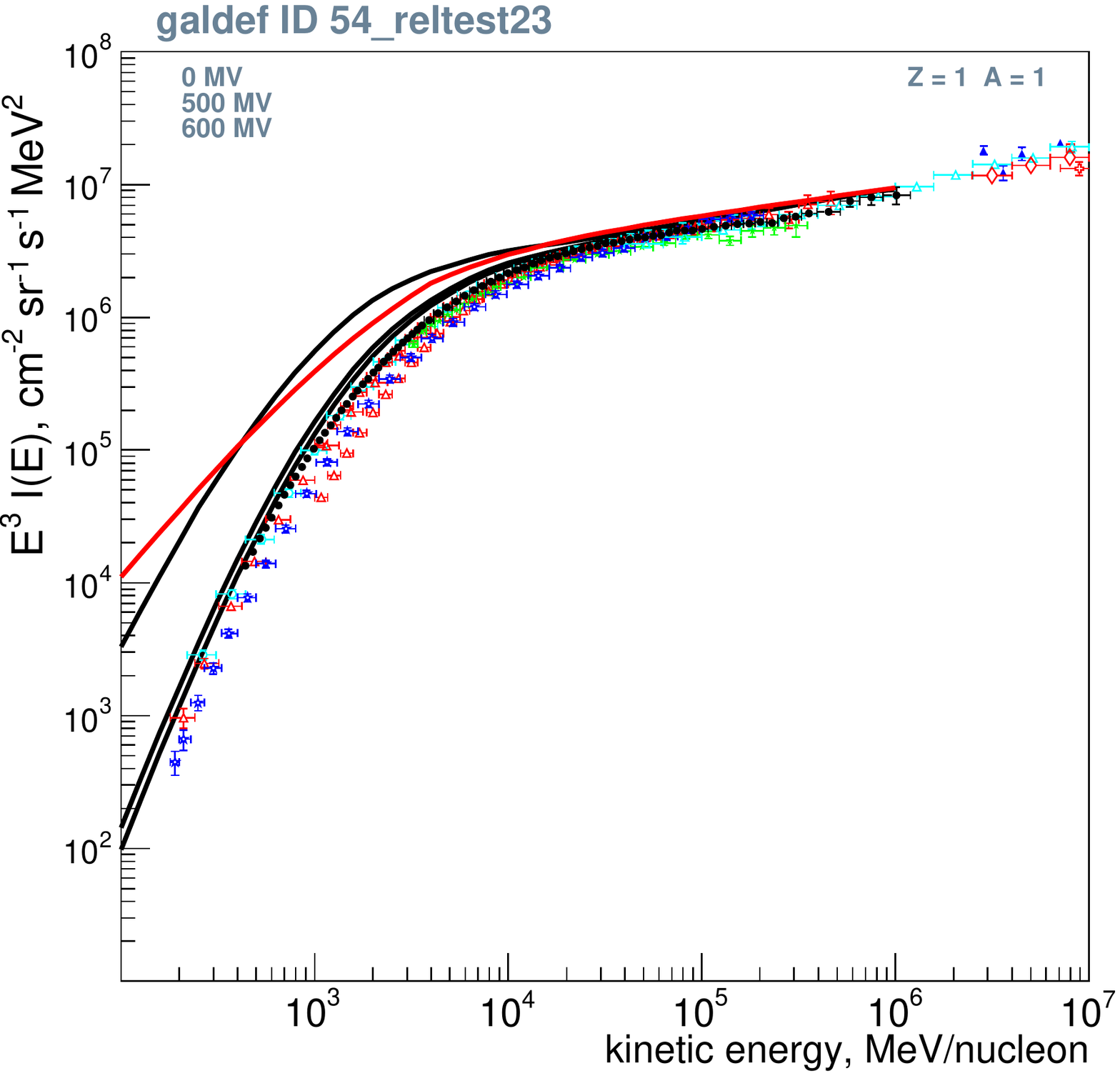}
              \includegraphics[width=0.5\textwidth]{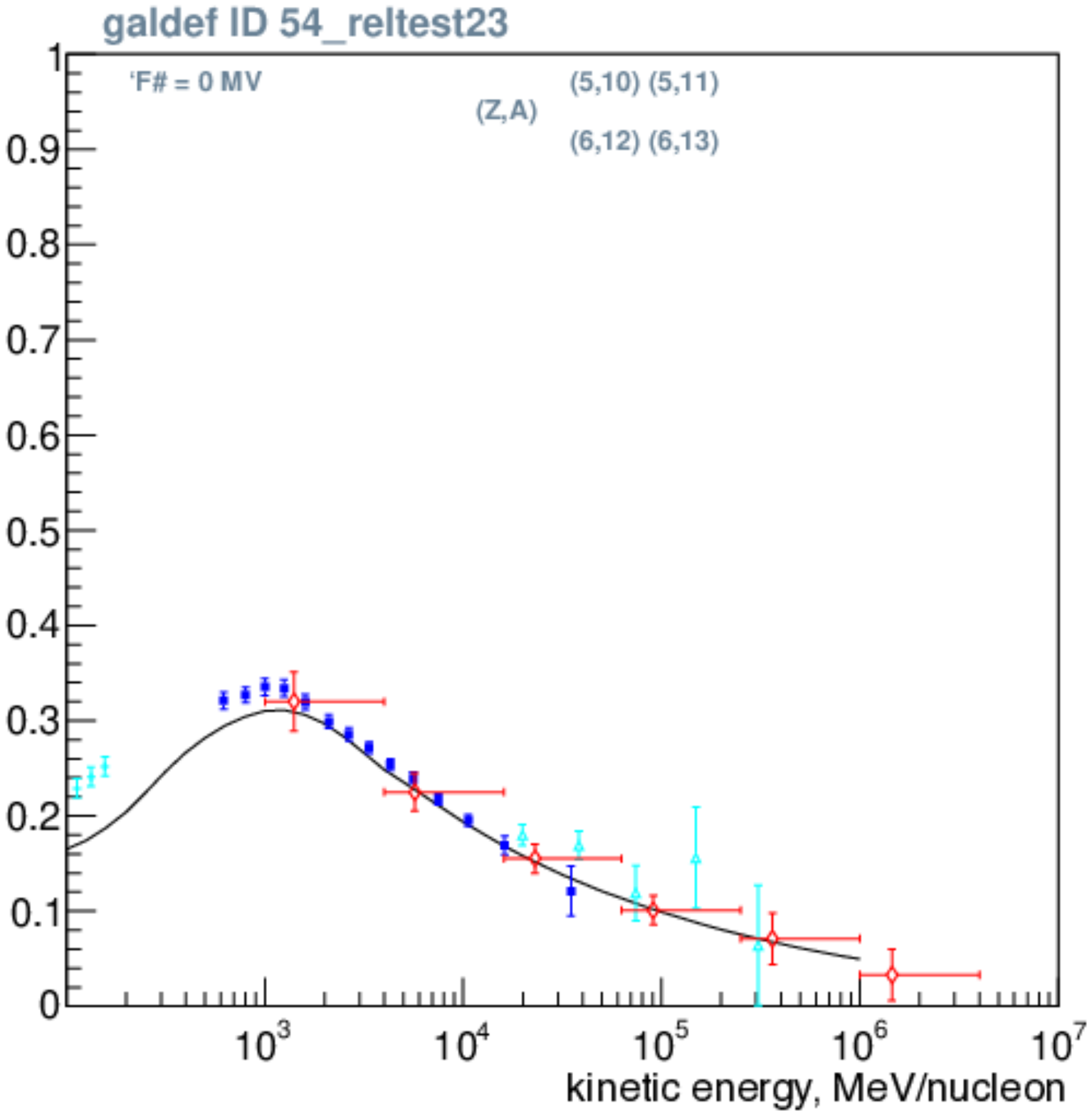}}
  \centerline{\includegraphics[width=0.5\textwidth]{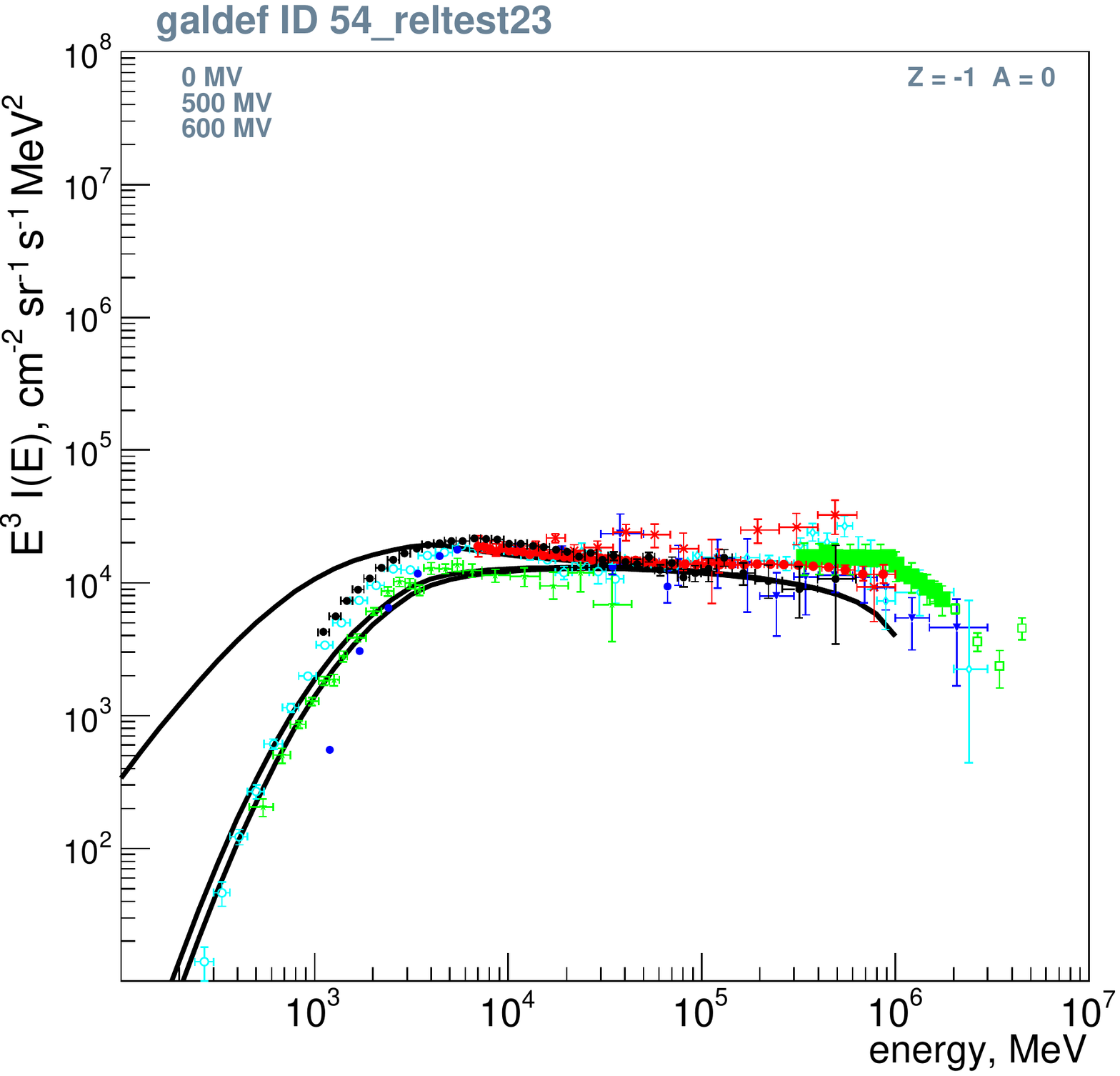}
              \includegraphics[width=0.5\textwidth]{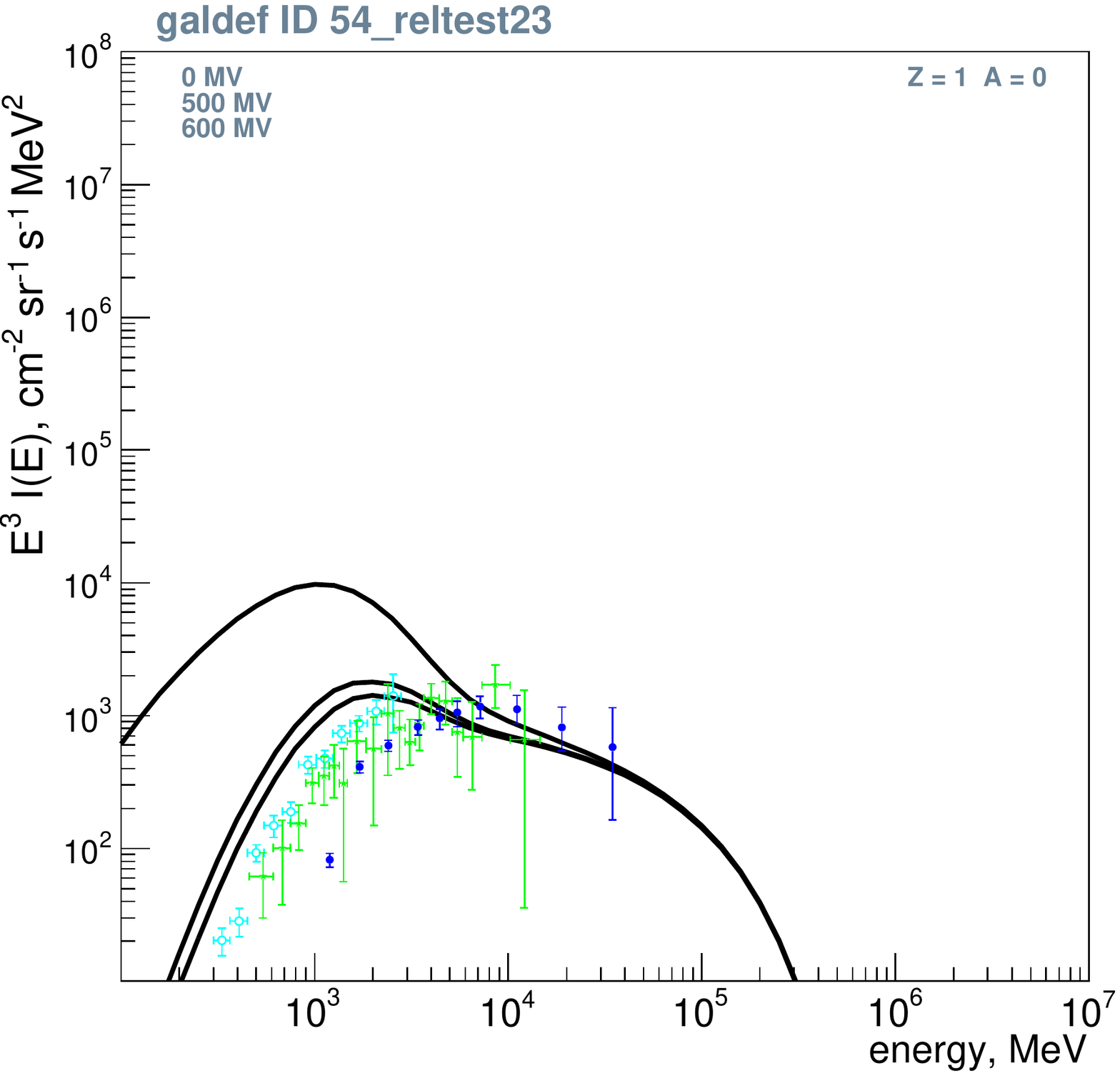}} 
\caption{Examples of GALPROP model predictions. Upper left: proton spectrum, upper right: Boron to Carbon ratio, lower left: electrons, lower right: positrons.
Top curves are interstellar at the solar position, and lower curves are solar-modulated with the parameters shown.
The plots were made with the GALPLOT package, available at  https://gitlab.mpcdf.mpg.de/aws/galplot.
}
\label{fig:galprop_CR_examples}
\end{figure*}

\begin{figure*}
  \centerline{\includegraphics[width=0.45\textwidth]{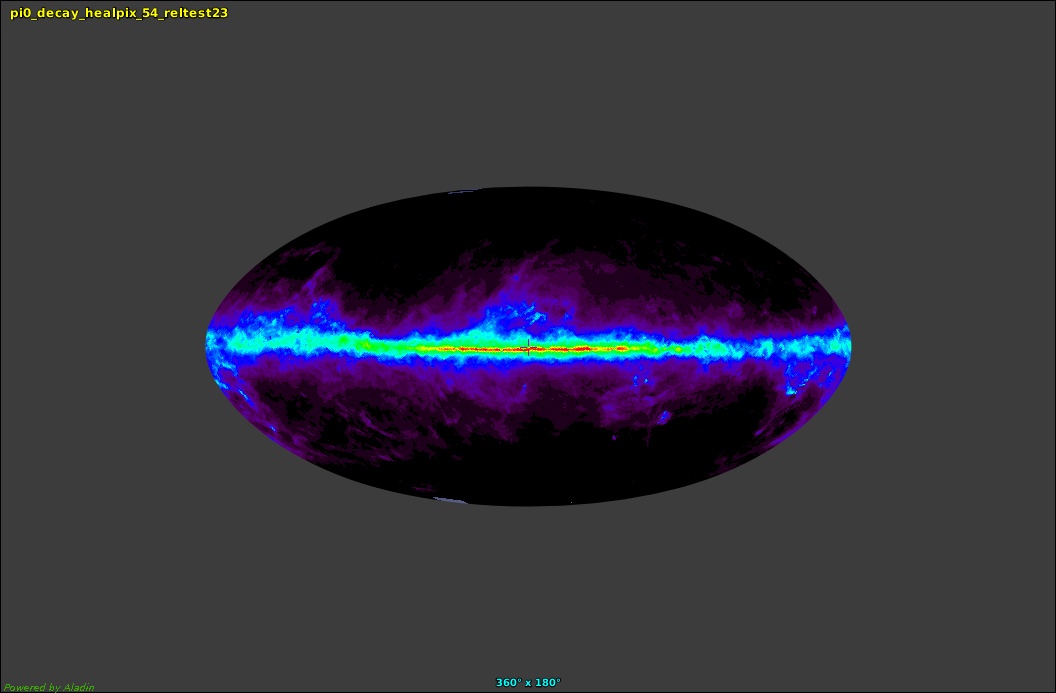}
              \includegraphics[width=0.45\textwidth]{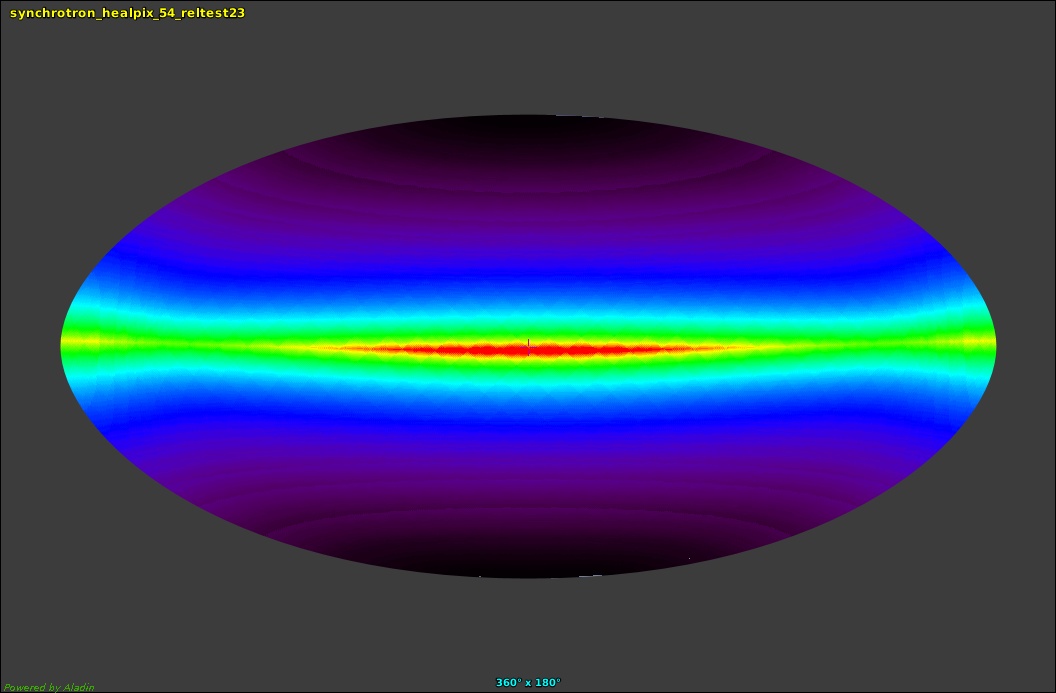}}
\caption{Examples of GALPROP model predictions. Left: gamma-ray skymap at 1 GeV, hadronic production via $\pi^o$-decay. Right: synchrotron skymap, 1 GHz,  Stokes I (total). The maps are in Galactic coordinates, Aitoff projection, centred on Galactic centre. The maps are GALPROP output in Healpix format, plotted with Aladin from CDS Strasbourg.
}
\label{fig:galprop_skymap_examples}
\end{figure*}

Fig.~\ref{fig:galprop_CR_examples} shows example predictions for a typical  GALPROP model:
protons, secondary-to-primary boron/carbon ratio, primary electrons and secondary positrons.
This model includes diffusive reacceleration, and breaks in both primary injection spectra and diffusion coefficient.
The halo height is 4 kpc.
The predictions are for the solar position in the Galaxy (though predictions are made for the entire Galaxy).
The parameters can be found in the GALPROP parameter file with name galdef\_54\_reltest23 supplied with the code.
Interstellar and example modulated results (simple force-field approximation) are shown.
Sample data are also shown for comparison; the fit is reasonable,  although no attempt has been made to use the latest available data, nor is any attempt made here to fit the data; 
this is covered in literature {\bf \citep{GrenierBlackStrong2015,2020ApJS..250...27B,2020ApJ...889..167B}.}

Fig.~\ref{fig:galprop_skymap_examples} shows sample all-sky maps from hadronic interactions with the interstellar gas, and synchrotron radiation from electrons and positrons on the magnetic field. The gamma-ray maps are generated by GALPROP using the computed p and He spectra over the Galaxy, and hadronic cross-sections and the GALPROP neutral, molecular and ionized hydrogen and Helium model, and integration over the line-of-sight from the solar position. The synchrotron maps use the electron and positron spectra over the Galaxy, the magnetic field model, and the full formula for synchrotron including the directional variation with respect to the magnetic field for the regular component. Such models can be directly compared to gamma-ray observations e.g. by the Fermi-LAT instrument and radio surveys by radiotelescopes and the Planck satellite.

\subsubsection{Diffusive reacceleration and alternatives}

Diffusive reacceleration has been a process frequently included since it explains B/C without an ad-hoc break in $D_{xx}(p)$ and is compatible with the Kolmogorov index 0.33 in this term.
It is simply diffusion in momentum due to the gain and loss of momentum  off moving scatterers; hence there is a basic relation between $D_{pp}$ and $D_{xx}$
(see e.g. \cite{StrongMoskalenkoPtuskin2007}). %
However the question remains of the actual importance of this process in the insterstellar medium. 
In fact in the models with large reacceleration, a significant amount of energy is being injected into CR from the ISM,
so that CR acceleration involves more than just the standard sources like SNR.
The source of this energy poses a problem. 
The issue has been recently addressed by 
\cite{2014MNRAS.442.3010T}. %
 This article has a useful and
clear derivation of the reacceleration formula and clarifies its relation to the original Fermi second-order mechanism. 
In future the energetics involved should be studied in more detail.
Reacceleration at the level often invoked seems incompatible with low-frequency synchrotron spectrum which is sensitive to the electron and positron spectra around a few GeV.

On the other hand, pure diffusion models without reacceleration do have a problem to reproduce B/C without either a very large break  in $D_{xx}(p)$
\citep{1998ApJ...509..212S} %
or an additional velocity dependence in $D_{xx}(p)$
\citep{2006ApJ...642..902P}; %
the latter paper also invokes dissipation of MHD waves as an alternative way to produce the required  $D_{xx}(p)$.
The use of B/C depends however critically  on the level of solar modulation used to convert the interstellar ratio to the observed one; a lower level of modulation than the value $\approx$500 MV often adopted would ease the tension.
\cite{2013ApJ...770..117L} %
compare  both reacceleration and pure diffusion models with B/C data including the latest ACE data at a few 100 MeV; a modulation potential of 250 MV is adopted there.
More likely convection plays a role in the sub-GeV/nucleon range  as well, since it gives an energy-independent escape time.

It is also possible to  include the effect of CR on
the diffusion coefficient \cite{2006ApJ...642..902P}. 
They studied the possibility that the nonlinear MHD cascade sets the power-law spectrum of turbulence that scatters charged
energetic particles. They found that the dissipation of waves due to the resonant interaction with cosmic-ray particles may
terminate the Kraichnan-type cascade below wavelengths $10^{13} \cm$. The effect of this wave dissipation has been
incorporated in the GALPROP numerical propagation code in order to asses the impact on measurable astrophysical
data. The energy dependence of the cosmic-ray diffusion coefficient found in the resulting self-consistent model may
explain the peaks in the secondary to primary nuclei ratios observed at about 1 GeV nucleon$^{-1}$.

\subsection{CRs in star formation}

Star formation occurs in the densest and coldest regions of the turbulent interstellar medium \citep{MacLowKlessen2004,McKeeOstriker2007,GirichidisEtAl2020b}. Typical densities of star forming cores exceed $10^6\,\mathrm{cm}^{-3}$ where the gas is as cold as $\sim10\,\mathrm{K}$. The spatial extents are small fractions of parsecs. The role of CRs in these regions includes several aspects. As CRs penetrate deeply into the protostellar cores, they deposit energy into regions which are opaque to electromagnetic radiation \citep[see e.g. recent review by][]{PadovaniEtAl2020}. This effectively sets a minimum temperature on the gas temperature in dense gas. As the Jeans mass scales with the temperature as $T^{3/2}$ the impact of CRs on the fragmentation might be dynamically relevant. The second aspect is the impact on the chemical composition. With their high energy, CRs are able to unbind and ionize many molecules. The changes in the chemical reactions alter the observable tracers and contain valuable information on the local conditions \citep[e.g.][]{BisbasEtAl2018}. In the context of star formation the main focus is on low-energy CRs because of their enhanced cross section with the thermal particles. Integrated, the CR energy density in the low-energy component is generally not comparable to the other energy densities. Consequently, the CR pressure is also not directly driving the motions of the gas and to first order CRs can be treated as tracer particles or a tracer fluid. 

Overall, the time scales for CR transport through star forming regions are much shorter than the dynamical times scales of the gas. This means that adiabatic gains and losses of CRs can be neglected contrary to galactic scales where advective and diffusive time scales are comparable. Given the strong magnetic field strengths of $\sim\mathrm{mG}$ \citep{Crutcher2012} and the low energy of CRs ($\lesssim\mathrm{GeV}$) the gyro-radii are smaller than the spatial extents of star forming cores and the CRs can be followed collectively following the magnetic field lines along the path $s$.

The distribution function is generally a function of the pitch angle, which depends on the local strength of the regular magnetic field, and CR scattering that occurs due to resonant CR interactions with magnetohydrodynamic (MHD) turbulence and gas nuclei. Depending on the degree of pitch angle scattering, there are two CR transport regimes -- free streaming and diffusive \citep{PadovaniEtAl2018b, SilsbeeIvlev2019, PadovaniEtAl2020}. The free-streaming approximation (also known as the continuous slowing-down approximation, CSDA, \citep[see e.g.][]{Takayanagi1973,PadovaniGalliGlassgold2009}) is the most common approach used to calculate the propagation of CRs in molecular clouds. Scattering processes are inefficient in this regime, so that the resulting mean squared deviation of the pitch angle along a CR track is small, so $\mu$ is conserved. The dominant regime of CR transport in regions surrounding dense cores embedded within molecular clouds is debated. MHD turbulence in these regions can resonantly scatter the pitch angles of penetrating CRs \citep{1969apj...156..445k}, leading to spatial diffusion. The spectrum of MHD turbulence determines the magnitude of the CR diffusion coefficient and its dependence on the particle energy\citep{SchlickeiserEtAl2016, SilsbeeIvlev2019}. However, MHD turbulence can also be driven by anisotropy in the CR distribution function \citep{SkillingStrong1976, MorlinoGabici2015, IvlevEtAl2018}, which arises in response to CR absorption in dense cores.

Observationally, the ionisation rate of CRs can be connected to observations of H$_3^+$, whose chemical formation chain is primarily regulated by CRs \citep[see, e.g.][]{IndrioloEtAl2007,IndrioloMcCall2012}. %

\subsection{CRs in the interstellar medium}

In this section we discuss the interactions of CRs with the interstellar medium. This covers the effects in the immediate surroundings of a supernova, the heating in the interstellar medium, and the dynamical impact within the ISM. The interactions on larger scales like the impact on galactic outflows are discussed further below. We will not cover the acceleration mechanism in SN remnant shocks as this involves the kinetic approach of CRs and is beyond the scope of this review.

Simulations of the interstellar medium allow to resolve the different thermal phases of the gas including the degree of ionisation. The spatial resolution also allows to resolve MHD shocks. On the one hand this allows to investigate in more detail the coupling of CRs with the thermal gas. On the other hand CRs can be injected at the position of the shock and allow a more consistent injection mechanism.

In the ISM the transport speeds of CRs and the local tubulent motions can be comparable, so the local conditions determine whether the CRs diffuse faster than they are advected or vice versa. \citet{CommerconMarcowithDubois2019} perfom simulations of the turbulent ISM and find that for most cases the effective diffusion speeds are dominating. As a result, no significant CR pressure gradient can build up and no dynamical effects are expected on scales of a few tens of parsecs.

\cite{2017MNRAS.465.4500P} extend the two-fluid shock tube by CR injection at the shock and applies an on-the-fly shock finder \citep{SchaalSpringel2015} in several idealised setups such as a Sedov explosion \citep{Sedov1959} using the \textsc{Arepo} code \citep{Springel2010}. \cite{paisetal2018} added an obliquity-dependent injection efficiency with respect to the magnetic field orientation and investigated the impact of CRs injected in the shock region on the Sedov explosion finding that the solution remains self-similar because the ellipticity of the propagating blast wave stays constant over time. Furthermore, their comparison to observed SN remnants suggests a lower injection efficiency of only 5 per cent rather than the canonical 10 per cent of the SN energy. This result is independent of the assumed magnetic coherence length.

\citet{2019A&A...631A.121D} use a similar approach to inject CRs at the loci of the MHD shocks with an on-the-fly shock finder in the larger boxes of the interstellar medium using \textsc{Ramses}. The efficiency is also coupled to the upstream magnetic obliquity. Their models include CR streaming as well as ansiotropic diffusion. Similar to \citet{paisetal2018} they find supernova bubbles with large polar caps in case of a homogeneous background magnetic field, and a patchy structure of the CR distribution in case of inhomogeneous background fields. The application in a turbulent box shows that the presence of shock-injected CRs significantly modifies the structure of the gas.

A more idealized study by \citet{2017MNRAS.467..646W} investigates  a stream of CRs interacting with cold clouds in a hot dilute environment. Using one-dimensional models including CR streaming they find that the bottleneck effect, which occurs when a population of CRs undergoes the streaming instability, will accelerate and heat the gas cloud. Corresponding two-dimensional models including radiative cooling of the clouds \citep{2019MNRAS.489..205W} confirm the acceleration of clouds by CRs, but reveal that the thermal impact might dominate. The cooling effect will keep the clouds intact to CR wave heating.

\subsection{CR driven large-scale instabilities of the ISM}

Observational data indicate that gas, magnetic fields and CRs appear in approximate energetic equipartition, which is interpreted   as  the effect of dynamical coupling of the diverse components of the interstellar medium. The stability of a system in vertical equilibrium
\begin{equation}
\frac{\dd}{\dd z}\left(P_\mathrm{therm} + P_\mathrm{CR} + \frac{B^2}{8\pi}\right) = -\rho g_z
\end{equation}
without CRs was analysed in \citet{Newcomb1961}; the inclusion of CRs dates back to \citet{1966ApJ...145..811P}. \citet{ZweibelKulsrud1975} generalized the model for a family equilibria. The dynamical role of CRs was first recognized by \cite*{1966ApJ...145..811P}, who noted that vertically stratified ISM consisting of thermal gas, magnetic fields and CRs is unstable due to buoyancy of the weightless components, i.e. the magnetic fields and the CRs. 

The physical  mechanism of the instability stems from the fact that two weightless components  -- magnetic fields and cosmic rays -- contribute to the total pressure of the gravitationally stratified interstellar medium. The thermal gas component is inflated by these nonthermal pressure contributions. 
The system is prone to a kind of convective instability tending to expel the weightless components from the system and therefore to reduce potential energy of the thermal gas. Linear stability analysis demonstrates that small-amplitude vertical corrugations  of magnetic field lines lead to gas motions down to valleys of magnetic fields unweighting  tops of the field lines and therefore enhancing the upward motion of the upper parts of the corrugations and a downward motion of gas into the valleys.

The instability, hereafter named  {\em Parker instability}, was recognized as a plausible mechanism leading to the formation of clouds of gas collecting in valleys of the large-scale galactic magnetic fields \citep{1967ApJ...149..517P,1980ApJ...238..148B}. %
\cite{1992ApJ...401..137P}  proposed, furthermore, that the instability enhanced by continues replenishment of CRs in supernova remnants may lead to an amplification of large-scale magnetic  fields in galaxies.

Numerous papers addressing the Parker instability in the ISM by means of linear stability analysis adopted a simplifying assumption that the diffusive propagation of CRs is fast enough to ensure a constant pressure of CRs along magnetic field lines.  \cite{RyuEtAl2003} performed linear a analysis of the Parker instability for the medium composed of thermal gas, horizontal magnetic field and  anisotropicaly diffusing CRs. The relevant set of equations  consisted of the set of ideal MHD equations and the diffusion-advection equation equivalent to Eqn.~(\ref{eq:e_all}) with the diffusion tensor in the form of Eqn.~(\ref{eq:diff-tensor}) describing the magnetic-field aligned anisotropic propagation of CRs. Their results have shown a strong dependence of the growth rate of the Parker instability on the value of the parallel diffusion coefficient. Lower diffusion coefficients turned out to reduce the growth rate of the instability with respect to the  case of constant CR pressure along magnetic field lines, representing the limit of an infinite diffusion coefficient.

The nonlinear evolution of the Parker instability with CRs was addressed by \cite{2003A&A...412..331H}, who extended the ZEUS 3D code with  a stable numerical algorithm combining anisotropic diffusion and advection of CRs and coupled the momentum-integrated  diffusion equation with the  MHD  system equations describing a thermal plasma (see Sect. \ref{subsect:two_fluid_models}). Numerical simulations have shown that the growth rate of the Parker instability depends essentially on the adopted value of the CR diffusion coefficient and that results differ from predictions made on the grounds of linear stability analysis. The  difference was  plausibly caused by a different triggering mechanism of the instability, which in \citep{RyuEtAl2003} relied on excitation of a specific 'idealized' eigenmodes,  while  in \citep{2003A&A...412..331H} the instability was excited by an instantaneous injection of CRs in a SN remnant.

The linear and nonlinear analysis of the effects of CR diffusion on the Parker instability was subsequently extended by \cite{2004ApJ...607..828K} who have shown that  the growth rate of the Parker instability becomes smaller if the coupling between CRs and thermal gas is stronger (i.e., if the CR diffusion coefficient is smaller). MHD simulations of the Parker instability with an appropriate perturbation confirmed this result. 
Similar conclusions were derived from studies of Parker--Jeans instability with anisotropic CR diffusion   \citep{2006ApJ...636..290K}. The system is less unstable when the CR diffusion coefficient is smaller (i.e., the coupling between the CRs and plasma is stronger) and if the CR pressure is larger. This conclusion is consistent with the fact that Jeans instability and Parker instability are less unstable when the pressure is higher. In the nonlinear regime the Parker--Jeans instability leads to the formation of gas filaments whose orientation depends on the parallel diffusion coefficient \citep{2020ApJ...899...72K}.
When the diffusion coefficient is large gas filaments form preferentially perpendicular to the magnetic field and when the diffusion coefficient is small the filaments are aligned parallel to the magnetic field. 

\cite{2016ApJ...816....2R}   examined the evolution of the Parker instability in galactic disks using 3D numerical simulations.  They have found that the instability develops a multimodal 3D structure, which cannot be quantitatively predicted from the earlier linearized studies. They calculated synthetic polarized intensity and Faraday rotation measure (RM) maps, and the associated structure functions. They suggested that correlation scales inferred from RM maps can be used to  probe the cosmic-ray content of galaxies. 

\cite{2018ApJ...860...97H}  performed a stability analysis of a stratified layer for three different cosmic-ray transport models: decoupled, corresponding to the classic Parker instability, coupled with $\gamma_c=4/3$ but not streaming,  named as modified Parker instability, and coupled with streaming at the Alfv\'{e}n speed. They  demonstrated that cosmic-ray heating of the gas is responsible for the destabilization of the system and concluded that  Parker instability with cosmic-ray streaming may play an important role in cosmic-ray feedback.  \cite{2020ApJ...891..157H}  expanded the work by including radiative cooling. Heating due to CR streaming has  a destabilizing effect which affects significantly the nonlinear regime of the Parker instability. While cooling depressurizes the dense gas, streaming CRs heat and inflate the diffuse extraplanar gas, greatly modifying the phase structure of the medium. The fastest growth affects typically the modes characterized by short wavelengths in the horizontal direction perpendicular to the background magnetic field.

\subsection{Cosmic ray driven galactic dynamo}
\note{0-application-dynamo.tex}

\par The idea of the CR-driven dynamo  was originally raised by \cite*{1992ApJ...401..137P}, who postulated that the buoyancy of CRs together with the Coriolis-force, galactic differential rotation and magnetic reconnection can lead to efficient amplification of galactic  magnetic fields.

Preliminary numerical experiments pursued with the aid of the thin flux-tube approximation \citep{2000ApJ...543..235H} have shown that the buoyancy of CRs injected in SN remnants may lead, in the presence of the Coriolis force, to efficient generation of the poloidal magnetic field component out of the initial  azimuthal magnetic field  (named as $\alpha$-effect) and to the diffusive transport of magnetic flux in the vertical direction. The CR-induced diffusive transport coefficients for the large mean magnetic field, plugged into the mean-field dynamo equation \citep{2003A&A...401..809L}, \citep[see also][]{2006A&A...445..915K} resulted in exponentially growing solutions, confirming the Parker's conjecture that CRs may efficiently drive the dynamo action on galactic scales.

First MHD numerical simulation models \citep{2004ApJ...605L..33H,2006AN....327..469H,2009A&A...498..335H,2010A&A...510A..97S} of the CR-driven dynamo were realized in a local, rectangular patch of galactic disk with shearing boundary conditions and rotational pseudo-forces (tidal and Coriolis forces) incorporated to study the magnetic field evolution in rotating galactic disks. 

The first global  model of the CR-driven dynamo in a Milky Way-type galaxy  \citep{2009ApJ...706L.155H} assumed that:
(1) Supernovae convert  $10\%$  of their explosion kinetic energy into cosmic rays.  
(2) A weak initial magnetic field of stellar origin was suppled to the system in selected  (plerionic type) SN remnants. 
(3) Differential rotation of the interstellar gas results from an assumed analytical model of an axisymmetric galactic gravitational potential \citep[e.g.][]{1991RMxAA..22..255A} or from a computational model of an N-body galactic disk \citep[][]{Woltanski-PhD-thesis}.
(4) The initial gas distribution follows the global Milky Way model by \cite*{1998ApJ...497..759F}.

\begin{figure*}
  \centerline{\includegraphics[width=0.45\textwidth]{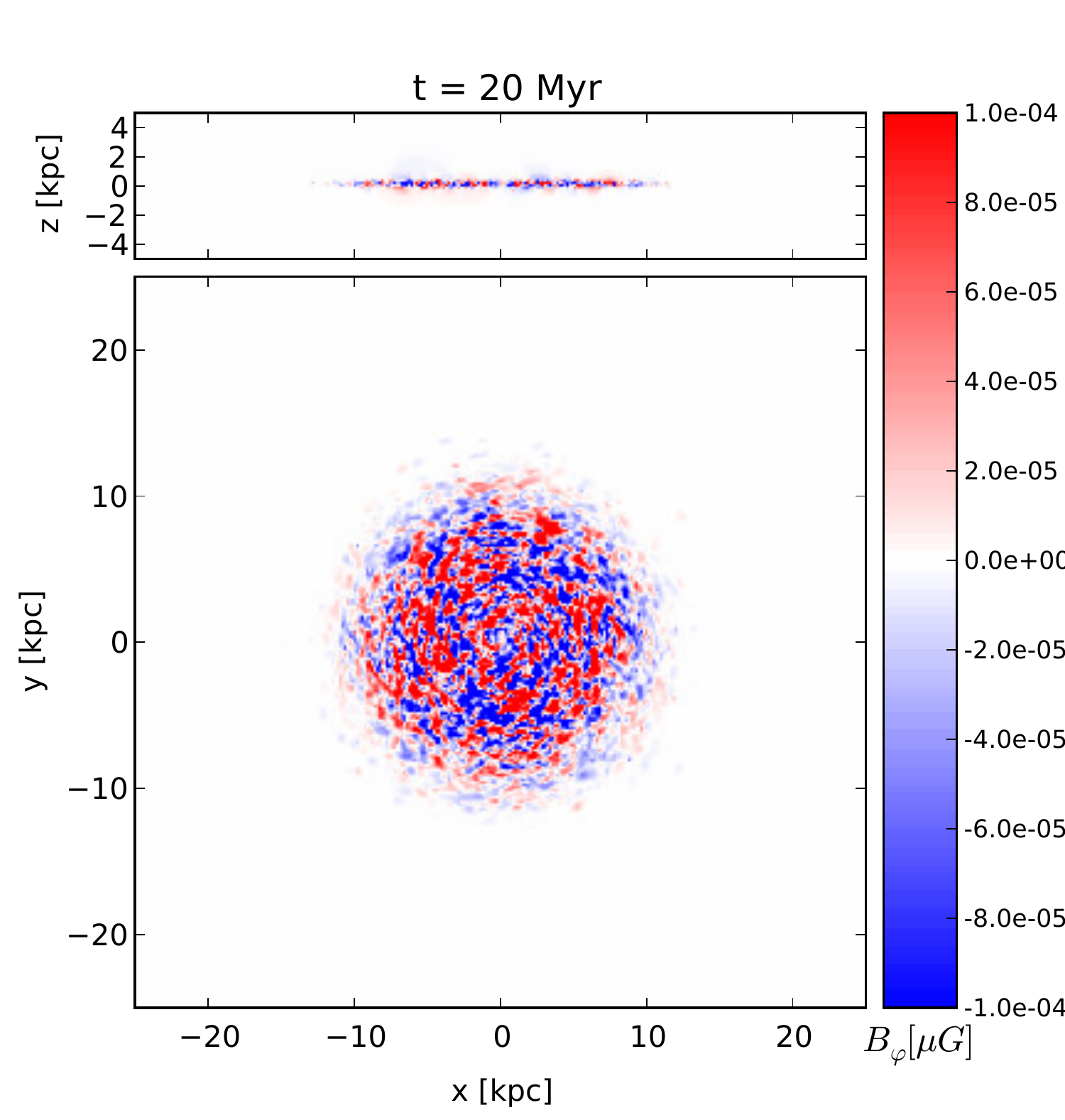}
              \includegraphics[width=0.45\textwidth]{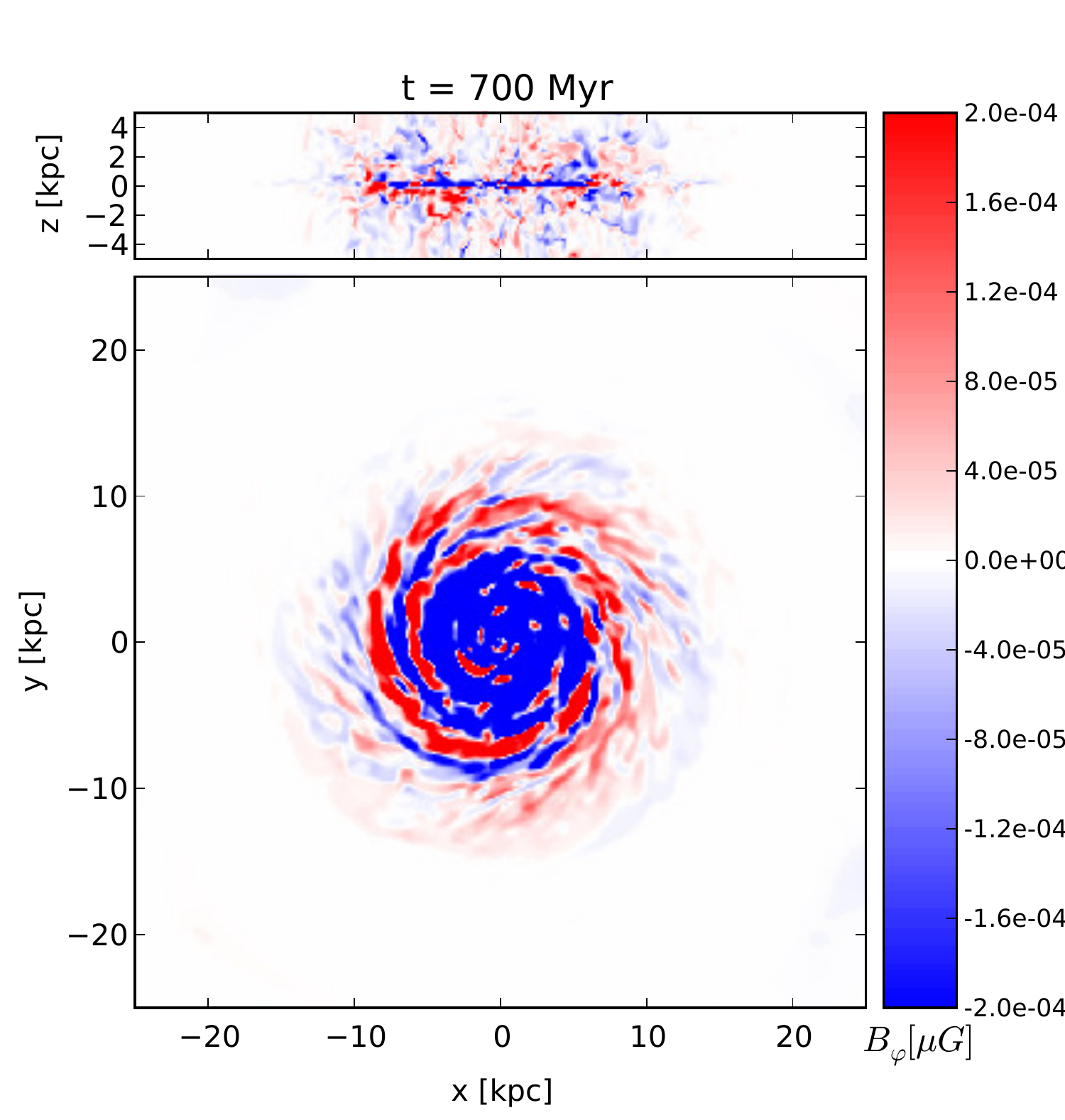}}
  \centerline{\includegraphics[width=0.45\textwidth]{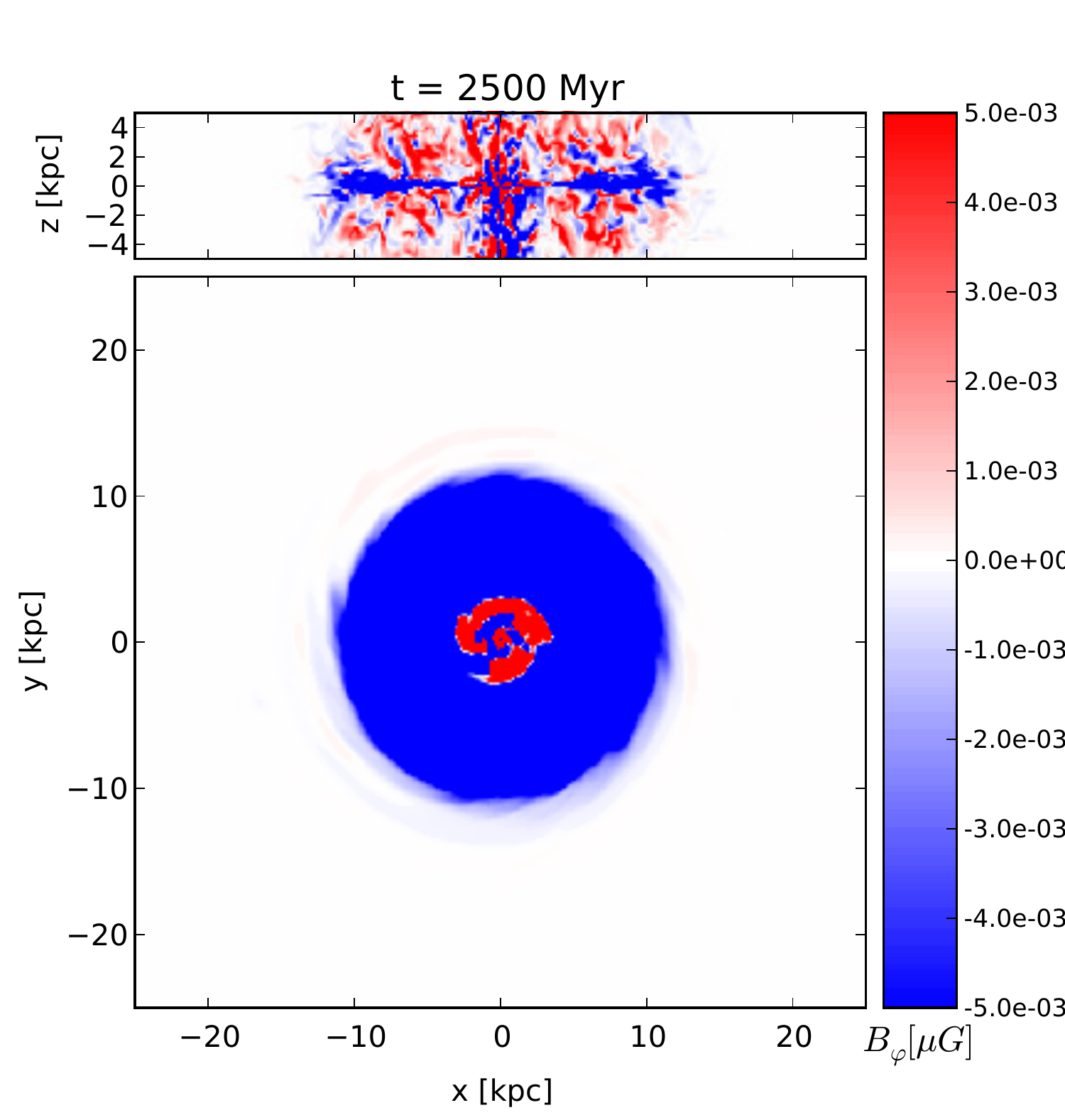}
              \includegraphics[width=0.45\textwidth]{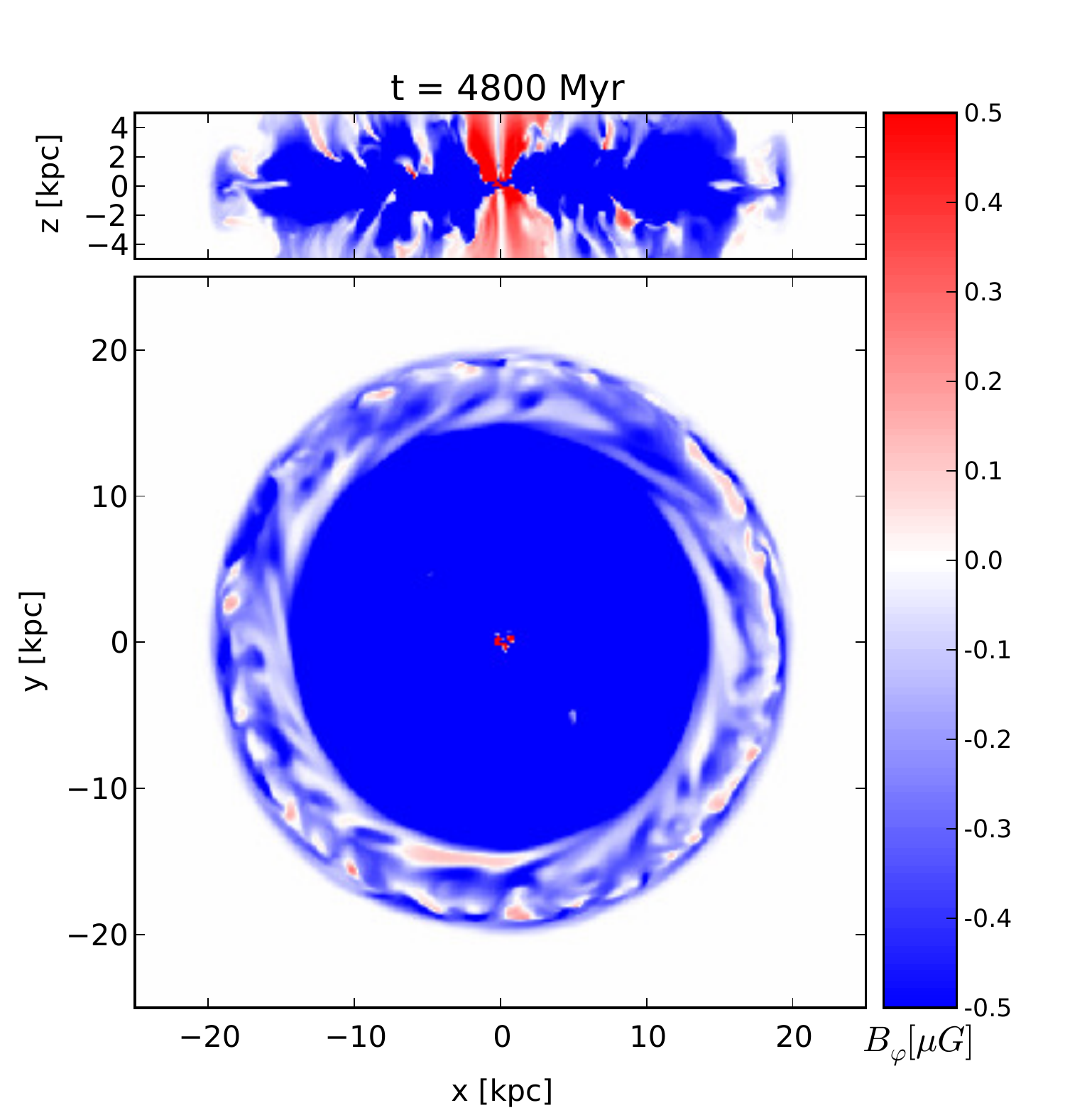}} 
\caption{Distribution of toroidal magnetic field at $t=20\,{\Myr}$ (top-left), $t=700\,{\Myr}$ (top-right), $t=2.5\,{\Gyr}$ (bottom-left), 
and $t =4.8\,{\Gyr}$ (bottom-right). Unmagnetised regions of the volume are white, while positive and negative toroidal magnetic fields are marked with red and blue respectively. Note that the colour scale in magnetic field maps is saturated to enhance weaker magnetic field structures in disk peripheries.
The maximum magnetic field strength are $5.9 \cdot 10^{-4}$, $4.4 \cdot 10^{-3}$, $1.5$ and $29\,\muG$ at $t=0.02$, $0.7$,  $2.5$ and $4.8\,\Gyr$ respectively \citep{2009ApJ...706L.155H}.}
\label{fig:slices-bb}
\end{figure*}				
\begin{figure}
\centerline{ \includegraphics[width=0.45\columnwidth]{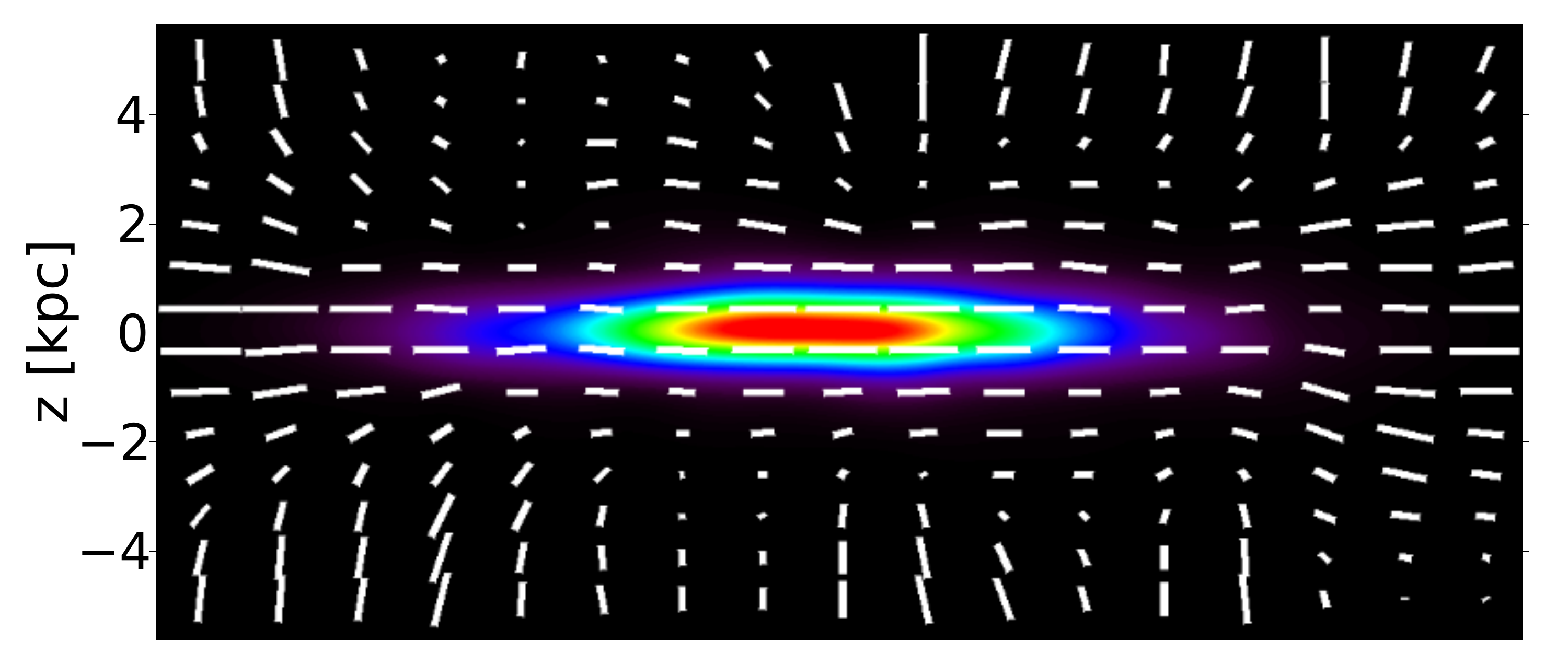}}
\centerline{ \includegraphics[width=0.45\columnwidth]{./hanasz_2009-f4a.pdf}}
\caption{Synthetic radio maps of polarised intensity (PI) of synchrotron emission, together with polarisation vectors are shown for the edge-on and face-on views of the galaxy at $t=4.8{\Myr}$. Vectors direction resembles electric vectors rotated by $90^{\circ}$, and their lengths are proportional to the degree of polarisation \citep{2009ApJ...706L.155H}.}
\label{fig:radiomaps}
\end{figure}
\par The amplification of the magnetic field originating from the small-scale, randomly oriented dipolar magnetic fields is apparent through the exponential growth by several orders of magnitude of both the magnetic flux and the magnetic energy. %
The growth of  magnetic field saturates at about $t = 4\,\Gyr$, reaching values of $3-5\,{\muG}$ in the disk. During the
amplification phase, magnetic flux and total magnetic energy grow by about 6 and 10 orders of magnitude, respectively. The average e-folding time of
magnetic flux amplification is approximately equal to $270\,{\Myr}$, corresponding to the orbital rotation period at the galactocentric radius ($\approx10{\,\kpc}$). The magnetic field originated from randomly oriented magnetic dipoles was initially chaotic, as shown in Fig.~\ref{fig:slices-bb}. Later on, the toroidal magnetic field
component formed a spiral structure revealing reversals in the plane of the disk. The magnetic field structure evolved gradually by increasing its correlation scale.
The toroidal magnetic field component became almost uniform inside the disk around $t=2.5\,\Gyr$. The volume occupied by the well-ordered magnetic field expanded continuously until the end of the simulation.
\par To visualise the magnetic field structure in a manner resembling radio observations of external galaxies,  synthetic radio maps of the synchrotron radio-emission were deduced in a simplified way from the distribution of CR protons. The polarised intensity of synchrotron emission  is shown  in Fig.~\ref{fig:radiomaps} together with polarisation vectors. Electric
vectors, computed on the basis of integrated Stokes parameters, are rotated by $90^{\circ}$ to reproduce the magnetic field direction projected onto the plane of sky.  Polarisation vectors, indicating the mean magnetic field direction, reveal a regular spiral structure in the face-on view, and the so-called \textit{X-shaped structure} in the edge-on view. A particular similarity can
be noticed between the edge-on synthetic radio map and the radio maps of observed edge-on galaxies such as NGC 891 \citep{2009RMxAC..36...25K}.

\cite{2011ApJ...733L..18K} added an analytical elliptical component to the axisymmetric gravitational potential. In the presence of the bar perturbation the CR-driven dynamo reveals new properties, such as the presence of a ring-like structure as well as a shift of the magnetic arms with respect to the crests of spiral density waves. \cite{Woltanski-PhD-thesis} used N-body simulations to compute gravitational field of a nonaxisymmetric disk with spiral density wave perturbations. Density waves were excited in the disk by addition of a small satellite galaxy. The amplification of magnetic flux by the CR-driven dynamo leads the large-scale magnetic field to the saturation level around  $10\,\muG$ with local maxima reaching  $20-30\,\muG$, located near the spiral crests of gas density.  Magnetic field vectors are parallel to the gaseous spiral arms as it is observed in real galaxies \citep{2011MNRAS.412.2396F}.  The strongest  magnetic fields are generated  in between gaseous spiral arms \citep{2011ApJ...733L..18K,2015A&A...575A..93K} similarly to magnetic   arms observed in some galaxies, such as: NGC 6946 \citep{2007A&A...470..539B} and IC 343 \citep{2015A&A...578A..93B}. 

The edge-on synchrotron radio maps  of the galaxy exhibit polarized synchrotron emission, extending several kiloparsecs above and below the disk.  Similar structures are common in radio-images  of edge on galaxies \citep[see e.g][]{2009RMxAC..36...25K,2011A&A...531A.127S,2013A&A...560A..42M}. The X-type structures are present in all global models of  CR-driven galactic dynamo \citep{2009ApJ...706L.155H,2011ApJ...733L..18K,2015A&A...575A..93K}.  Global-galactic winds accompanying the magnetic field amplification are present in practically all numerical realizations of  the CR-driven dynamo. 
The commonly present CR-driven winds leaving galaxies  imply magnetic field  transport out of the disk and effective magnetisation of intergalactic space. This process is particularly efficient in dwarf galaxies \citep{2014A&A...562A.136S}. 

An interesting observation has been made by \cite{2017ApJ...843..113B} who performed numerical simulations of an isolated galaxy with a stellar feedback prescription in which thermal energy and magnetic energy are supplied by supernova explosions, but CRs are not included. 
They noted that, similarly to the model of the CR-driven dynamo, the magnetic field reaches equipartition levels over gigayear timescales and 
 raised the question of the relative importance of the two primary types of energy supplied by supernovae: 
  turbulence driven by the thermal expansion of the remnants themselves, or energy released as CRs, 
 since either by itself appears to be sufficient.

\subsection{Cosmic ray driven galactic winds}

As supernovae drive strong shocks into the interstellar medium (ISM), some fraction of the explosion energy is consumed to accelerate ionised particles to relativistic energies, which are then injected into the ISM \citep{Krymskii1977,Bell1978,BlandfordOstriker1978}. This
relativistic fluid is coupled to the galactic magnetic field and, in particular the hadronic component, is less prone to energy losses than the gaseous
component of the ISM.

The idea of CR  wind driving has been proposed by \cite{1975ApJ...196..107I} and developed by numerous authors including  \cite{1991A&A...245...79B,1993A&A...269...54B,1996A&A...311..113Z,1997A&A...321..434P,2002A&A...385..216B,2012MNRAS.423.2374U,2012A&A...540A..77D},  who find that CRs together with magnetic fields and thermal pressure can contribute to the galactic winds phenomenon. \cite{EverettEtAl2008} applied a wind model, driven by combined cosmic-ray and thermal-gas pressure, to the Milky Way, and shown that the observed Galactic diffuse soft X-ray emission can be better explained by a wind than by previous static gas models. They find that cosmic-ray pressure is essential to driving the observed wind.

CRs are strongly coupled to magnetic fields and their mutual interaction should be followed in a self-consistent
way. \citet{2004ApJ...605L..33H,2009ApJ...706L.155H}, \citet{2010A&A...510A..97S}, \citet{2011ApJ...733L..18K} and  \cite{2018ApJ...860...97H}  have shown that CRs induce buoyancy effects in the ISM, leading to the break-out of magnetic fields from galactic disks  \citep{1992ApJ...401..137P}. Plausibly, such processes are also relevant for star-forming galaxies at high redshift which are observed to have significant magnetic fields at the level of tens of~$\muG$ \citep{2008Natur.454..302B}. Recent observations even demonstrate the existence of large magnetic fields up 50~kpc away from the galaxy, indicating strong large-scale magnetised winds \citep{2013ApJ...772L..28B}.

\subsubsection{1D models}
An early popular model for investigating the impact of CRs on galactic winds are one-dimensional models. Usually, they focus on large scales compared to the details of galactic substructure, such that details of the interstellar medium and substructures inside the galaxy are neglected. The infinitesimally thin disc is modelled as the source of CRs. The simplified geometry does not allow for details of the magnetic field or turbulence to be included, but under the assumption of external confinement of CRs by isotropic turbulent motions or a magnetic field model the radial dynamics of a wind including CRs can be modelled.
The first application of CRs hydrodynamics in the context of galactic winds was performed by \citet{1991A&A...245...79B} assuming a steady state flow along magnetic flux tubes. They found that CRs can drive supersonic mass loss for a wide range of parameters. Exploring the parameter space of thermal and CR pressure on the mass loss has been done by \citet{EverettEtAl2008} finding that in the Milky Way CRs and thermal pressure contribute equally to the outflow. Extensions of the models to time-dependent winds \citet{2012A&A...540A..77D} confirm asymptotic solutions of previous steady state problems and also offer an explanation to the observed galactic CR spectrum. More recent models include a full CR spectrum in a semi-analytic model of advective and diffusive CR transport \citep{2016MNRAS.462L..88R,2016MNRAS.462.4227R,2017MNRAS.470..865R}. Whereas the qualitative results of winds are reproduced, the inferred spectra are in tension with observed ones.

\subsubsection{3D models}

\begin{figure}
\includegraphics[width=\textwidth]{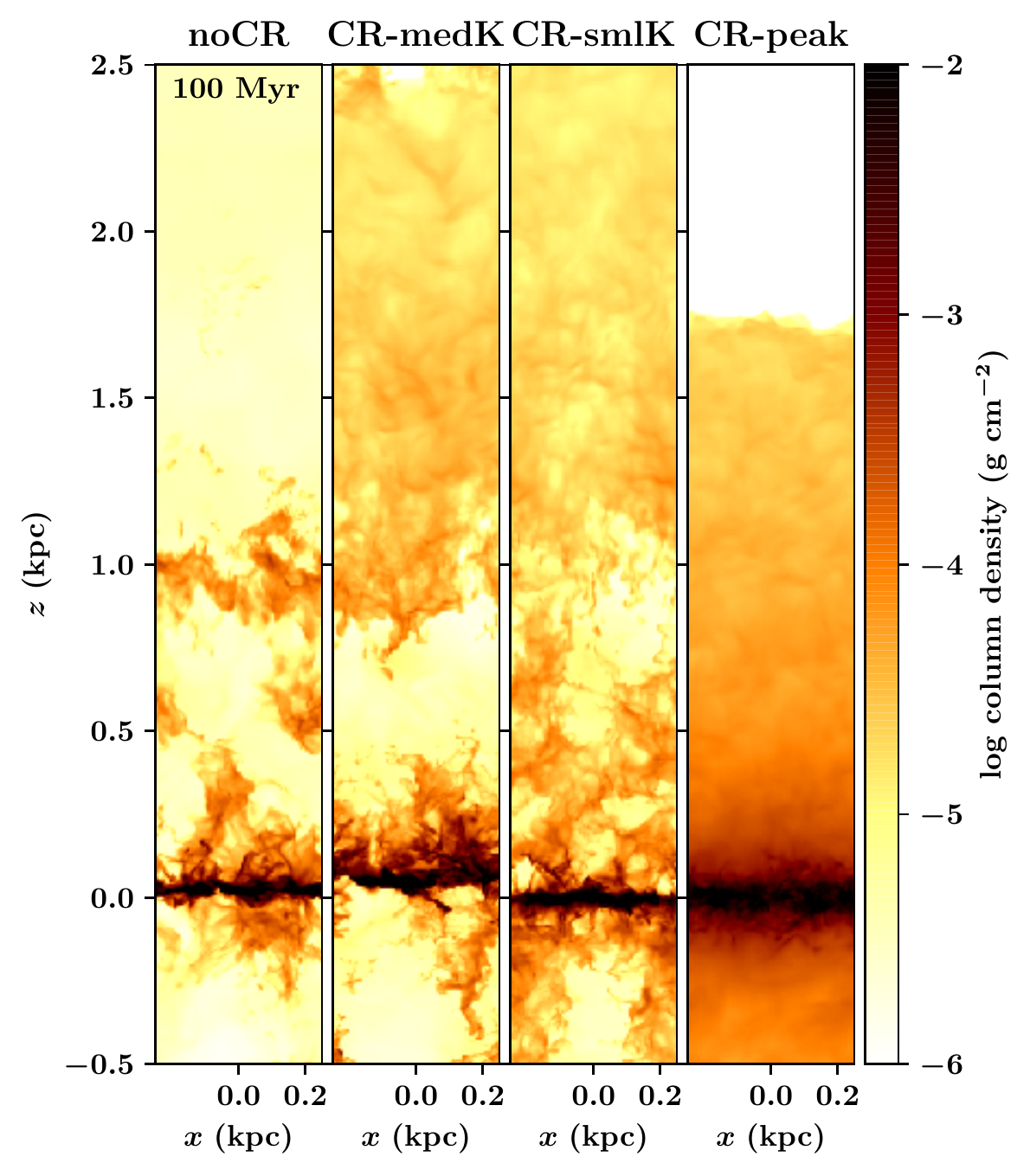}
\caption{Comparison of the column density structure in simulations of the interstellar medium. The left-hand column shows a setup with only thermal energy injection. The three other simulations include CR injection with an efficiency of 10\% of the SN energy. The two central panels show two different diffusion coefficients. In the right-hand panel the SNe all explode in dense regions, which limits their thermal impact. This illustrates that CRs alone are able to lift gas out of the galactic disc \citep{GirichidisEtAl2018a}}
\label{fig:stratbox-comparison-girichidis}
\end{figure}

Three dimensional models of the interstellar medium and galaxies including CRs allow for a more natural combination of turbulent motions, a dynamical magnetic field evolution and the interactions with CRs. However, the numerical cost only allows for reduced CR physics and a small parameter range to be explored. 
\citet{2008A&A...481...33J}
conduct simulations of isolated galaxies including the feedback of the diffusive CR component included in the GADGET code. They find that CRs can significantly reduce the star formation efficiencies of small galaxies, with virial velocities below $\sim 80 \km \s^{-1}$, an effect that becomes progressively stronger towards low-mass scales. 
\citet{2012ApJ...761..185Y} simulate the onset of a CR-driven outflow comparable to the Milky Way Fermi bubbles using anisotropic CR diffusion. \citet{2012MNRAS.423.2374U} use SPH simulations of entire isolated galaxies to test the wind dynamics including CRs including CR heating due to streaming. Without magnetic fields, the simulations approximated the Alfv\'{e}n velocity by the sound speed and only focus on isotropic transport. Similar approaches have been employed by \citep{BoothEtAl2013, SalemBryan2014} using isotropic CR diffusion. Both studies find that CRs are able to drive winds, but only for some fraction of the probed parameter space. 
\citet{BoothEtAl2013}  note that the effects of CR-driven winds are much larger in dwarf galaxies compared to their Milky Way models, which is still affected by CRs.
More recent simulations by \citet{DashyanDubois2020} investigate in more detail the impact of CR on dwarf systems with different CR transport mechanisms. They report a stronger impact on the different transport mechanism than the effective diffusion coefficient. 

In a similar framework \citet{2016MNRAS.456..582S} 
investigate the impact of CRs on a forming $10^{12} \Msun$ halo.
 They find that CR-inclusive runs, contrasted with a run with star formation and energetic feedback, but no CRs, substantially enrich the circumgalactic medium with metals due to robust and persistent outflows from the disk. The CR-inclusive models reveal more diffuse gas at lower temperatures, down to $10^{4} \K$, than the non-CR cases. The CR inclusion leads to a better match of HI, SiIV, CII, and OIV line intensities than the case without CRs. Comparison of gamma-ray luminosity to  observational data favor CR  diffusion coefficients close to the Milky Way canonical values $3\cdot 10^{28} \cm^2 \s^{-1}$.

Galaxy simulations including magnetic fields and anisotropic CR diffusion are performed by \citet{2013ApJ...777L..38H}. Their models are not only able to drive galactic winds with mass loading factors of order unity but also sustain a CR-driven dynamo with magnetic field strength at a several $\mathrm{\mu G}$ level. \citet{2016ApJ...824L..30P} compare isotropic and anisotropic diffusion models finding that anisotropic diffusion leads to more realistic magnetic field strengths than in the isotropic case. Including CR streaming in galactic models was investigated by \citet{WienerPfrommerOh2017,2017ApJ...834..208R}.
\cite{2018ApJ...868..108B} 
compare the role of isotropic and anisotropic diffusion and streaming mechanisms for shaping the structure of circumgalactic medium (CGM). They find that all three transport mechanisms result in strong, metal-rich outflows but differ in the temperature and ionization structure of their CGM. Isotropic diffusion results in a spatially uniform, warm CGM that underpredicts the column densities of low ions. Anisotropic diffusion develops a reservoir of cool gas that extends farther from the galactic center, but disperses rapidly with distance. CR streaming projects cool gas out to radii of 200 kpc, supporting a truly multiphase medium. 

A systematic parameter study of the impact of CR for different halo masses was performed by \citet{JacobEtAl2018} finding that CRs only have a significant effect on the launching of a wind for halo masses below $10^{12}\,\mathrm{M}_\odot$. The combined effects of AGN and CRs has been simulated by \citet{WangRuszkowskiYang2020} in isolated galaxies, where the authors argue that CR streaming and the corresponding heating has to be included in order to prevent too efficient cooling of the circumgalactic medium.

Besides the isolated disc setups, recent models include CR transport in cosmological zoom simulations, in which individual galaxies are evolved with high resolution. \citet{BuckEtAl2019} rerun simulations from the AURIGA project including CRs. They limit the transport coefficients to the Alfv\'{e}n speed and include the corresponding heating effect, which changes the structure of the outflow. The modified structure of the circumgalactic medium results in a different distribution of angular momentum and thereby alters the stellar and gaseous disk. In a similar cosmologial setup \citet{HopkinsEtAl2020b} compare a large variety of subgrid models for CR confinements, which are connected to the MHD equations using effective transport coefficients in the diffusion, the streaming as well as combined unified CR transport. Besides the moderate effect that CRs have on the galactic winds, their models favour effective diffusion coefficients that range from $10^{29}-10^{33}\,\mathrm{cm}^2\,\mathrm{s}^{-1}$,  which is based on the comparison with gamma-ray emission.
The details of how CR winds are driven and in particular the relative importance and dynamical interplay between SN-driven motions and CR dynamics cannot be investigated in full galactic models at the current resolutions. Instead simulations of a smaller fraction of the galaxy are needed. \citet{2016ApJ...816L..19G}, \citet{2015ApJ...813L..27P} and \citet{GirichidisEtAl2018a} compare thermally and CR driven winds in more sophisticated model of the ISM, which includes a chemical evolution to accurately model the phases of the ISM. The models confirm that CRs alone are able to drive and support an outflow with mass loading of order unity. The CR driven outflows are denser, cooler and smoother than their thermally driven counterparts. Figure~\ref{fig:stratbox-comparison-girichidis} illustrates the difference between outflows that are only driven by thermal energy injection by SNe (left-hand panel) and the outflows driven 10\% of the energy injected as CRs (three right-hand panels). The two central panels show two different diffusion coefficients. The column density on the right-hand side depicts the structure of the ISM if all SNe explode in dense regions. The resulting strong overcooling illustrates that CRs alone are able to lift gas out of the disk. Similar models by \citet{SimpsonEtAl2016} compare simulations with advection only and advection plus diffusion. They conclude that diffusion is needed in order to drive an outflow. The decoupling of CR and the thermal gas in a neutral environment leads to a broader spatial distribution of cosmic rays and higher wind speed compared to the uniformly applied advection-diffusion approach \citep{FarberEtAl2018}. A more sophisticated, locally determined coupling between CRs and the gas confirms these results \citep{HolguinEtAl2019}.

\subsection{Cosmic rays in galaxy clusters}

CRs are also expected to affect galaxy clusters. One particularly important aspect is the possible heating effect provided by CR protons from AGN. Simulations including CR protons produced at shocks reveal that they could provide a substantial additional pressure in the intracluster medium \citep{2001ApJ...559...59M}. A follow-up study included CR electrons as an additional non-thermal component \citep{2001ApJ...562..233M}. In both cases the CRs are transported with the gas and include cooling effects. Simulations including diffusion of CR protons that are injected in shock regions close to the AGN find large X-ray cavities and radio lobes in the hot diffuse gas of the central regions of galaxy clusters. The cavities are long-lived if the diffusion coefficient does not exceed $10^{28}\,\mathrm{cm^2\,s^{-1}}$ \citep{2007ApJ...660.1137M,2008ApJ...676..880M,2008MNRAS.383.1359R}. A range of galaxy clusters and shock related CR injection models has been performed by \citet{2007MNRAS.378..385P} omitting CR diffusion. They find that CRs can be efficiently accelerated in strong structure formation shocks such as accretion and merger shocks. The high relative fraction of CR pressure in the central region increases the compressibility and allows for more star formation. A detailed discussion on the differences between CR advection, diffusion and streaming by \citet{2011A&A...527A..99E} suggest that the details of the transport mechanisms are essential to understand the non-thermal signatures of clusters. Hydrodynamical simulations including CR streaming and the resulting heating effect have been run by \citet{2017ApJ...844...13R} find that the CR can efficiently heat the gas and provide a viable channel for the AGN energy thermalization.
More recently, \citet{EhlertEtAl2018} couple CRs to an AGN jet model and simulate galaxy clusters including a simplified streaming approach. They introduce an effective CR diffusion coefficient $\kappa_\mathrm{cr}\sim l_\mathrm{cr}v_\mathrm{A}$ with the CR gradient length $l_\mathrm{cr}$ and the Alfv\'{e}n speed $v_\mathrm{A}$, which emulates the combined effects of streaming and spatial diffusion \citep{SharmaEtAl2009, WienerPfrommerOh2017}. They conclude that the CR-induced Alfv\'{e}n heating matches the CR heating rates needed to solve the cooling flow problem. 
 \citet{VazzaEtAl2012} model CRs in galaxy clusters by injecting them in numerically identified shocks and advect them with the gas flow. The simulations reveal a small but not dominant impact of CRs on the evolution with changes at the percent level for the temperature, pressure and density distribution.

One of the first attempts to investigate the spectrum of relativistic electrons in cosmological shocks has been made by \citep{2003ApJ...585..128K} who studies the radiation emitted due to inverse Compton process by shock-accelerated electrons in hydrodynamic cosmological simulations of a Lambda cold dark matter ($\Lambda$CDM) universe. They performed  a posteriori detection and analysis of shocks forming in cosmological simulations  to predict the spectrum of CR electrons and  the spectrum of $\gamma$-ray photons emitted from cosmological shocks. They subsequently constructed all-sky maps of $\gamma$-ray emission from the nearby universe.

\cite{2006MNRAS.367..113P} derived an analytic solution to the one-dimensional Riemann shock tube problem for a composite plasma of CRs and thermal gas. They applied their solution to study the properties of structure formation shocks in high-resolution hydrodynamic simulations of the $\Lambda$CDM universe. They found that most of the energy is dissipated in weak internal shocks with Mach numbers $M \sim 2$. They recognized the dynamical importance of shock-injected CRs which is comparatively large in the low-density, peripheral halo infaling regions, buts is less important for the weaker flow shocks ocuring in central high-density regions of haloes. They raised  questions of cosmological implications of the CRs component and for observational signatures of this radiation.

In the context of AGN feedback \citet{SijackiEtAl2008} include CRs together with thermal feedback. The two-component fluid with an effectively softer equation of state and the less efficient cooling for CRs compared to the thermal gas allows CR supported bubbles to rise to the outskirts of the clusters and leak into the surrounding intergalactic medium. However, neither the accretion rates onto the black hole nor the star formation rate are significanty affected by the non-thermal component.

\section{Conclusions and outlook} 

We have reviewed the numerical treatment of CRs assuming a fluid description of this high-energy component. We identified three main approaches in treating CRs numerically. The first is a spectrally resolved approach with a focus on the physical processes of CRs like the primary CRs and the production of secondaries. The most prominent code that represents this approach is \textsc{Galprop} with more recent frameworks \textsc{Dragon} and \textsc{Picard}. The second numerical model focuses on the dynamical coupling of CRs and the thermal gas in (magneto)-hydrodynamical simulations. Here, CRs take the role of dynamical drivers of hydrodynamical flows. In the context of the interstellar medium and galaxy formation CRs are an important additional reservoir of energy. The numerical complexity in this approach is more in the interaction between the two fluids. Simple linear diffusion models are challenged by more complex models that account for the CR streaming instability and couple CRs to the gas including more plasma effects. A third approach tries to spectrally resolve CRs and at the same time include the dynamical impact onto the gas dynamics. The numerical and computational complexity forces this approach to be limited in both spectral complexity and multiple species of CRs and in the details of plasma interactions of relativistic and thermal gas.

Depending on the application the currently developed models still face major limitations. In the interstellar medium and for the dynamics of galaxies all three approaches would need to be merged and coupled. For the connection with observations -- in particular for a comparison with the Milky Way -- a detailed modeling of secondaries is of major importance. Similarly, a spectrally resolved modeling of CR electrons as tracers of local thermal and CR properties are key to understand the galaxies. The structure of the magnetic field in galaxies is strongly connected to the dynamics of CRs. Magnetohydrodynamical simulations naturally include an evolving magnetic field that follows and/or shapes spiral structures and interacts with dynamos and local instabilities of the gas like Parker loops. The combined CR-MHD models using a single CRs fluid approach reveal that CRs can be dynamically important for shaping the magnetic field, structuring the interstellar medium, and driving galactic outflows and winds. However, the relative importance of CRs compared to other processes in galaxies as well as which details of CR physics are the crucial ones to resolve, include and simulate are still strongly debated. Currently, there is converging consensus that CRs -- in particular the dynamically more relevant CR protons -- should be coupled to the thermal gas and treated in a time-dependent manner. The missing physical aspects in this approach like a spectral CR description and the inclusion of multiple species and secondaries are on the agenda for the near future.

\begin{acknowledgements} 

MH acknowledges support of the (Polish) National Science Centre through the grant 2015/19/ST9/02959. PG acknowledges funding from the European Research Council under ERC-CoG grant CRAGSMAN-646955.\\

\end{acknowledgements}

\end{document}